\documentclass[11pt,a4,twoside,longbibliography]{article}
\usepackage[dvips,a4paper,left=2.0cm,right=2.0cm,top=2.0cm,bottom=2.0cm,headsep=1em]{geometry}
\usepackage{fancyhdr}
\usepackage{titlesec}
\usepackage[latin1]{inputenc}
\usepackage{amsmath,amsfonts,amssymb,array,amsthm,mathrsfs}
\usepackage{rotating}
\usepackage[font={small}]{caption}
\usepackage{lmodern,slantsc}
\usepackage{multirow}
\usepackage{chngcntr}
\usepackage{placeins}
\usepackage{caption}
\usepackage{enumitem}
\usepackage{cite}
\usepackage[sort&compress,numbers]{natbib}
\def\bibsep{6pt}
\def\bibfont{\small}
\usepackage[breaklinks, bookmarks=true,
colorlinks=true, linkcolor=blue, citecolor=blue, urlcolor=blue, filecolor=blue,
pdfborder={0 0 0}, pdfpagelabels]{hyperref}
\newcommand{\PRLsep}{\noindent\makebox[\linewidth]
{\resizebox{0.3333\linewidth}{1pt}{$\bullet$}}\bigskip}
\usepackage{tikz}
\newcommand{\ExternalLink}{%
\tikz[x=1.2ex, y=1.2ex, baseline=-0.05ex]{%
\begin{scope}[x=1ex, y=1ex]
\clip (-0.1,-0.1) --++ (-0, 1.2) --++ (0.6, 0) --++ (0, -0.6) --++
(0.6, 0) --++ (0, -1); \path[draw, line width = 0.7, rounded
corners=0.5] (0,0) rectangle (1,1);
\end{scope}
\path[draw, line width = 0.7] (0.5, 0.5) -- (1, 1); \path[draw,
line width = 0.7] (0.6, 1) -- (1, 1) -- (1, 0.6);}}
\let\oldabstract\abstract
\let\oldendabstract\endabstract
\makeatletter
\renewenvironment{abstract}
{%
{\list{}{\addtolength{\leftmargin}{-0.2em} 
\listparindent 1.5em%
\itemindent\listparindent%
\rightmargin\leftmargin%
\parsep\z@\@plus\p@}%
\item\relax}%
{\endlist}%
\oldabstract}
{\oldendabstract}
\makeatother
\def\vx{\mathbf x}

\def\vr{\mathbf r}

\def\v0{\boldsymbol{0}}

\def\vlambda{\boldsymbol{\lambda}}

\def\vphi{\boldsymbol{\phi}}

\def\L{\langle}
\def\R{\rangle}
\newlength{\FigureHeight}
\newlength{\FigureHeightHalf}

\numberwithin{equation}{section}
\titleformat{\section}
{\large\bfseries}{\thetitle.}{0.5em}{}
\titleformat{\subsection}
{\normalfont\bfseries}{\thetitle.}{0.5em}{}
\titleformat{\subsubsection}
{\normalfont\itshape}{\normalfont\thetitle.}{0.5em}{}
\newcommand{\symfootnote}[1]{%
\let\oldthefootnote=\thefootnote%
\stepcounter{mpfootnote}%
\addtocounter{footnote}{-1}%
\renewcommand{\thefootnote}{\fnsymbol{mpfootnote}}%
\footnote{#1}%
\let\thefootnote=\oldthefootnote%
}
\begin{document}

\title{A closer look at predicting turbulence statistics of arbitrary moments when based on a non-modelled symmetry approach}

\author{Michael Frewer$\,^1$\thanks{Email address for correspondence:
frewer.science@gmail.com}$\:\,$ \& George Khujadze$\,^2$ \\[1.0em]
\small $^1$ Heidelberg, Germany\\
\small $^2$ Chair of Fluid Mechanics, Universit\"at Siegen, 57068
Siegen, Germany}
\date{{\small\today}}
\clearpage \maketitle \thispagestyle{empty}

\setcounter{mpfootnote}{6}

\begin{abstract}

\noindent
A recent Letter by Oberlack {\it et al.} [\href{https://doi.org/10.1103/PhysRevLett.128.024502}{Phys.~Rev.~Lett.~{\bf 128},~024502~(2022)}]
claims to have derived new symmetry-induced solutions of the non-modelled statistical Navier-Stokes equations of turbulent\linebreak[4] channel flow. A high accuracy match to DNS data for all streamwise moments up to order 6
is presented, both in the region of the channel-center and in the inertial sublayer close to the wall. Here we will show that the findings and conclusions in that study are highly misleading, as they give the impression that a significant breakthrough in turbulence research has been achieved. But,\linebreak[4] unfortunately, this is not the case. Besides trivial and misleading aspects, we will demonstrate that even basic turbulence-relevant correlations as the Reynolds-stress cannot be fitted to data using the proposed symmetry-induced scaling laws. $\!$The Lie-group symmetry method as used by Oberlack {\it et~al.}\linebreak[4] cannot bypass the closure problem
of turbulence. It is just another assumption-based method that requires modelling and is not, as claimed, a first-principle method that leads directly to solutions.\symfootnote{A concluding review
to a Reply by Oberlack {\it et al.} is provided in \href{https://arxiv.org/abs/2302.04853}{arXiv:2302.04853}.}
Next to PRL, two more papers by Oberlack {\it et al.} are called out for correction or a retraction.

\vspace{0.5em}\noindent

\end{abstract}

\newpage

\thispagestyle{empty}
\noindent
{\bf\large Contents}

\vspace{2em}\noindent
{\bf 1$\;\;\!$ A brief synopsis of what will be shown and analyzed\hfill\hyperref[Sec1]{1}}

\vspace{1em}\noindent
{\bf 2$\;\;\!$ The trivial and misleading aspects of \href{https://doi.org/10.1103/PhysRevLett.128.024502}{{\small PRL$\;$(2022)}}\hfill\hyperref[Sec2]{3}}

\vspace{1em}\noindent
{\bf 3$\;\;\!$ The incorrect aspects of \href{https://doi.org/10.1103/PhysRevLett.128.024502}{{\small PRL$\;$(2022)}} about symmetries and solutions\hfill\hyperref[Sec3]{6}}

\vspace{1em}\noindent
{\bf 4$\;\;\!$ A physically consistent symmetry-based modelling approach\hfill\hyperref[Sec4]{8}}
\\[0.25em]
\phantom{4$\;\;$} 4.1$\;\;$ Implementation of a realizable statistical scaling invariance in channel center\hfill$\!$. . . . . . .\hfill\hyperref[Sec41]{13}
\\[0.25em]
\phantom{4$\;\;$} 4.2$\;\;$ Reiterating the causality principle for statistical symmetries\hfill\hspace{-0.4mm}. . . . . . . . . . . . . . . . .\hfill\hyperref[Sec42]{14}

\vspace{1em}\noindent
{\bf A$\;\;\!$ On the usefulness of a Lie-group symmetry analysis in turbulence\hfill\hyperref[SecA]{16}}

\vspace{1em}\noindent
{\bf B$\;\;\!$ Parameter values for the figures shown\hfill\hyperref[SecB]{18}}
\\[0.25em]
\phantom{B$\;\;$} B.1$\;\;$ Indicator functions to Fig.$\,$2\hfill$\!$. . . . . . . . . . . . . . . . . . . . . . . . . . . . . . . . . . . .\hfill\hyperref[SecB1]{20}
\\[0.25em]
\phantom{B$\;\;$} B.2$\;\;$ The unnatural scaling in \hypersetup{urlcolor=black}\href{https://doi.org/10.1103/PhysRevLett.128.024502}{{\small PRL$\;$(2022)}}\hypersetup{urlcolor=blue}
for the Reynolds-stress $\overline{u_1^2}$ in the log-layer\hfill. . . . .\hfill\hyperref[SecB2]{21}

\vspace{1em}\noindent
{\bf C$\;\;\!$ On the DNS accuracy and uncertainty in the fluctuation moments\hfill\hyperref[SecC]{22}}
\\[0.25em]
\phantom{C$\;\;$} C.1$\;\;$ An experiment quantifying the uncertainty of $R_3$\hfill. . . . . . . . . . . . . . . . . . . . . . .\hfill\hyperref[SecC1]{24}

\vspace{1em}\noindent
{\bf D$\;\;\!$ System of moment equations for statistically stationary flow in channel center\hfill\hyperref[SecD]{26}}

\vspace{1em}\noindent
{\bf E$\;\;\!\!$ The non-existent universal log-law in \href{https://doi.org/10.1017/jfm.2014.98}{{\small JFM$\;$(2014)}}\hfill\hyperref[SecE]{27}}

\vspace{1em}\noindent
{\bf F$\;\;\!$ The importance for a correction, or if necessary, for a retraction of \href{https://doi.org/10.1017/S0022112000002408}{{\small JFM$\;$(2001)}}\hfill\hyperref[SecF]{29}}
\\[0.25em]
\phantom{F$\;\;$} F.1$\;\;$ Introduction\hfill$\!$ . . . . . . . . . . . . . . . . . . . . . . . . . . . . . . . . . . . . . . . . . . . . .\hfill\hyperref[SecF1]{29}
\\[0.25em]
\phantom{F$\;\;$} F.2$\;\;$ Proof that result (3.16) in \hypersetup{urlcolor=black}\href{https://doi.org/10.1017/S0022112000002408}{{\small JFM$\;$(2001)}}\hypersetup{urlcolor=blue}
is in error and cannot be reproduced\hfill$\,$. . . . . . .\hfill\hyperref[SecF2]{30}
\\[0.25em]
\phantom{F$\;\;$} F.3$\;\;$ Proof that result (3.17) in \hypersetup{urlcolor=black}\href{https://doi.org/10.1017/S0022112000002408}{{\small JFM$\;$(2001)}}\hypersetup{urlcolor=blue}
is in error\hfill\hspace{-1mm}. . . . . . . . . . . . . . . . . . . . . . . .\hfill\hyperref[SecF3]{32}
\\[0.25em]
\phantom{F$\;\;$} F.4$\;\;$ Further points for correction in \hypersetup{urlcolor=black}\href{https://doi.org/10.1017/S0022112000002408}{{\small JFM$\;$(2001)}}\hypersetup{urlcolor=blue}$\;\;$
\hfill\hspace{-3.5mm}. . . . . . . . . . . . . . . . . . . . . . . . . . .\hfill\hyperref[SecF4]{33}
\\[0.25em]
\phantom{F$\;\;$} F.5$\;\;$ List of all false statements made in \hypersetup{urlcolor=black}\href{https://doi.org/10.1017/S0022112000002408}{{\small JFM$\;$(2001)}}\hypersetup{urlcolor=blue}$\;\;$
\hfill\hspace{-4.3mm}. . . . . . . . . . . . . . . . . . . . . . . . .\hfill\hyperref[SecF5]{34}



\newpage
\pagenumbering{arabic}\setcounter{page}{1}

\label{p1}
\section{A brief synopsis of what will be shown and analyzed\label{Sec1}}

The present review will unequivocally demonstrate that no breakthrough or any progress in \cite{Oberlack22} has been made
{\it ``to overcome the problem related to the closure problem of turbulence"} [p.$\,$5]. Instead, the paper \cite{Oberlack22} is seriously misleading and makes false promises.

In a nutshell, the key results of \cite{Oberlack22} are: Invariant turbulent scaling laws are derived that can be near-perfectly matched to DNS data, where these laws are claimed to be true solutions of the statistical Navier-Stokes equations obtained from first principles alone, i.e., without making any prior assumptions or using any modelling approaches. Then, two so-called `symmetries' are highlighted and claimed to be a measure of the intermittent and non-Gaussian behaviour of turbulence. But, as we will show and demonstrate in the upcoming sections, these results are either trivial or misleading or even not~true~at~all.

First of all, the scaling behaviour shown in Figs.$\,$1-3 in \cite{Oberlack22} is clearly dominated by the mean flow~and says nothing about the fluctuating part. Due to the overwhelming dominance of the mean flow in the two flow regions considered, the $n$-th full-field correlation in the streamwise direction $\overline{U_1^n}$ is close to equal to the $n$-th power of the mean field itself, $\overline{U}_{\!1}^{\,n}$, i.e., the near-perfect fitting results all reveal only the trivial aspect that $\overline{U_1^n}\approx\overline{U}_{\!1}^{\,n}$. This is also independently confirmed when employing an {\it unbiased} indicator function, proving that the scaling of the full-field correlations is simply driven by the scaling of the mean flow, regardless of the specific scaling (power or log law) used.\footnote{Important to note here is that the so-called indicator-plot in \cite{Oberlack22}, Fig.$\,$1(b), is based on a {\it biased} indicator function, and not on an unbiased one, as is usually the case. This is because the indicator function Eq.$\,$(18) in \cite{Oberlack22} is not free of parameters. It contains the modelling parameters $B_n$ and therefore modifies the data in bias towards the function being used. Therefore, it does not have the same significance as an unbiased or model-free indicator function. All this is discussed in detail in Sec.$\,$\ref{SecB1}.}

Therefore, it is not surprising that only a few parameters are needed to fit the data up to high order. In fact, with the symmetry method used in \cite{Oberlack22}, it is easy to show that even fewer parameters are already sufficient to achieve the same quality in all fitting results. Thus, once the lowest order $n=1$ (the mean field) is matched, all full-field correlations of higher order $n>1$ will follow suit, because in the end it is effectively just the exponentiated mean velocity that is being matched. Said differently, once the mean field has been fitted to data, the scaling behaviour of the full-field correlations as shown in Figs.$\,$1-3 in \cite{Oberlack22} is trivial~to~predict.

The perfect scaling of the full-field moments shown in \cite{Oberlack22} is therefore highly misleading, since it does not lead to a correct prediction of turbulent scaling beyond the mean flow. What we see in~\cite{Oberlack22}
is just the coarse structure of the mean flow and its various powers, which ultimately constitutes only a trivial result once the scaling of the fitted mean field is known, while all the interesting and relevant fine structure of turbulence cannot be predicted. Hence, no conclusions about non-Gaussianity and intermittency of turbulence can be drawn from the scaling of the full-field moments, as wrongly done in~\cite{Oberlack22}, simply because
the mean flow in the full-field moments just dominates, thus turning these moments into coarse quantities not sensitive enough to resolve the necessary fine structure of turbulence in order to approach the issues of non-Gaussianity and intermittency.

It is well-known that in statistical physics the finding insight or the gain in knowledge lies in the study of the fluctuation correlations and not in the full-field correlations which include and involve the mean field. Already more than a century ago, Osborne Reynolds was among the first to have realized\linebreak[4] that it is necessary to separate the fluctuating part from the mean flow when trying to analyze and\linebreak[4]
understand the intricate nature of turbulence.\footnote{In \hspace{-0.15mm}particular, \hspace{-0.15mm}when \hspace{-0.15mm}studying \hspace{-0.0mm}the \hspace{-0.15mm}intermittent \hspace{-0.15mm}behaviour \hspace{-0.15mm}of \hspace{-0.0mm}turbulence \hspace{-0.0mm}driven \hspace{-0.15mm}by \hspace{-0.15mm}a \hspace{-0.15mm}strong \hspace{-0.15mm}mean \hspace{-0.15mm}flow, \hspace{-0.15mm}at \hspace{-0.15mm}which \hspace{-0.15mm}\cite{Oberlack22} \hspace{-0.15mm}aims \hspace{-0.15mm}at,\linebreak[4] it is a clear mistake not to separate the fluctuations from the mean flow. This is because intermittency, such as observed in fully developed turbulent channel flow, is encoded in the fine structure of the (higher order) fluctuation correlations, and not in the coarse full-field moments, which are essentially insensitive to the detection of intermittent~effects.}
Notwithstanding this, the authors of \cite{Oberlack22} {\it ``depart here from the usual approach"} [p.$\,$2] and instead of analyzing and showing the scaling behaviour of the turbulence-relevant fluctuation correlations, they only show the trivial scaling relation of the full-field correlations. By doing so, they claim to have finally found the prediction rule for the scaling of turbulence in two regions of turbulent channel flow, the inertial and center region.

The findings asserted in \cite{Oberlack22} are not restricted to pure channel flow. From the first author's group\linebreak[4] it\hfill is\hfill claimed\hfill for\hfill example\hfill that\hfill they\hfill have\hfill also\hfill found\hfill
the\hfill passive\hfill temperature\hfill scaling\hfill in\hfill turbulent\hfill jet

\newgeometry{left=2.0cm,right=2.0cm,top=2.0cm,bottom=1.75cm,headsep=1em}

\noindent
flow~\cite{Sadeghi21}, but here again only the misleading near-perfect scaling of the full-field correlations is shown. The relevant scaling of the fluctuation correlations of the velocity field, however, was shown elsewhere in an earlier publication, but had to be corrected in \cite{Sadeghi20}, with the result that their new scaling theory is no better and partially even worse than the well-known classical one.

Now, let's turn to and focus on the authors' general argument, which is supposed to justify their approach of analyzing only the full-field correlations. Their argument is that once the scaling relation of the full-field correlations  is determined, the corresponding scaling relation of the fluctuation correlations can be retrieved and derived from it by just using the unique Reynolds decomposition,\footnote{This argument is not explicitly mentioned in \cite{Oberlack22}, but can be read elsewhere, e.g., most recently in \cite{Sadeghi21} on p.$\,$8: {\it ``there is a unique relation between the instantaneous and the fluctuation approach, which simply allows us to change from one approach to the other."}} which, for the streamwise components as studied in \cite{Oberlack22}, is given iteratively by
\begin{equation}
\overline{u_1^n}=\overline{U_1^n}-\overline{U}_{\!1}^{\,n}-\sum_{k=1}^{n-2}\binom{n}{k}\,\overline{U}_{\!1}^{\,k}\,\overline{u_1^{n-k}}\, .
\label{220122:1231}
\end{equation}
This argument that the scaling of the fluctuation correlations $R_{1,n}:=\overline{u_1^n}$ can be derived from the scaling relation of the full-field correlations $H_{1,n}:=\overline{U_1^n}$ when only using the above
mapping \eqref{220122:1231}, is theoretically valid with perfect (infinitely accurate) data and with knowledge of the {\it exact} full-field scaling, but practically this argument is neither sound nor valid.

The reason for this is that such a process is very ill-conditioned: When comparing or fitting data to the full moments and then use \eqref{220122:1231} to determine the corresponding fit of the fluctuation moments, it automatically develops the error of suppressing or removing the fluctuations from the comparison due to the large difference in magnitudes, and this error amplifies the stronger the mean flow is, exactly as it is the case for the two particular flow regions considered in \cite{Oberlack22}.\footnote{\label{fn4}Important to note here is that when using relation \eqref{220122:1231} to only extract the fluctuation data from the full-field data itself, then such a process is\hspace{0.05mm} {\it not}\hspace{0.35mm} ill-conditioned, as analyzed and discussed in Sec.$\,$\ref{SecC}. The problem with using \eqref{220122:1231} is when\linebreak[4] curve-fitting the moments, i.e., when trying to determine the scaling exponents from the data, where a best fit in the full-field moments then does not automatically lead to a best fit in the fluctuation moments when \eqref{220122:1231} is used. In other words, to achieve a best fit for the fluctuation moments, they have to be fitted separately from the full-field moments.} Therefore to avoid such ill-conditioned comparisons, the usual practice in statistical physics is to consider and work with correlations where the background or the bulk motion has been subtracted, i.e., if possible, one should always co-move with the phenomenon being considered.

However, this ill-conditioning problem, namely that a best-fit for $\overline{U_1^n}$ does not automatically lead to a best-fit for $\overline{u_1^n}$ when using relation \eqref{220122:1231}, is not the only problem that the publication \cite{Oberlack22} faces.\linebreak[4] A~major key problem is that the scaling laws for the relevant {\it fluctuation} correlations $\overline{u_1^n}$, induced by\linebreak[4] Eqs.$\,$(15-20) through~\eqref{220122:1231}, cannot be fitted to the provided data at all. For the center region of the channel, this failure already occurs for the streamwise Reynolds stress $(n=2)$, while for the inertial or so-called log-region, the mismatch starts at the next higher order $(n=3)$, and worsens for higher orders in both regions. To note is that although the streamwise Reynolds stress in the log-region can be more or less fitted
in \cite{Oberlack22} by Eq.$\,$(16), it fits very unnaturally, since a quantity which only varies by order $1$ has to be fitted with parameters of order 100.

The reason for this failure is that the proposed scalings in \cite{Oberlack22} are not solutions to the statistical Navier-Stokes equations, as incorrectly claimed, simply because these scalings are based on two non-physical invariances (Eqs.$\,$8-9) that violate the classical principle of cause and effect between the fluctuations and the mean fields \cite{Frewer15,Frewer16.1,Frewer17,Sadeghi20,Frewer21.1,Frewer21.4}. This violation is suppressed and therefore not visible when analyzing the symmetry-based scaling of the full-field correlations, but becomes measurable and clearly visible when analyzing the corresponding fluctuation correlations.

To note is that this symmetry-based scaling failure is not specific to channel flow. It can also be clearly seen in other flow configurations when turbulence-relevant moments are explicitly matched to data~\cite{Sadeghi20,Frewer16.2,Frewer16.3,Frewer14.1}, and simply stems from the fact that this failure is methodologically rooted in the symmetry-based scaling approach as developed by Oberlack {\it et al.} over the last two decades.

Finally, another issue that \cite{Oberlack22} faces is that of {\it ``identical gradients"} [p.$\,$4], as emphasized and shown in Fig.$\,$3, which is not a pure turbulence phenomenon. It is a trivial universal aspect that can also be obtained by simply exponentiating the laminar profile, as we will now show and discuss in the next section as our first point to clarify and put the results of \cite{Oberlack22} into the right perspective.

\restoregeometry

\newpage

\label{p3}
\section{The trivial and misleading aspects of \cite{Oberlack22}\label{Sec2}}

\begin{figure}[t!]
\centering
\begin{minipage}[c]{.49\linewidth}
\includegraphics[width=.90\textwidth]{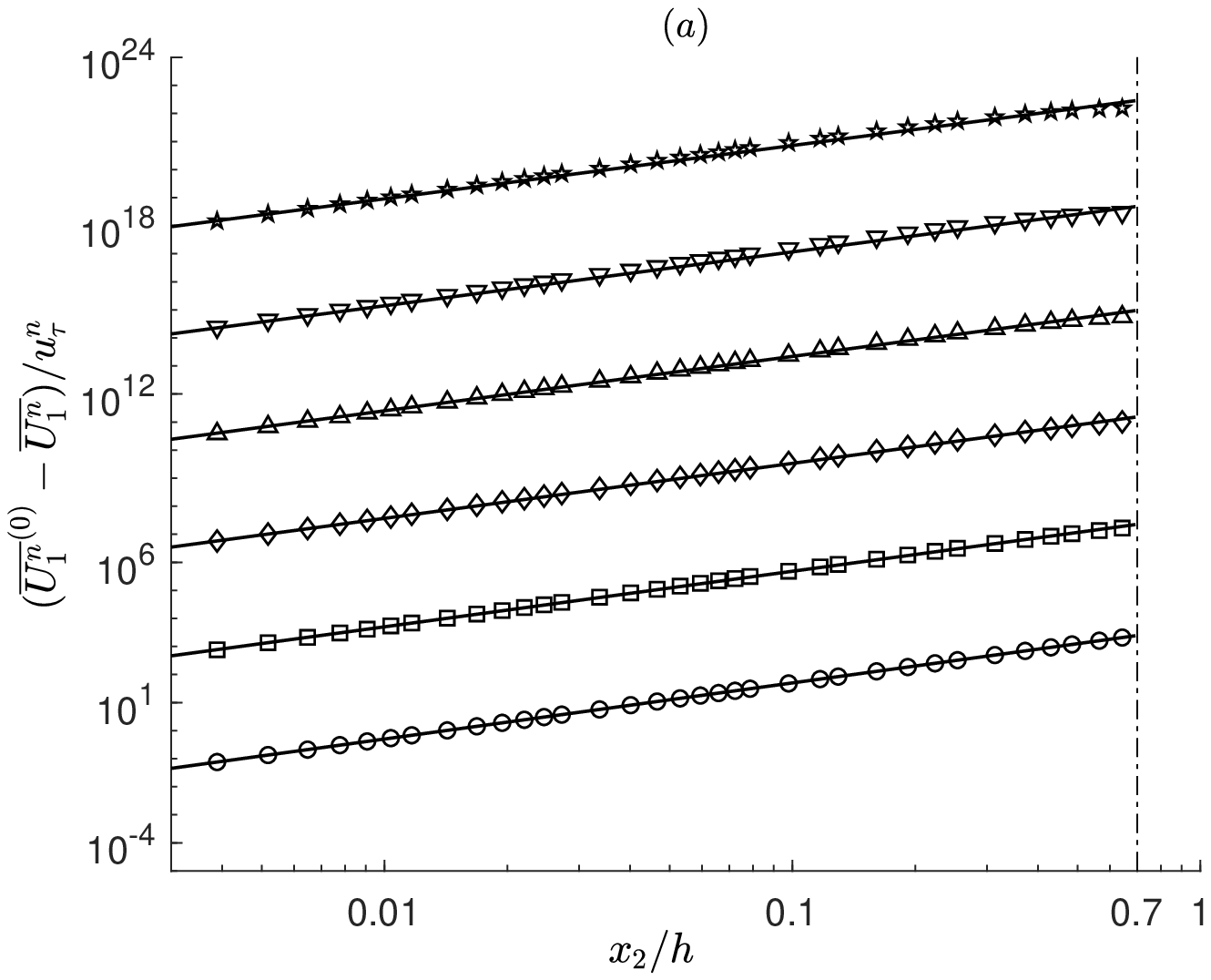}
\end{minipage}
\hfill
\begin{minipage}[c]{.49\linewidth}
\includegraphics[width=.90\textwidth]{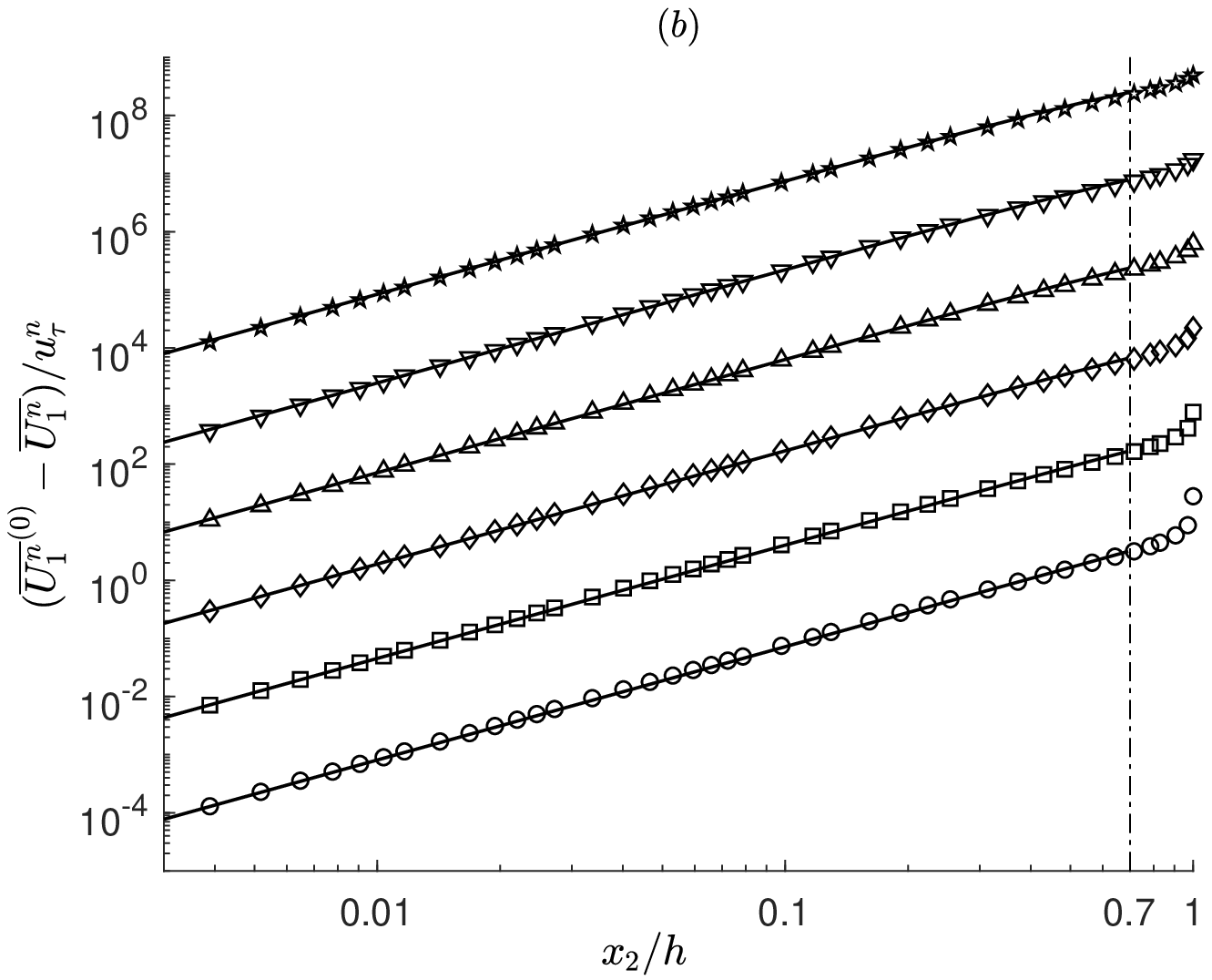}
\end{minipage}
\caption{
(a) Corresponding laminar case of Fig.$\,$3 in \cite{Oberlack22}, where $\overline{U_1^n}=\overline{U}_{\!1}^{\,n}$ holds trivially. The symbols display the different powers of the laminar quadratic profile (in deficit form) up to order $n=6$ (from bottom to top); the solid lines are the best-fit using the turbulent scaling law Eqs.$\,$(19-20) from \cite{Oberlack22}.
(b) Comparison plot to Fig.$\,$3~in~\cite{Oberlack22}. The symbols display the turbulent full-field correlations $\overline{U_1^n}$ in deficit form, exactly as in \cite{Oberlack22}, while the solid lines show the $n$-th power of the mean velocity field $\overline{U}_{\!1}^{\,n}$ (in deficit form) up to order $n=6$ (from bottom to top). Here, only the mean velocity ($n=1$, bottom curve) was fitted according to the turbulent scaling law Eqs.$\,$(19-20). Therefore, instead of four $(\sigma_1,\sigma_2,\alpha^\prime,\beta^\prime)$, only a reduced set of two parameters $(\sigma_1,C_1^\prime)$ is needed to fit all full-field moments as shown. The data for the full-field moments was extracted and digitized from the plots in \cite{Oberlack22} using the software \cite{Rohatgi21} and then compared to the database provided, with no visible difference between them. The values of the fitted parameters used in both figures are listed in Appendix \ref{SecB}.
}
\label{Fig1}
\vspace{-0.75em}
\end{figure}

The matching result in channel-center, shown in Fig.$\,$3 in \cite{Oberlack22}, is a trivial result, since the same profile structure of parallel lines and the same fitting result can also be obtained with the corresponding laminar flow. Fig.$\,$\hyperref[Fig1]{1(a)} shows the corresponding laminar case for the same flow parameters as used in Fig.$\,$3 in~\cite{Oberlack22}: The set of symbols displays the powers of the laminar quadratic profile (in deficit form) up to order $n=6$ (from bottom to top). The solid lines show the best fit to these profiles, based on the turbulent scaling law Eqs.$\,$(19-20) in \cite{Oberlack22}, with the same number of fitting parameters ($\sigma_1,\sigma_2,\alpha^\prime,\beta^\prime)$ and the same fitting-range in $x_2/h$ as in Fig.$\,$3 for turbulent flow.
From this simple finding, three conclusions can already be drawn that refute \cite{Oberlack22}:

\begin{itemize}[leftmargin=2.5em]

\vspace{-1em}
\item[(1)]
The structure of parallel lines as emphasized and shown in \cite{Oberlack22} is not a feature of turbulence.
It's a universal result, irrespective of whether the flow is turbulent or not. Important is only to represent the flow fields in deficit form and to normalize appropriately, in order to get such a structure of uniformly distributed parallel lines in a log-log-plot. The origin of this structure lies trivially in the leading quadratic term when asymptotically expanding around the global maximum in channel-center.\footnote{\label{fn5}In fact, for any function $f(x)$ whose limit at zero is finite and non-zero, $\lim_{x\to 0}f(x)=f_0\neq 0$, the set of functions in deficit form $\lvert f^n-f_0^n\rvert$ will appear parallel when shown in a log-log-plot in the range $0<x<\epsilon<1$, where the parallel alignment is of course to be understood in an asymptotic sense: the smaller the range $\epsilon$, the better this alignment, and the higher the exponentiation $n$ gets, the smaller will be the range $\epsilon$. The $n$-independent and thus constant slope $m$ for this set of parallel lines is given by $m=\lim_{x\to 0}\log(\lvert f(x)^n-f_0^n\rvert)/\log(x)$, for all $n>1$.} When reducing the upper fitting range of $x_2/h$ from 0.7 already to slightly lower values, the two scaling exponents rapidly converge to the same trivial value $\sigma_1=\sigma_2=2$, both for the laminar as well as for the turbulent case. The claim in \cite{Oberlack22} that~$\sigma_1\approx\sigma_2$ is the result of  anomalous scaling due to {\it ``intermittent behavior"}~[p.$\,$5] is therefore neither plausible nor given.\footnote{The key word ``intermittency" is used several times in \cite{Oberlack22}, although it is not clear from the context what exactly~is meant by it. Nowhere in the text a definition or explanation is given to what type or kind of intermittency they refer~to.\linebreak[4] For example, to attribute in fully developed turbulence the globally constant scaling invariance Eq.$\,$(9) as a measure of intermittency is not plausible. As is well known, intermittency breaks global scaling in fully developed turbulence and instead gets replaced by local and non-constant scaling, which can be modelled by using e.g. the idea of multi-fractals~\cite{Frisch95}.} Further illustrative examples can be taken from \cite{Frewer22}.

\vspace{-0.7em}
\item[(2)]
Since the result shown in Fig.$\,$3 in \cite{Oberlack22} is universal and since Eqs.$\,$(19-20) scales in the laminar case just as well as in the turbulent case, it makes the derivation of Eqs.$\,$(19-20) based on new symmetries superfluous and unnecessary. In other words, the so-called ``statistical symmetries"~in~\cite{Oberlack22} are not essential to predict this universal scaling behaviour. The classical scaling symmetries of the Navier-Stokes equation for the mean profile ($n=1$) are already fully sufficient to scale all higher {\it full-field} moments in the streamwise direction, due to the overwhelming dominance of the mean field in those type of correlations.

\vspace{-0.7em}
\item[(3)]
In a hierarchy of profiles obtained by exponentiation of a base-profile, it is clear that the base-profile, i.e. the lowest-order profile~$n=1$, dictates the scaling behaviour of all higher-order profiles~$n>1$.
In laminar flow this base profile is the laminar field, while in turbulence it's the mean field, which will be shown and proven next.

\end{itemize}

\vspace{-0.5em}\noindent
The channel flow considered in \cite{Oberlack22} is driven by a strong mean velocity field $\overline{U}_{\!1}$ that significantly exceeds the magnitude of the fluctuations in the system. Hence, when matching to a full-field correlation $\overline{U_1^n}$ that is fully aligned in the streamwise direction ($i=1$), as done in \cite{Oberlack22}, where $\overline{U_1^n}$ then contains $n$~times the mean velocity field $\overline{U}_{\!1}$, the matching of $\overline{U_1^n}$ simply degenerates to a matching of the mean field itself to the power $n$, i.e. to $\overline{U}_{\!1}^{\,n}$. This is shown in Fig.$\,$\hyperref[Fig1]{1(b)}, where the symbols show the full-field correlations $\overline{U_1^n}$ ($n\geq 1)$ from \cite{Oberlack22}, while the solid lines show the mean field and its powers $\overline{U}_{\!1}^{\,n}$ up to order $n=6$ (from bottom to top), where, and this is important, only the lowest order $n=1$, that is, only the mean field $\overline{U}_{\!1}$ has been fitted.

Fig.$\,$\hyperref[Fig1]{1(b)}, which corresponds to Fig.$\,$3 in \cite{Oberlack22}, shows the universal result of parallel profiles in channel-center, but now for the turbulent case. Instead of two scaling
$(\sigma_1,\sigma_2)$ and two shifting $(\alpha^\prime,\beta^\prime)$ para\-meters, as used in \cite{Oberlack22}, only a single scaling exponent $\sigma_1$ with a single shift parameter
$C_1^\prime$ was needed in Fig.$\,$\hyperref[Fig1]{1(b)} to obtain for all higher-order moments qualitatively the same fitting result as shown in~\cite{Oberlack22}. A not surprising result for such a simple structure of parallel lines, where the scaling of the whole set is already encoded in the lowest-order moment ($n=1$), i.e. in the mean velocity profile~$\overline{U}_{\!1}$, and with all higher-order moments then automatically given by its $n$-th power $\overline{U}_{\!1}^{\,n}$.\footnote{It is trivially clear that the more parameters for the fitting process are used, the better the result. A natural extension of the parameter-set is achieved by exploiting the fact that the full-field moment equations Eq.$\,$(4) in \cite{Oberlack22} are linear, thus allowing to superpose symmetries and its induced invariant functions. For example, in channel-center this will yield the higher orders in the asymptotic expansion $\sum_{n\geq 2} c_n (x_2/h)^{\sigma_n}$ of the mean field $\overline{U}_{\!1}$ around the global maximum, where a reduction to a polynomial power series ($\sigma_n=n$) already appears to be sufficient. The fit of the higher-order full-field moments $H_{1,n\geq 2}$ are then again automatically obtained by taking the corresponding powers $\overline{U}_{\!1}^{\,n}$ of the fitted mean field.}

Important to note here is that since the two scaling exponents in channel center are defined in \cite{Oberlack22} through the group parameters as $\sigma_1=(1-a_{St}/a_{Sx})+a_{Ss}/a_{Sx}$
and $\sigma_2=2\cdot(1-a_{St}/a_{Sx})+a_{Ss}/a_{Sx}$,~the successful reduction to only $\sigma_1$, as shown in Fig.$\,$\hyperref[Fig1]{1(b)}, implies that $a_{Ss}$ is redundant, thus proving that their so-called ``intermittency~symmetry"~(Eq.$\,$9) is not needed or not of any relevance to scale the {\it full}\hspace{0.5mm}-field correlations. The two classical scaling symmetries (Eq.$\,$6), with the group parameters\linebreak[4] $a_{St}$ and $a_{Sx}$, already prove to be fully sufficient. In the next sections, we even demonstrate that this ``intermittency symmetry" Eq.$\,$(9) has to be discarded, because only when excluding it, the fluctuation correlations can be matched to the data, otherwise not. This just proves again that this scaling invariance Eq.$\,$(9) is nonphysical, as has been proven already several times before \cite{Frewer14.1,Frewer15,Frewer16.1,Frewer16.2,Frewer17,Frewer18.2,Sadeghi20,Frewer21.1,Frewer21.4}.
The~same is true also for the nonphysical translation invariance Eq.$\,$(8).

\begin{figure}[t!]
\centering
\begin{minipage}[c]{.49\linewidth}
\includegraphics[width=.90\textwidth]{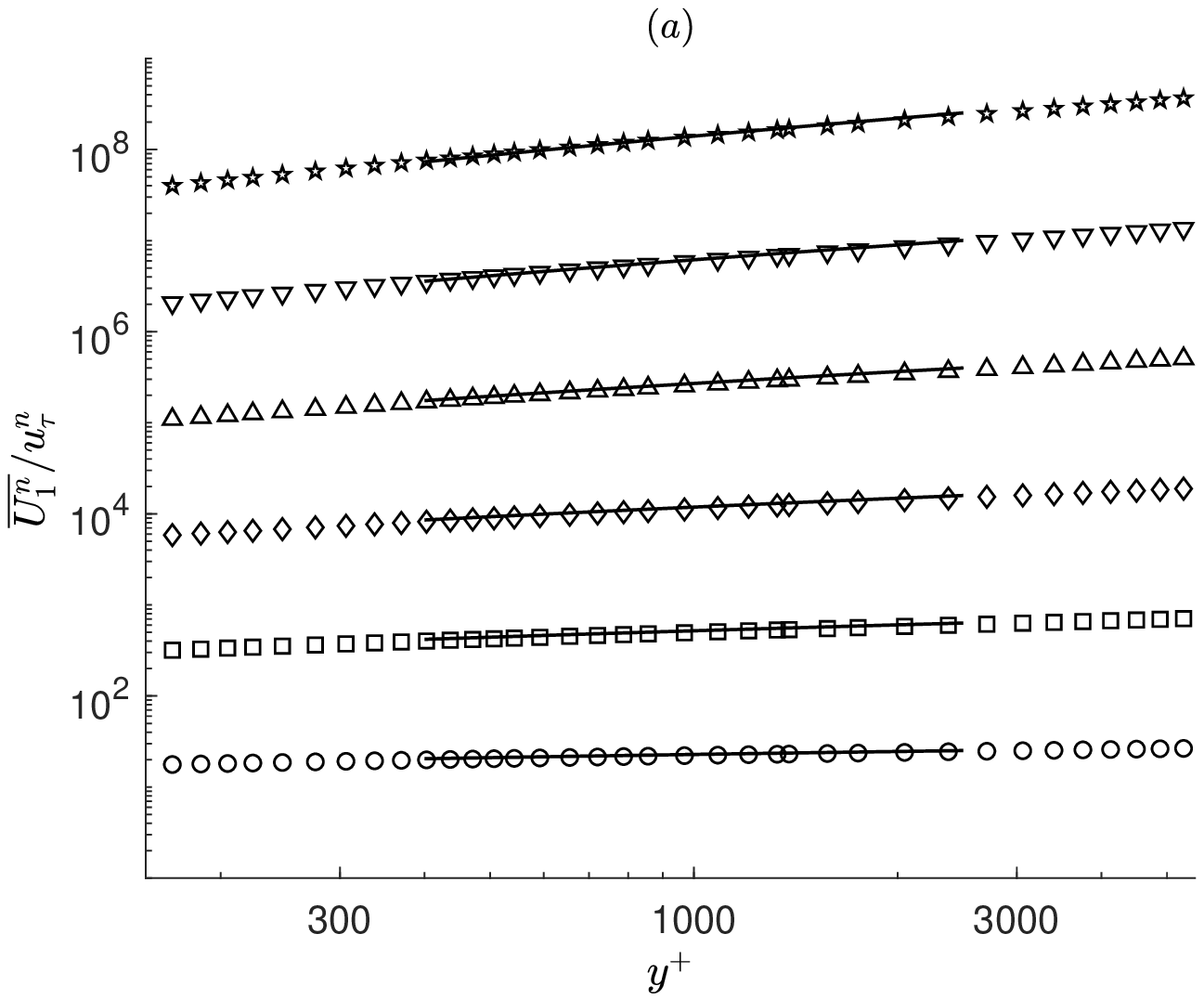}
\end{minipage}
\hfill
\begin{minipage}[c]{.49\linewidth}
\includegraphics[width=.90\textwidth]{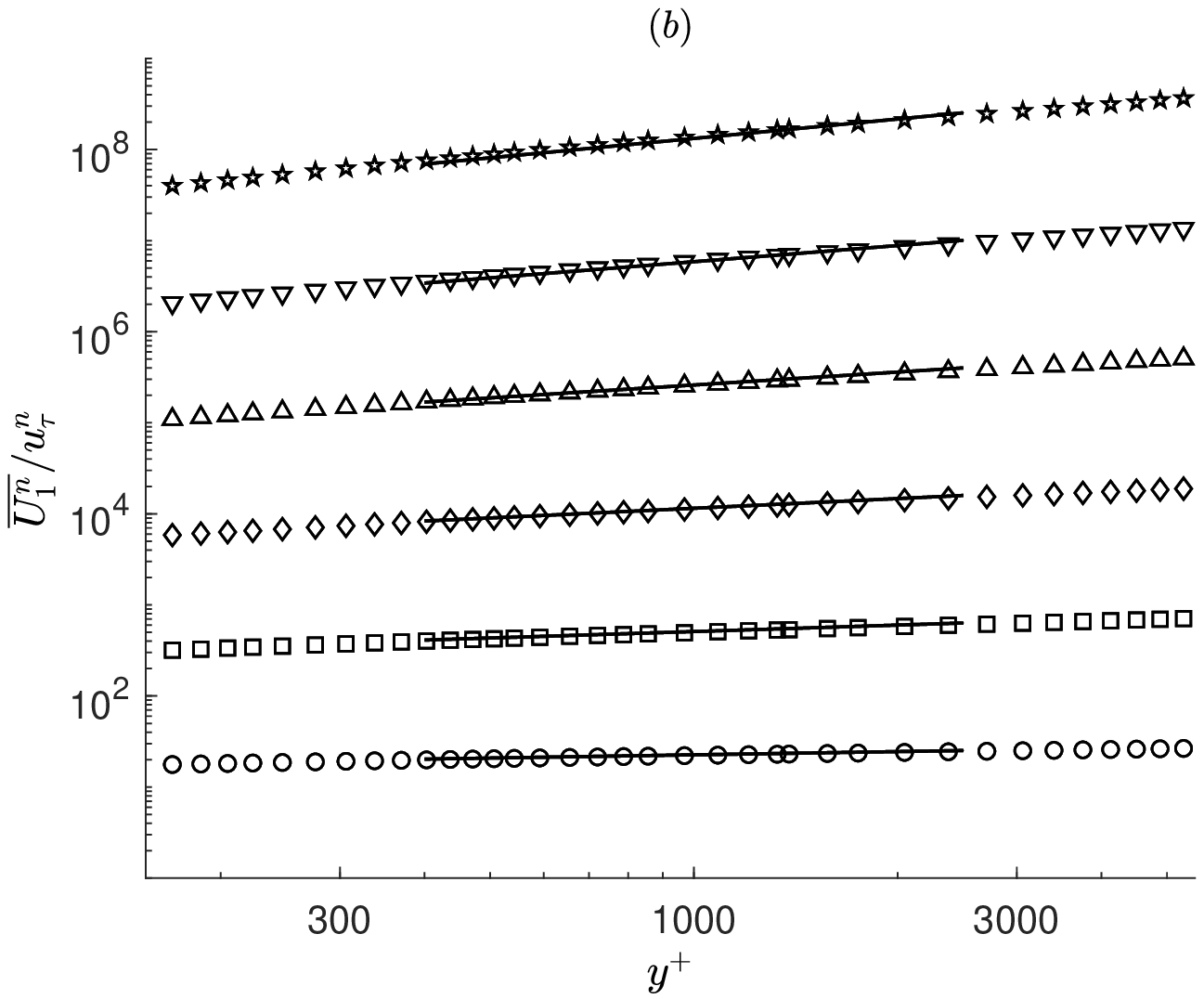}
\end{minipage}
\caption{
Comparison plots to Fig.$\,$1(a) in \cite{Oberlack22}. The symbols display the turbulent full-field correlations $\overline{U_1^n}$ in\linebreak[4] the inertial region, exactly as in \cite{Oberlack22}, while the solid lines show $\overline{U}_{\!1}^{\,n}$, the $n$-th power of the mean velocity field up to order $n=6$ (from bottom to top). In both plots only the mean velocity itself ($n=1$, bottom curve) was fitted: in (a) according to the turbulent scaling law Eq.$\,$(15) and in (b) according to Eqs.$\,$(16-17), but for $n\geq 1$, which also is a valid symmetry-induced scaling law for the mean velocity field in the inertial range (see text). That is, in the inertial sublayer the mean velocity field is once modelled as a log-law (a) and once as a power law (b). In both cases, however, since only a single field is fitted again, a reduced set of only two parameters is needed again to fit all full-field moments as shown: instead of the seven parameters $(\kappa,B,\omega,\alpha,\beta,\tilde{\alpha},\tilde{\beta})$ as used in \cite{Oberlack22}, only $(\kappa,B)$ were used in (a), and $(\omega,C_1)$ in (b). Again, the data for the full-field moments was extracted and digitized from the plots in \cite{Oberlack22} using the software \cite{Rohatgi21} and then compared to the database provided, with no visible difference between them. The values of the fitted parameters used in both figures are\linebreak[4] listed in Appendix~\ref{SecB}. The corresponding indicator functions are shown and discussed in Sec.$\,$\ref{SecB1}.
}
\label{Fig2}
\vspace{-1em}
\end{figure}

Now, due to the overwhelming dominance of the mean velocity field considered, a structure of parallel profiles is not necessary to successfully apply this reduced full-field matching process also to other flow regions of the channel. Fig.$\,$\hyperref[Fig2]{2}, which corresponds to Fig.$\,$1(a) in \cite{Oberlack22}, shows the best-fit of the mean profile $\overline{U}_{\!1}$ for the inertial sublayer, once modelled as a log-law (Fig.$\,$\hyperref[Fig2]{2(a)}) and once modelled as a power law (Fig.$\,$\hyperref[Fig2]{2(b)}). The matching to all higher-order moments $H_{1,n>1}=\overline{U_1^n}$ (from bottom to top) is then automatically obtained again by just taking the $n$-th power of the fitted mean field~$\overline{U}_{\!1}$. Instead\linebreak[4] of seven fitting parameters $(\kappa,B,\omega,\alpha,\beta,\tilde{\alpha},\tilde{\beta})$, as used in \cite{Oberlack22}, only two parameters were used here again to qualitatively obtain the same result again as shown in \cite{Oberlack22}: For the log-law modelled version in Fig.$\,$\hyperref[Fig2]{2(a)}, these two parameters are $(\kappa,B)$, while for the power-law modelled version in Fig.$\,$\hyperref[Fig2]{2(b)}, they are $(\omega,C_1)$.

Important to note here is that the used log-law for the mean velocity field Eq.$\,$(15) in \cite{Oberlack22} is an empirical assumption made by Oberlack {\it et al.}, and not a necessary condition that results from theory.
A properly performed Lie-group invariance analysis does not restrict to a particular scaling function~in

\newgeometry{left=2.0cm,right=2.0cm,top=2.0cm,bottom=1.20cm,headsep=1em}

\noindent
the inertial sublayer, as incorrectly claimed in \cite{Oberlack22} --- for a detailed account, see e.g.~\cite{Frewer14.2}. Also the claim on p.$\,$3 that the wall-friction velocity is symmetry breaking, with the effect that $a_{Sx}-a_{St}+a_{Ss}=0$,\linebreak[4] is a plain assumption and not dictated by Lie-group theory. It's a reverse-engineered ansatz just to get the log-law as a ``symmetry-induced" result.\footnote{It should be clear that we do not exclude the log-scaling as the more appropriate scaling for the mean velocity profile in the inertial sublayer. Here we only want to stress that when applying the method of Lie-groups in turbulence correctly, it does not result into a specific scaling function as claimed in \cite{Oberlack22}. In fact, an infinite number of different scaling functions can be obtained when performing a symmetry analysis without any prior assumptions (see Sec.$\,$\ref{SecA}), simply due to that the statistical equations of turbulence are unclosed, making their admitted set of symmetries thus
also unclosed\linebreak[4] --- see e.g. \cite{Frewer14.2,Frewer16.3,Frewer18.2} for a full and complete Lie-group invariance analysis to certain turbulent flow configurations.}
In fact, a correct analysis shows again that
when putting the translation group parameter for the mean flow to zero, the wall-boundary condition $\overline{U}_{\!1}=0$ can be well implemented in Eq.$\,$(11) without breaking the invariance transformation in the mean velocity field.\linebreak[4] This is simply because with $\overline{U}_{\!1}^*=e^{a_{Sx}-a_{St}+a_{Ss}}\overline{U}_{\!1}$ in its unshifted form now, the wall-condition $\overline{U}_{\!1}=0$ gets~mapped to $\overline{U}_{\!1}^*=0$, i.e.~invariantly to the same boundary condition again, irrespective of whether $a_{Sx}-a_{St}+a_{Ss}$ is zero or not. Hence, the wall-friction velocity $u_\tau$ need not to be symmetry breaking as claimed, with the effect therefore that the Lie-group method also allows a power scaling for the mean velocity profile in the inertial sublayer, $\overline{U}_{\!1}^{\,+}=C_1(y^+)^\omega$
(Eq.$\,$(10) for $a_{Sx}-a_{St}+a_{Ss}\neq0$, $a_{1{\scriptscriptstyle \{1\}}}^H=0$),\linebreak[4]
as successfully shown in Fig.$\,$\hyperref[Fig2]{2(b)}.

Finally it is to note that for all turbulence-based results shown in this section through Fig.$\,$\hyperref[Fig1]{1(b)} and Figs.$\,$\hyperref[Fig2]{2(a)-(b)}, only classical symmetries were used:
the two inviscid scaling symmetries (Eq.$\,$6~in~\cite{Oberlack22}), the translation symmetry in wall-normal direction (not shown but used in \cite{Oberlack22}), and the Galilean boost symmetry in the streamwise direction (not used in \cite{Oberlack22}), where the latter symmetry serves as the correct substitute for the nonphysical field-translation symmetry Eq.$\,$(8). Since for the full-field correlations $\overline{U_1^n}$ up to order $n=6$ in both channel center and inertial sublayer only the mean velocity field~$(n=1)$\linebreak[4] needs to be fitted, we naturally obtained a 2-parameter scaling model as shown in the figures above. Using Occam's razor, a 2-parameter model is then to be preferred over the unnecessary multi-parameter scaling models Eqs.$\,$(15-17) and Eqs.$\,$(19-20), thus making all the new ``statistical symmetries" in \cite{Oberlack22} dispensable, let alone the fact they are not even symmetries but only nonphysical equivalences that violate the classical principle of cause and effect and therefore should be discarded in the first place, as was already shown and proven before in \cite{Frewer14.1,Frewer15,Frewer16.1,Frewer16.2,Frewer17,Frewer18.2,Sadeghi20,Frewer21.1,Frewer21.4}, and here once again shown by the next section's Fig.$\,$\hyperref[Fig3]{3},
which will be discussed next.\vspace{-0.5em}\pagebreak[4]

\restoregeometry
\newgeometry{left=2.0cm,right=2.0cm,top=2.0cm,bottom=1.2cm,headsep=1em}

\label{p6}
\section{The incorrect aspects of \cite{Oberlack22} about symmetries and solutions\label{Sec3}}

Simply put, the Lie-group symmetry method cannot bypass the closure problem of turbulence, since it only shifts the closure problem of equations to a closure problem of symmetries. All results obtained by this method are assumption-based results and not first-principle results as claimed by Oberlack~{\it et~al}. When applying this method to turbulence, major assumptions are made that are not visible to the reader who is not familiar with Lie-groups (see the discussion in Appendix \ref{SecA} for further details). It~is an {\it ad~hoc} method too, not free of assumptions. In the end, the Lie-group method in turbulence is effectively no different to the classical invariance method of von K\'arm\'an and Prandtl. Not {\it solutions} to the statistical Navier-Stokes equations are produced, but only possible candidate functions are obtained that either are useful or not to describe turbulence data to a certain degree of accuracy. However, in contrast to the classical invariance method, the Lie-group symmetry method in turbulence is also able to produce a large set of nonphysical invariances, which cannot be realized by simulation or experiment. This leads us to the ``solutions" Eqs.$\,$(15-20) in \cite{Oberlack22}.

Although the symmetry analysis in \cite{Oberlack22} is set up for arbitrarily orientated correlations in turbulent shear flow, the analysis always only leads to isotropic scaling results, and this irrespective of the length scale considered. That is, also when generalizing the three streamwise scaling symmetries Eqs.$\,$(6-7,9) to symmetries of the governing equation (Eq.$\,$4) for arbitrary directions $i$, their scaling exponents will always remain independent of the coordinate direction. Consequently, all differently orientated correlations will scale identically in \cite{Oberlack22}, which obviously is a non-realistic result in a highly anisotropic flow as channel flow. This circumstance could be one of the reasons why in \cite{Oberlack22} only the scaling of the full-field correlations in the streamwise direction is shown. Because for all other moments when they are not fully aligned in the streamwise direction, the symmetry-induced scaling of \cite{Oberlack22} fails. The less mean fields a full-field correlation carries, the worse the failure, in particular for all pure fluctuation correlations $R_{i,n}=\overline{u_i^n}$, where the mean field in all components has been subtracted, the failure is most pronounced. For example in the spanwise direction $i=3$, the symmetry-induced results are such that they even lead to a contradictive scaling \cite{Zimmerman21}, a finding not cited and shared with the reader in \cite{Oberlack22}.

\begin{figure}[t!]
\centering
\begin{minipage}[c]{.49\linewidth}
\includegraphics[width=.90\textwidth]{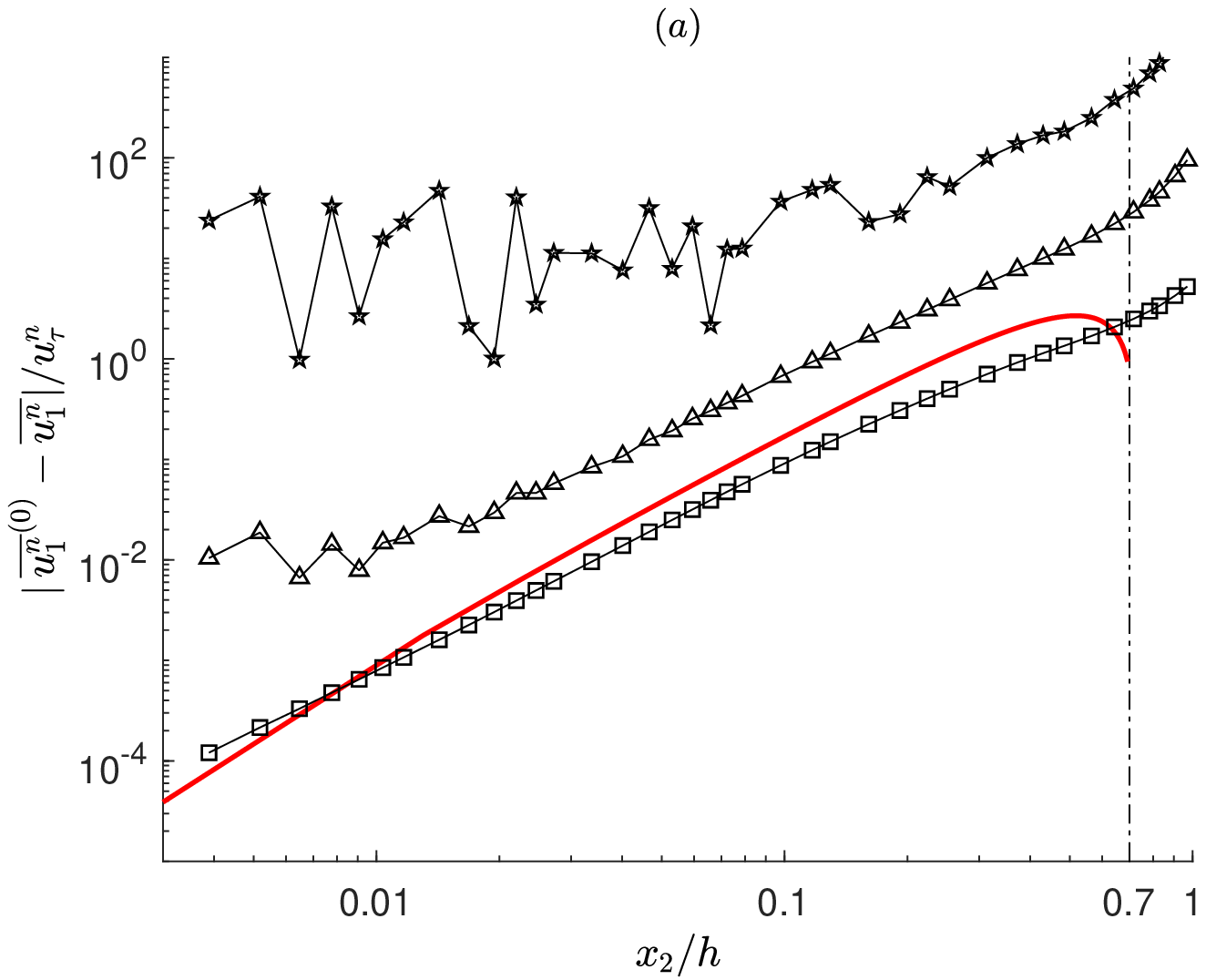}
\end{minipage}
\hfill
\begin{minipage}[c]{.49\linewidth}
\includegraphics[width=.90\textwidth]{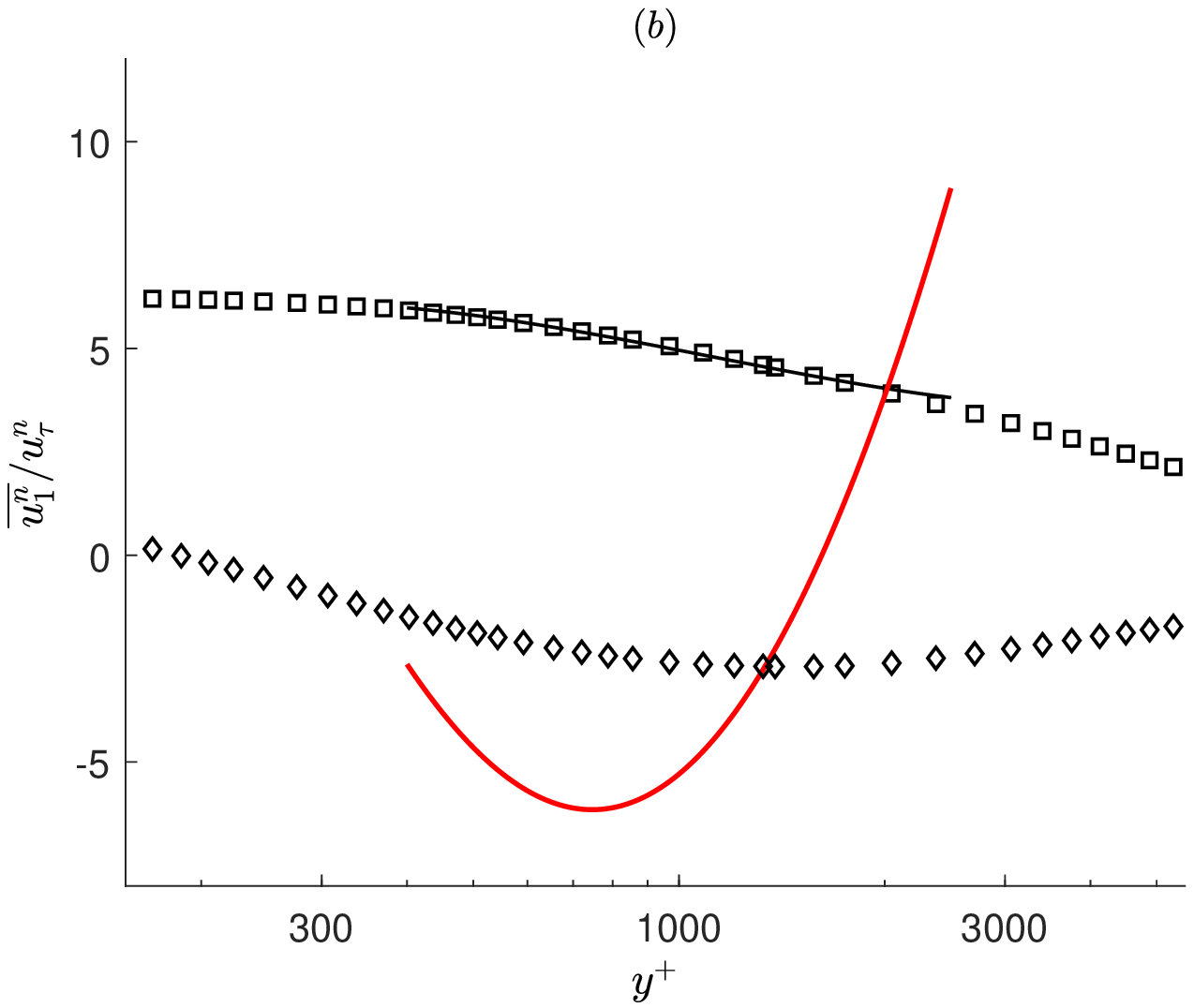}
\end{minipage}
\caption{
(a) Comparison plot to Fig.$\,$3 in \cite{Oberlack22}. The symbols display the fluctuation correlations $\overline{u_1^n}$ (in deficit form) of the three even orders $n=2,4,6$ (from bottom to top).\protect\footnotemark $\,$
The data points were obtained by extracting the full-field moments from the corresponding figure in \cite{Oberlack22} and then transforming them into the representation of the fluctuation moments using the unique relationship \eqref{220122:1231}. Hence, the discrete points shown (connected with a thin line) correspond exactly to those points shown in Fig.$\,$3 in~\cite{Oberlack22} when transforming from the full-field to the fluctuation correlations. Obviously, the higher moments are not yet fully converged, indicating that the performed DNS in \cite{Oberlack22} either needs to be run longer or spatially resolved better, or both. Therefore, only the lowest fluctuation moment $n=2$ was fitted according to its associated scaling law Eq.$\,$(19) in~\cite{Oberlack22} (when transformed accordingly). The result is shown by the red solid line. A clear failure, proving that the scaling law cannot be matched to turbulence-relevant data already at the lowest level $n=2$, and thus further proving that it cannot be a solution to the statistical Navier-Stokes equations as claimed in \cite{Oberlack22}.
\\
(b) Comparison plot to Fig.$\,$1(a) in \cite{Oberlack22}. The symbols display the fluctuation correlations $\overline{u_1^n}$ in wall-units for $n=2$ (squares) and $n=3$ (diamonds). The data points were obtained again as described in (a). The black solid line shows the best-fit to the second moment and the red line to the third moment, each according to their~associated scaling law Eq.$\,$(16) in~\cite{Oberlack22} (when transformed accordingly). While the second moment can be well fitted, though very unnaturally, over the whole range $400<y^+<2500$ specified by \cite{Oberlack22}, the fitting of the third moment fails. Thus, the scaling law Eq.$\,$(16) for the inertial sublayer too cannot be a solution.
\\
It should be noted that the failure in (a) and (b) is quite robust, i.e. even when fitting on significantly smaller ranges, it results in the same ill-shaped profiles as shown above by the red line. Ironically, the failure stays invariant under scaling of the fitting domain. For all details on the fitting process in both (a) and (b), see Appendix \ref{SecB}.
The extracted and then transformed data used in the above plots were carefully compared with the directly transformed data from the provided database of \cite{Oberlack22}, with no difference in the end result visible.
}
\label{Fig3}
\vspace{-1em}
\end{figure}

Another reason why in \cite{Oberlack22} only the full-field and not the turbulence-relevant fluctuation correlations are shown, is that when changing the representation in the streamwise direction from the full-field
$H_{1,n}=\overline{U_1^n}$ to the fluctuating field $R_{1,n}=\overline{u_1^n}$, they cannot be matched to the data anymore. For the channel-center region this failure already starts at the level of the Reynolds stresses ($n=2$), as shown in Fig.$\,$\hyperref[Fig3]{3(a)}, while for the region of the inertial sublayer it starts at the next higher order ($n=3$), as\linebreak[4] shown in Fig.$\,$\hyperref[Fig3]{3(b)}
--- although $n=2$ can be fitted in this layer, it fits very unnaturally (see Sec.$\,$\ref{SecB2}).

To generate Fig.$\,$\hyperref[Fig3]{3}, we proceeded as follows: the unique mapping \eqref{220122:1231} was used to equivalently rewrite the scaling laws Eqs.$\,$(15-17) and Eqs.$\,$(19-20) in \cite{Oberlack22} from the full-field to the fluctuation corre\-lations. Since that unique mapping only acts on the left-hand side of those scaling laws, while their right-hand side, i.e. the modelling side, remains untouched, the correct and consistent approach would therefore be to take for the fluctuation correlations the fitted values of the full-field correlations. But, when using those fitted parameter values from \cite{Oberlack22}, it inevitably leads to an even larger discrepancy\textsuperscript{$\hookrightarrow$ \hspace{-0.5mm}\ref{fn4}} than what is already shown in Fig.$\,$\hyperref[Fig3]{3}. Hence, we had to use a {\it new} best set of values for the fitting parameters, and what is shown here is the ultimate best-fit that results from comparing even different norm functions such as the Euclidean vs. the infinite norm. Thus, no better fit can be found in this ill-conditioned setting --- even when considering significantly smaller fitting domains.

This failure in Fig.$\,$\hyperref[Fig3]{3} clearly shows that neither Eqs.$\,$(15-17) nor Eqs.$\,$(19-20) is a set of solutions of the moment equations, as misleadingly claimed in \cite{Oberlack22}. Because, if they would be true solutions of the statistical Navier-Stokes equations, then a match of the full-field correlation $H_{1,n}$ will result to a corresponding match of the associated fluctuation correlation $R_{1,n}$ and vice versa,
simply because this change is based on an analytical one-to-one map \eqref{220122:1231} that maps solutions to solutions and therefore does not change the solution manifold (up to numerical stability issues associated with this map).\footnote{The numerically most stable result will of course be obtained by applying the unique mapping \eqref{220122:1231} only after all small-magnitude correlations (fluctuation correlations) have been fitted. In other words, when fitting to data, the mapping $H_{1,n}\rightarrow R_{1,n}$ is ill-conditioned,\textsuperscript{$\hookrightarrow$ \hspace{-0.5mm}\ref{fn4}} while the inverse mapping $R_{1,n}\rightarrow H_{1,n}$ is not.}
But, for Eqs.$\,$(15-17) and Eqs.$\,$(19-20) the solution manifold does change when applying this unique mapping, as can be clearly seen, respectively, from the mismatch in Fig.$\,$\hyperref[Fig3]{3(a)} and Fig.$\,$\hyperref[Fig3]{3(b)} already in the lowest moments and for which no better fits can be found,\footnote{Note, since no better fits can be found, the failure in Fig.$\,$\hyperref[Fig3]{3} is structural and no longer a numerical stability~issue.} hence, they cannot be solutions.

\restoregeometry
\newgeometry{left=2.0cm,right=2.0cm,top=2.0cm,bottom=1.35cm,headsep=1em}

\footnotetext{To avoid a too busy plot, we only show the situation of the even moments. The odd moments $(n>1)$ are no better.}

This failure, that all scaling laws in \cite{Oberlack22} are not solutions of the statistical Navier-Stokes equations is not a singular case due to channel flow. It has also been shown and proven for zero-pressure-gradient (ZPG) turbulent boundary layer flow \cite{Frewer14.1} (see Sec.$\,$5). Therein it is even shown that the fitting process improves by several orders of magnitude and becomes well-defined again for all higher-order moments as soon as one excludes from the modelling process all nonphysical symmetries, which without reason or proof were declared in \cite{Oberlack22} to be a measure of ``non-Gaussianity and intermittency''. The same experience was also made in turbulent jet flow \cite{Sadeghi20}, but subsequently obscured again in \cite{Sadeghi21} by showing the misleading ``nicely collapse" of the full-field correlations again.

Also, in the case of turbulent channel flow with wall-transpiration, the claim in \cite{Avsarkisov14} of a universal log-law for different transpiration rates  is not true. The authors provided a correction \cite{Avsarkisov21}, but it's still flawed. The correction is even worse than the original version, as DNS data that was previously considered consistent has now been changed into inconsistent data. For a varying transpiration rate at fixed Reynolds number, the mean velocity profiles do not collapse, as incorrectly claimed and shown in Fig.$\,$9(a) in \cite{Avsarkisov14}, and again in the falsely corrected Fig.$\,$1(a) of \cite{Avsarkisov21}$\,$\footnote{Since Fig.$\,$1(a) in \cite{Avsarkisov21} is not a correction to Fig.$\,$9(a) in \cite{Avsarkisov14}, and since the latter figure cannot be reproduced 
from the original DNS data provided, the pressing question is: How did the authors manage to create this perfect Fig.$\,$9(a) in their original article \cite{Avsarkisov14}$\,$? The same question also applies to the construction of Fig.$\,$7 in \cite{Sadeghi18}, another figure of Oberlack {\it et al.}\linebreak[4] that cannot be reproduced from the data provided. 
In this case, too, exactly the same problem: Although a correction is given by \cite{Sadeghi20}, it does not provide a correction to the original Fig.$\,$7. Instead, a completely new figure based on new results is presented and therefore unrelated to the original one. Hence, also in this case, the same central question: How and with what tools did the authors manage to do the original Fig.$\,$7 in \cite{Sadeghi18}$\,$?} --- not even approximately do

\restoregeometry

\noindent
they collapse --- thus invalidating their universal log-law. All details and the proof to our claims can be found in Appendix \ref{SecE}.
Therein we also provide a simple physical explanation for why the profiles cannot collapse.

Returning to \cite{Oberlack22}, finally note that Fig.$\,$\hyperref[Fig3]{3(a)} also reveals the separate problem that the higher-order moments generated by the DNS in \cite{Oberlack22} cannot be relied upon to show the correct scaling. The noise in the correlations for $n>2$ is still significant and can only be attributed to the fact that the performed DNS in \cite{Oberlack22} is not well and sufficiently resolved, either in time or in space, or both.

\label{p8}
\section{A physically consistent symmetry-based modelling approach\label{Sec4}}

In this section we will demonstrate how the scaling behaviour of the turbulence-relevant fluctuation correlations $R_{1,n}=\overline{u_1^n}$ can be modelled by using a physically consistent Lie-group-based symmetry approach.
However, as we will show at the end of this section, a consistent modelling approach does not necessarily mean or guarantee that the results obtained thereby also match the data. Like any other analytical method when used in turbulence research, the Lie-group-based symmetry approach is also just another trial-and-error approach.

To keep the approach simple and concise, we will not generalize the idea here to arbitrary correlation orders, but will limit ourselves only up to order $n=3$. The generalization to higher orders $n\geq4$ and even to correlations other than the streamwise direction is then more complicated, but not impossible. It is the idea of this section we want to bring across, not the technical details. To~note is that the following analysis is a symmetry-based {\it modelling} approach and not a symmetry-based {\it solution} approach to turbulence. All invariant functions determined below are only possible but not guaranteed solutions of the statistical Navier-Stokes equations, in complete contrast, of course, to what is claimed in \cite{Oberlack22}, where all invariant functions are declared as ``solutions".

The first modelling assumption we make is to take the inviscid Euler equations ($\nu=0$) as the governing equations
\begin{equation}
\partial_t U_i+ U_k\partial_k U_i=-\partial_i P,\qquad \partial_k U_k=0,
\label{220125:2229}
\end{equation}
with its two classical scaling symmetries (as in \cite{Oberlack22})
\begin{equation}
\left.
\begin{aligned}
\mathsf{S_1}\!:&\quad t^*=e^{a_1}t,\quad x_i^*=x_i,\quad U_i^*=e^{-a_1}U_i,\quad P^*=e^{-2a_1}P,
\\[0.5em]
\mathsf{S_2}\!:&\quad t^*=t,\quad x_i^*=e^{a_2}x_i,\quad U_i^*=e^{a_2}U_i,\quad P^*=e^{2a_2}P,
\end{aligned}
~~~\right\}
\label{220125:2134}
\end{equation}
as the basis to model the scaling in the inertial sublayer and the center-region of turbulent channel flow. In \cite{Oberlack22}, it is incorrectly claimed that the viscous Navier-Stokes equations {\it ``in the
limiting case~$\nu\to 0$ possess two scaling symmetries, i.e., in principle exactly like the Euler equation"} [p.$\,$2]. The intention of this statement is clear: it should give the impression that for high-Reynolds-number turbulent flows the validity of the two symmetries \eqref{220125:2134} is not a model assumption, but a fact, analytically derived by Oberlack on {\it ``the basis of a multiscale expansion in the correlation space $\vr$ in Ref.$\,$[10]"}~[p.$\,$2]. However, the multiscale analysis in that cited paper, and presented in more detail e.g. in \cite{Oberlack02}, is misleading and not justified because the scales were just naively separated in a reverse-engineered form instead of performing a correct singular asymptotic analysis as it should have been done.

Fact is, the symmetries \eqref{220125:2134} are not admitted by the Navier-Stokes equations, not even to leading order within a symmetry perturbation for small viscosities $\nu\to 0$. In other words, the Navier-Stokes equations cannot be converted into the Euler equations by just employing any exact or approximate Lie-group invariance transformation.

To model the scaling of the mean velocity field in channel flow, we also consider the Galilean boost symmetry in the streamwise direction admitted by \eqref{220125:2229}
\begin{equation}
\mathsf{G}\!:\quad t^*=t, \quad x_i^*=x_i+c_1\delta_{i1}\cdot t,\quad U_i^*=U_i+c_1\delta_{i1},\quad P^*=P.
\label{220126:0950}
\end{equation}
The next step is to consider the 1-point moment equations of the governing model-equations~\eqref{220125:2229},

\newpage

\noindent
where we truncate the infinite hierarchy of moments at order $n=3$:
\begin{equation}
\left.
\begin{aligned}
&\partial_t \overline{U}_{\!i}+\partial_k\overline{U_iU_k}=-\partial_i \overline{P},\qquad \partial_k \overline{U}_{\!k}=0,
\\[0.5em]
&\partial_t\overline{U_iU_j}+\partial_k\overline{U_iU_jU_k}=-\overline{U_i\partial_jP}-\overline{U_j\partial_iP},
\\[0.5em]
&\partial_t\overline{U_iU_jU_k}+\partial_l\overline{U_iU_jU_kU_l}=-\overline{U_iU_j\partial_kP}-\overline{U_jU_k\partial_iP}-\overline{U_kU_i\partial_jP}.
\end{aligned}
~~~\right\}
\label{220126:0948}
\end{equation}
Now, since we are interested in the scaling of the streamwise fluctuation correlations $\overline{u_1^n}$ of statistically stationary turbulent parallel shear flow, we can now Reynolds decompose the above symmetries and the moment equations  via $U_i=\overline{U}_{\!i}+u_i$. Since the wall-normal coordinate $y=x_2$, and, for the pressure, also the streamwise coordinate $x_1$, are the only statistically relevant coordinates in this flow configuration, the above moment equations \eqref{220126:0948} for the dynamics of the streamwise fields, i.e., the equations for $\partial_t\overline{U}_{\!1}$, $\partial_t\overline{u_1^2}$ and $\partial_t\overline{u_1^3}$, then respectively reduce to
\begin{equation}
\left.
\begin{aligned}
&\partial_{2\,}\overline{u_1u_2}=-\partial_1\overline{P},
\\[0.5em]
&\partial_2\!\left(\overline{u_1^2u_2}+2\overline{U}_{\!1\,}\overline{u_1u_2}\right)
=-2\left(\overline{U}_{\!1}\partial_1\overline{P}+\overline{u_1\partial_1p}\right)\!,
\\[0.5em]
&\partial_2\!\left(\overline{u_1^3u_2}+3\overline{U}_{\!1\,}\overline{u_1^2u_2}+3\overline{U}_{\!1\,}^{2\,}\overline{u_1u_2}\right)
=-3\left(\overline{U}_{\!1}^2\partial_1\overline{P}+\overline{u_1^2}\partial_1\overline{P}+2\overline{U}_{\!1}\overline{u_1\partial_1p}+\overline{u_1^2\partial_1p}\right)\!,
\end{aligned}
~~~\right\}
\label{220126:1019}
\end{equation}
which, when inserting each lower-order equation into the next higher-order one, can be simplified to
\begin{equation}
\left.
\begin{aligned}
&\partial_{2\,}\overline{u_1u_2}=-\partial_1\overline{P},
\\[0.5em]
&\partial_{2\,}\overline{u_1^2u_2}+2\,\overline{u_1u_2}\,\partial_2\overline{U}_{\!1\,}
=-2\,\overline{u_1\partial_1p},
\\[0.5em]
&\partial_{2\,}\overline{u_1^3u_2}+3\,\overline{u_1^2u_2}\,\partial_2\overline{U}_{\!1\,}
=-3\left(\,\overline{u_1^2\partial_1p}-\overline{u_1^2}\,\partial_2\overline{u_1u_2}\right)\!,
\end{aligned}
~~~\right\}
\label{230112:0939}
\end{equation}
where the mean streamwise pressure gradient is a constant: $-\partial_1\overline{P}=K>0$, where $K=u_\tau^2/h$, with $u_\tau$ the friction velocity and $h$ the half-height of the channel.
Next to the symmetries \eqref{220125:2134} and \eqref{220126:0950}, which in Reynolds-decomposed form read
\begin{equation}
\left.
\begin{aligned}
\mathsf{S_1}\!:&\quad t^*=e^{a_1}t,\quad x_i^*=x_i,\quad \overline{U}_{\!i}^*=e^{-a_1}\overline{U}_{\!i},\quad \overline{P}^*=e^{-2a_1}\overline{P},\quad u_i^*=e^{-a_1}u_i,\quad p^*=e^{-2a_1}p,
\\[0.5em]
\mathsf{S_2}\!:&\quad t^*=t,\quad x_i^*=e^{a_2}x_i,\quad \overline{U}_{\!i}^*=e^{a_2}\overline{U}_{\!i},\quad \overline{P}^*=e^{2a_2}\overline{P},\quad u_i^*=e^{a_2}u_i,\quad p^*=e^{2a_2}p,
\\[0.5em]
\mathsf{G}\!:&\quad t^*=t, \quad x_i^*=x_i+c_1\delta_{i1}\cdot t,\quad \overline{U}_{\!i}^*=\overline{U}_{\!i}+c_1\delta_{i1},\quad \overline{P}^*=\overline{P},\quad u_i^*=u_i,\quad p^*=p,
\end{aligned}
~~~\right\}
\label{220126:2221}
\end{equation}
and which indeed, as can be easily verified, are symmetries also of the reduced moment equations~\eqref{230112:0939}, we further include in our analysis the following statistical invariance of \eqref{220126:0948} (up to the order of the fields appearing therein)
\begin{equation}
\left.
\begin{aligned}
\!\!\!\!\!\mathsf{T}\!:&\quad t^*=t, \quad x_i^*=x_i,\quad \overline{U}_{\!i}^*=\overline{U}_{\!i},\quad \overline{P}^*=P,
\\[0.5em]
&\quad\overline{u_iu_j}^{\,*}=\overline{u_iu_j}+c_2\delta_{i1}\delta_{j1},\quad \overline{u_iu_ju_k}^{\,*}=\overline{u_iu_ju_k}+c_3\delta_{i1}\delta_{j1}\delta_{k1},
\quad \overline{u_i\partial_jp}^{\,*}=\overline{u_i\partial_jp},
\\[0.5em]
&\quad\overline{u_iu_ju_ku_l}^{\,*}=\overline{u_iu_ju_ku_l}+c_2T_{ijkl}+c_4\delta_{i1}\delta_{j1}\delta_{k1}\delta_{l1},
\quad \overline{u_iu_j\partial_kp}^{\,*}=\overline{u_iu_j\partial_kp},
\\[0.5em]
&\quad\text{with}\;\; T_{ijkl}=\overline{u_iu_j}\delta_{k1}\delta_{l1}+\overline{u_ju_k}\delta_{l1}\delta_{i1}+\overline{u_ku_l}\delta_{i1}\delta_{j1}+\overline{u_lu_i}\delta_{j1}\delta_{k1}
+\overline{u_iu_k}\delta_{j1}\delta_{l1}+\overline{u_ju_l}\delta_{i1}\delta_{k1}.
\end{aligned}
~\right\}
\label{220126:2319}
\end{equation}
This transformation will model the statistical streamwise-translational invariance in the governing equations \eqref{220125:2229}. By this we first mean that \eqref{220126:2319} is an invariance admitted by the subsystem of\linebreak[4] moment equations \eqref{220126:0948} in the streamwise direction, which, when Reynolds-decomposed, are given by~\eqref{230112:0939} and, hence, is a valid invariance of stationary turbulent channel flow in the streamwise direction we are interested in here --- for a more general translation invariance, a more general ansatz than \eqref{220126:2319} must of course be sought.
Secondly, since this invariance is not induced by an underlying symmetry of the governing equations \eqref{220125:2229} and further, since these equations are statistically unclosed, the invariance~\eqref{220126:2319} thus only corresponds to a statistical {\it equivalence}\hspace{0.3mm}\footnote{\label{fn13}An equivalence transformation acts in a weaker sense than a symmetry transformation. While a symmetry maps solutions to solutions of the same equation, an equivalence only maps equations to different equations of the same class. However, if the equations mapped by an equivalence differ only in the existing parameters and not in some unclosed functions, then an equivalence transformation can also map a solution of one equation to a corresponding solution of another equation, but otherwise not. For more details, see e.g. Sec.$\,$2 in \cite{Frewer14.1} and the last two footnotes in \cite{Frewer18.1}.} transformation of \eqref{220125:2229}, exactly as the two statistical invariances Eqs.$\,$(8-9) proposed in~\cite{Oberlack22}.
But, unlike~to~\cite{Oberlack22}, where those two invariances are nonphysical and cannot be realized by any transformation of the governing equations because of violating the classical principle of cause and effect between the fluctuations and the mean fields \cite{Frewer15,Frewer16.1,Frewer17,Sadeghi20,Frewer21.1,Frewer21.4},
we shall now prove that the above proposed invariance \eqref{220126:2319} is fundamentally different to the similar-looking statistical translation invariance Eq.$\,$(8) in \cite{Oberlack22}, for the single but important reason that \eqref{220126:2319} can be realized by an appropriate transformation of the governing equations.

If in the governing equations \eqref{220125:2229} the fluctuating fields $u_i,\, p$ (which are obtained by $U_i-\overline{U}_{\!i}$, $P-\overline{P}$) get transformed as
\begin{equation}
t^*=t,\quad x_i^*=x_i,\quad u_i^*=u_i+\eta\cdot\delta_{i1},\quad p^*=p,
\label{220129:1727}
\end{equation}
where $\eta=\eta(t,\vx)$ is an arbitrary random space-time field with zero mean, such that to all orders it is statistically independent of the governing fluctuation fields for velocity $u_i$ and pressure gradient
$\partial_j p$,\linebreak[4] i.e., such that
\begin{equation}
\overline{\eta}=0, \qquad \overline{\eta^m\hspace{0.3mm} u_{i_1}\cdots u_{i_n}\partial_jp}=\overline{\eta^m}\cdot \overline{u_{i_1}\cdots u_{i_n}\partial_jp}, \;\;\; \forall n,m\geq 1,
\label{220410:2051}
\end{equation}
then the fluctuation correlations will transform as (up to order $n=4$)
\begin{equation}
\left.
\begin{aligned}
&\overline{u_iu_j}^{\,*}=\overline{u_iu_j}+\overline{\eta^2}\hspace{0,3mm}\delta_{i1}\delta_{j1},\quad \overline{u_iu_ju_k}^{\,*}=\overline{u_iu_ju_k}+\overline{\eta^3}\hspace{0,3mm}\delta_{i1}\delta_{j1}\delta_{k1},
\quad \overline{u_i\partial_jp}^{\,*}=\overline{u_i\partial_jp},
\\[0.5em]
&\overline{u_iu_ju_ku_l}^{\,*}=\overline{u_iu_ju_ku_l}+\overline{\eta^2}\hspace{0,5mm}T_{ijkl}+\overline{\eta^4}\hspace{0,3mm}\delta_{i1}\delta_{j1}\delta_{k1}\delta_{l1},
\quad\overline{u_iu_j\partial_kp}^{\,*}=\overline{u_iu_j\partial_kp},
\end{aligned}
~~~\right\}
\label{220209:0934}
\end{equation}
from which, if compared with the invariance \eqref{220126:2319}, yields the following realizability conditions
\begin{equation}
c_2=\overline{\eta^2},\qquad c_3=\overline{\eta^3},\qquad c_4=\overline{\eta^4}.
\label{220128:1624}
\end{equation}
Now, since the parameters $c_2$, $c_3$, and $c_4$ are defined as space-time constants, the simplest generating random field $\eta(t,\vx)$ to yield such constant moments is to let the field be an uncorrelated process
in both space and time (white-noise process)
\begin{equation}
\overline{\eta(t,\vx)\eta(t^\prime,\vx^\prime)}=\overline{\eta^2}\cdot \delta(t-t^\prime)\delta(\vx-\vx^\prime).
\label{220129:1728}
\end{equation}
Since we further want to specify these three translation parameters $c_2$, $c_3$, and $c_4$ independently of~each other, we need a non-Gaussian probability density function (PDF)\footnote{A PDF-independent method can also be used to generate a non-normal univariate random variable with pre-specified moments. For example, the polynomial approach of \cite{Fleishman78}, expanding the targeted non-normal variable $\eta$ into powers of~a normal (Gaussian) variable $z$ with zero mean and unit variance, i.e., $\eta=\sum_{m=0}^n a_m z^m$, with the $m$-th moment of the standard normal distribution: $\overline{z\vphantom{.}^{m}}=2^{m/2-1}(1+(-1)^m)\Gamma(m/2+1/2)/\sqrt{\pi}$.
The coefficients $a_m$ are determined to match the values of the targeted moments $\overline{\eta\vphantom{.}^n}$, which then leads to a system of nonlinear equations that can be solved numerically.\linebreak[4]
As shown in \cite{Vale83}, this method extends to the multivariate case $\vphi=(\phi_1,\ldots,\phi_n)$, to then comply with a pre-specified  covariance matrix $C_{ij}=\overline{\phi_i\phi_j}$, using the technique of matrix decomposition. The Cholesky decomposition is usually used to map uncorrelated to correlated normal random variables, but it can also be used to map non-normal ones~\cite{Foldnes16}.}
to generate and extract the random numbers for $\eta(t,\vx)$, which then gets assigned to each space-time point independently. Hence, the realization conditions \eqref{220128:1624} constrains the random field $\eta=\eta(t,\vx)$ to be a non-Gaussian white-noise
process.\footnote{To extract the random numbers for $\eta$ from a Gaussian PDF is obviously not adequate. There, only the first two moments can be specified independently. All higher-order moments are expressible by these two, simply because a Gaussian is fully determined by its first two moments.}

The construction of a PDF from its moments is not uniquely determined.\footnote{For example, the PDF by Heyde
$f_{\alpha,k}(x)=(x\sqrt{\pi})^{-1}\exp(-\ln(x)^2)(1+\alpha\sin(4k\pi\ln(x)))$, $|\alpha|\leq 1$, $k\in\mathbb{N}$, $x\geq0$, taken from \cite{Heyde63},
illustrates very effectively the nonuniqueness problem of the moments: Although the PDF depends on the parameters $\alpha$ and $k$, all its moments do not depend on them. Therefore, even if the moments are known to all orders, they do not uniquely determine the underlying PDF.} To obtain a unique result, further fundamental constraints have to be placed (see e.g. \cite{Frontini94,Bandyopadhyay05}). However, for the PDF-construction of $\eta$ from the three moments \eqref{220128:1624}, the nonuniqueness problem is not of concern here, since the only issue here is just to find at least one realization for $\eta$. Quite the contrary, the more realizations exist, the easier and more effective the construction of $\eta$.

\newpage

It is rather the other issue of the moment problem we need to solve here, namely to find a non-negative function $f(x)\geq 0$ from a given but finite set of moments$\,$\footnote{When referring to $\eta$ in
\eqref{220129:1727}-\eqref{220129:1728}, the only given moments are: $\mu_0=1$,$\:$ $\mu_1=\overline{\eta}=0$,$\:$ $\mu_k=\overline{\eta^k}=c_k$, $k=2,3,4$.}
\begin{equation}
\mu_k=\int_{-\infty}^\infty x^k f(x) dx,\quad k=0,\ldots,M,
\end{equation}
on the infinite interval $-\infty <x<\infty$. By using the maximum entropy approach~(e.g.~as in \cite{Frontini94}), the~functional structure of the unknown density $f(x)$ is restricted to
\begin{equation}
f(x)=\exp\left(-\sum_{k=0}^{M^\prime}\lambda_k x^k\right),\quad  M^\prime>M,\mod\!(M^\prime,2)=0,
\label{220129:1640}
\end{equation}
supplemented by the condition that the first $M+1$ moments be given by $\mu_k$,
\begin{equation}
\int_{-\infty}^\infty x^k f(x)dx = \mu_k, \quad k=0,\ldots,M,
\end{equation}
being $M+1$ nonlinear equations for the unknown Lagrange multipliers $\lambda_k$ in \eqref{220129:1640}, which then can be solved numerically. The remaining Lagrange multipliers in \eqref{220129:1640}, i.e.
all $\lambda_k$ for $k=M+1,\ldots,M^\prime$, are left arbitrary and can vary freely in order to stabilize the numerical search algorithm --- only the highest Lagrange multiplier $\lambda_{M^\prime}$ (where $M^\prime$ is even) is restricted and chosen to be positive $\lambda_{M^\prime}>0$ and larger in value than all lower multipliers, $\lambda_{M^\prime}>|\lambda_{k}|$, for all $k<M^\prime$, to ensure convergence.

Hence, since we found a realization of $\eta$ \eqref{220129:1727}, we have also found a cause for the statistical invariance \eqref{220126:2319}. However, it should be clear that the causal transformation \eqref{220129:1727} itself is not a symmetry of the fluctuation equations of the governing equations \eqref{220125:2229}. It is a non-invariant transformation that maps the unclosed fluctuation equations of \eqref{220125:2229} to a new and different set of fluctuation equations,\linebreak[4] but in such a way that when taking statistical averages, it emerges as an invariance \eqref{220126:2319} of the induced moment equations \eqref{220126:0948},
considered here in their reduced form \eqref{230112:0939}.
In other words, on the coarse-grained (averaged)~level, the statistical invariance~\eqref{220126:2319} emerges as an effect from a non-invariant cause~\eqref{220129:1727} on the fine-grained (fluctuating) level.\footnote{This fact, that the cause itself need not to be a symmetry in order to induce a symmetry as an effect, can also be illustrated very nicely by the example of the diffusion equation: Its underlying fine-grained discrete random walk does not admit the variable transformation $t^*=c^2\cdot t$,$\,$ $x^*=c\cdot x$ as a symmetry transformation; only when coarse-graining this stochastic process, to yield the continuous and diffusive Fokker-Planck equation, it will turn into a scaling symmetry, resulting, however, from the cause of a non-invariantly transformed or changed random walk. For more details, see \cite{Frewer16.1}.}

Combining now all the aforementioned invariances $\mathsf{S}_1$, $\mathsf{S}_2$, $\mathsf{G}$ \eqref{220126:2221} and $\mathsf{T}$ \eqref{220126:2319} to determine the corresponding invariant scaling functions for the mean velocity $\overline{U}_{\!1}$ and the moments of the streamwise fluctuations $\overline{u_1^n}$ in the statistically stationary and fully developed regime of turbulent channel flow, we~arrive (up to order $n=3$) at the following characteristic system$\,$\footnote{How to generate a characteristic system from Lie-group symmetries and equivalences,
see e.g. \cite{Ovsiannikov82,Stephani89,Bluman89,Fushchich93,Olver93,Ibragimov94,Andreev98,Hydon00,Cantwell02}.}
\begin{equation}
\frac{dx_2}{a_2x_2}
=\frac{d\overline{U}_{\!1}}{(a_2-a_1)\overline{U}_{\!1}+c_1}
=\frac{d\overline{u_1^2}}{2(a_2-a_1)\overline{u_1^2}+c_2}
=\frac{d\overline{u_1^3}}{3(a_2-a_1)\overline{u_1^3}+c_3}.
\label{220130:1025}
\end{equation}
The general solution of \eqref{220130:1025} can generate two types of invariant functions, either a log function (for $a_2=a_1$) or a power function (for $a_2\neq a_1$). In the following we will only consider the latter case, once by solving \eqref{220130:1025} in the deficit-form representation for the center region of the channel
\begin{equation}
\frac{\overline{U}_{\!1}^{(0)}-\overline{U}_{\!1}}{u_\tau}=C^\prime_1\left(\frac{x_2}{h}\right)^\sigma,\qquad \frac{\overline{u_1^n}^{(0)}-\overline{u_1^n}}{u^n_\tau}=C^\prime_n\left(\frac{x_2}{h}\right)^{n\sigma},
\;\; n=2,3,
\label{220131:1155}
\end{equation}
and once in the wall-units representation for the inertial sublayer
\begin{equation}
\overline{U}_{\!1}^{\,+}=C_1\big(y^{+}\big)^\gamma+B_1,\qquad \overline{u_1^n}^{\,+}=C_n\big(y^{+}\big)^{n\gamma}+B_n,
\;\; n=2,3,
\label{220131:1156}
\end{equation}
with the $C$-parameters all being integration constants, and the group constants all comprised in the remaining parameters.\footnote{Note that although the scaling laws \eqref{220131:1155} and \eqref{220131:1156}
are similar to those in \cite{Oberlack22}, there is a decisive difference between them. Here they directly apply to the fluctuation correlations and not through a transformation over the full-field correlations, as in \cite{Oberlack22}.}

\newpage

\newgeometry{left=2.0cm,right=2.0cm,top=2.0cm,bottom=1.5cm,headsep=1em}

\begin{figure}[t!]
\centering
\begin{minipage}[c]{.49\linewidth}
\includegraphics[width=.90\textwidth]{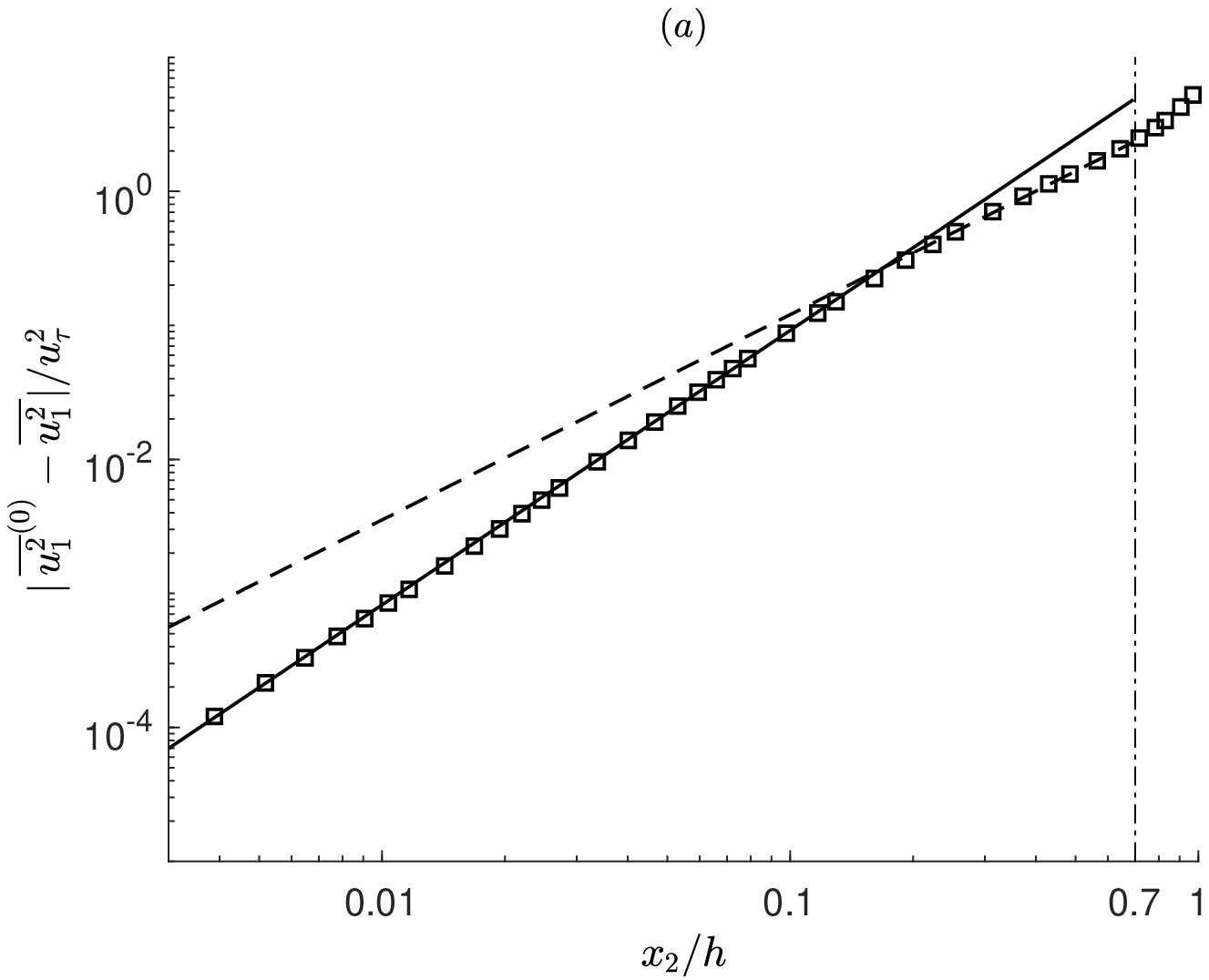}
\end{minipage}
\hfill
\begin{minipage}[c]{.49\linewidth}
\includegraphics[width=.90\textwidth]{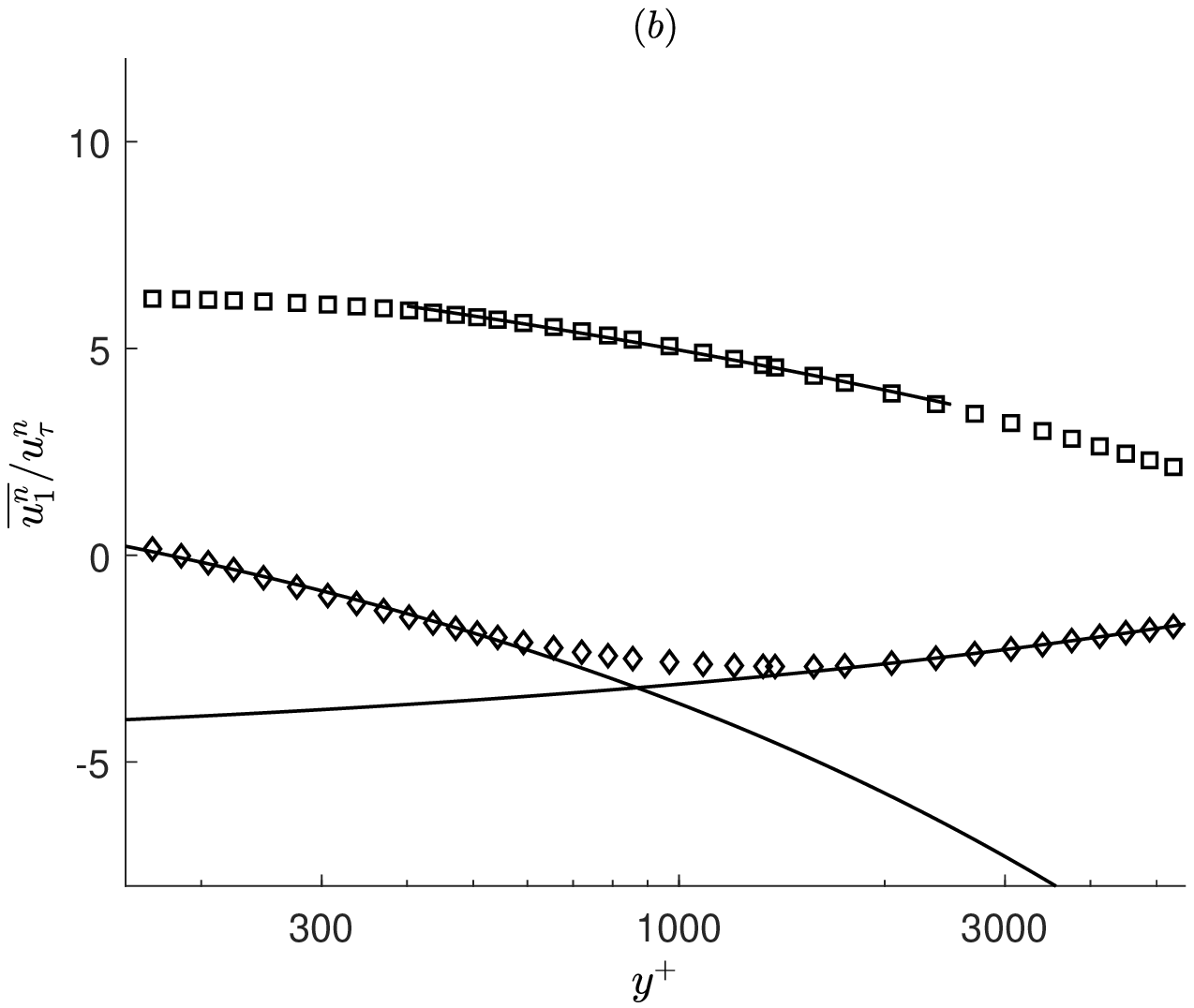}
\end{minipage}
\caption{
(a) The symbols show the second moment of the streamwise fluctuation field in deficit form, presented in the same arrangement as before in Fig.$\,$\hyperref[Fig3]{3(a)}. The solid line shows the best-fit according to the scaling law~\eqref{220131:1155} in the range $0<x_2/h<0.15$, leading to the exponent $2\sigma\approx 2$. The dashed line shows the best-fit for the same scaling law, but for the range
$0.15<x_2/h<0.7$, resulting in $2\sigma\approx 1.5$. However, since the best-fit for the mean~velocity field yields the exponent $\sigma\approx 2$ over the whole range $0<x_2/h<0.7$, the scaling law \eqref{220131:1155} turns inconsistent over this whole range.
\\
(b) The symbols show the second moment (squares) and the third moment (diamonds) of the streamwise fluctuation field in wall units, presented in the same arrangement as before in Fig.$\,$\hyperref[Fig3]{3(b)}. Based on the best-fit of the mean velocity field $(n=1)$ according to the scaling law \eqref{220131:1156}, which yields $\gamma\approx 0.12$ in the whole range $400<y^+<2500$, the solid line through the square symbols shows the best-fit of the second moment according to the same scaling law \eqref{220131:1156}, but for $n=2$. The left solid line through the diamond symbols shows the best-fit of the third moment $(n=3)$ in the fitted range $400<y^+<600$, based on the already fixed scaling exponent $\gamma\approx 0.12$. As shown in the figure, the fitted range can then be extended to the lower end $150<y^+<600$.\linebreak[4] The right solid line through the diamond symbols shows the best-fit of the third moment $(n=3)$ in the fitted range $1500<y^+<2500$, based again on the same fixed scaling exponent $\gamma\approx 0.12$. As shown in the figure, the fitted range can then be extended to the upper end $1500<y^+<5000$. Thus, in contrast to \eqref{220131:1155}, the scaling law \eqref{220131:1156} can be consistently applied. All fitted values can be taken again from Appendix \ref{SecB}.
}
\label{Fig4}
\vspace{-1em}
\end{figure}

In Fig.$\,$\hyperref[Fig4]{4(a)} the best-fit of \eqref{220131:1155} for $n=2$ is shown. As can be seen, there is no global power-law scaling for the second moment. The scaling is rather divided into two regions: A region very close to channel center $0<x_2/h<0.15$, with the trivially predictable scaling exponent $2\sigma\approx 2$ (see Section~\ref{Sec2}), and an adjacent region further away from it, $0.15<x_2/h<0.7$, with a different scaling of $2\sigma\approx 1.5$.

However, this subdivision is not the problem of the scaling law \eqref{220131:1155}, since each region can still be fitted by it --- the specified range in \cite{Oberlack22} is simply too long to exhibit global scaling behaviour for~the fluctuation moments. The core problem lies in the consistency of the scaling exponent $\sigma$, namely that when fitting the {\it mean velocity} according to \eqref{220131:1155}, we trivially get an exponent also close to $2$ (due~to the universal aspect discussed in Section~\ref{Sec2}, particularly in footnote \hyperref[fn5]{5}), and since $\sigma$ is the same symbol
in \eqref{220131:1155} for both moments, we get the inconsistent relation $2\sigma=\sigma$ for the range $0<x_2/h<0.15$, and the mismatch between $\sigma\approx 2$ and $\sigma\approx0.75$ for the range $0.15<x_2/h<0.7$.

From this we can conclude that the symmetry-based modelling assumptions made in this section leading to the scaling law \eqref{220131:1155} are not applicable to the center region of turbulent channel flow. Other invariances and different arguments have to be found to generate a consistent scaling law in this region, which will be done in the next section, Sec.$\,$\ref{Sec41}.

Based on the best-fit of the first moment, in Fig.$\,$\hyperref[Fig4]{4(b)} the best-fit of the scaling law \eqref{220131:1156} for the second and third moment in the inertial sublayer is shown. The squares refer to the data of the second moment $(n=2)$, and the diamonds to the third moment $(n=3)$, while the solid lines display the best-fit for each moment in the region $400<y^+<2500$, for the same range as specified in~\cite{Oberlack22}.

As before for the center region, we face again a global scaling problem, but now at higher order: While the second moment can be well fitted over the whole range according to \eqref{220131:1156}, this is not possible for the third moment. We chose to split the region into two separately reduced regions,\footnote{Important to note here is that such a split is not possible for the full-field scaling law in \cite{Oberlack22}. A best-fit fails even on such smaller regions. For any region, the result will always be the ill-shaped parabolic-like profile as shown in  Fig.$\,$\hyperref[Fig3]{3(b)}.} $400<y^+<600$\hfill and\hfill $1500<y^+<2500$,\hfill for\hfill two\hfill
reasons:\hfill First,\hfill the\hfill range\hfill of\hfill the\hfill inertial\hfill sublayer\hfill for

\restoregeometry

\noindent
higher moments may be smaller than for lower moments and may shorten either towards the lower or the higher end. Therefore, it is not necessary that the second and the third moment have the same length of fitting range. Second, the used DNS results of \cite{Oberlack22} cannot be relied too much upon for moments beyond the Reynolds-stress $n=2$, since obviously the higher moments are not yet fully converged, as was revealed in Fig.$\,$\hyperref[Fig3]{3(a)}. Thus, the third moment may suffer of a not yet fully developed inertial sublayer, as can also be independently seen from the small oscillations present in the data even in the reduced ranges $400<y^+<600$ and $1500<y^+<2500$.

However, unlike the scaling \eqref{220131:1155} proposed for channel center, the scaling law \eqref{220131:1156} for the inertial sublayer does not lead to any inconsistencies. The best-fit result for the mean velocity
yields~the~scaling exponent $\gamma\approx 0.12$, which remains valid also for the second and third moment. Interestingly, as shown in Fig.$\,$\hyperref[Fig4]{4(b)}, the fitted scaling law for the reduced ranges $400<y^+<600$ and $1500<y^+<2500$ can be extended to the longer ranges $150<y^+<600$ and $1500<y^+<5000$, respectively.

Hence, we can conclude that the symmetry-based modelling assumptions made so far in this section are more applicable to the inertial sublayer than to the center region of turbulent channel flow. However, due to the infinite degree of freedom in choosing an invariance in unclosed systems, this drawback in channel center can be easily resolved, as will be shown next.

\label{p13}
\subsection{Implementation of a realizable statistical scaling invariance in channel center\label{Sec41}}

In order to model the universal structure of parallel lines in channel center with invariant functions, as shown in Fig.$\,$\hyperref[Fig1]{1} for the full-field correlations $\overline{U_1^n}$, which trivially will also hold for the fluctuation moments $\overline{u_1^n}$, due to being all non-zero in channel center (see footnote \hyperref[fn5]{5}), we need a third scaling invariance to fix the scaling exponent of the set of moments \eqref{220131:1155} to the universal value $2$. That~it's here in this case exactly the value $2$, and no other value, is rooted in the trivial fact that in channel center all $\overline{u_1^n}$ have a non-zero local extremum that start off quadratically, thus leading to a trivial quadratic power-law scaling $(x_2/h)^2$ for all $\overline{u_1^n}$ close to channel center. Sure, the further away we move from channel center, the less it will be a pure quadratic scaling, since the higher-order Taylor terms slowly start to get relevant --- for more illustrative examples, see also Figs.$\,$1-3 in \cite{Frewer22}.

Two points should be noted here: First, that a third scaling invariance is necessary to model equal scaling for moments of different order is based on the idea of \cite{Oberlack22}, though redundant for instantaneous moments (see Sec.$\,$\ref{Sec2}), it will here be consistently implemented now with a different statistical scaling invariance being physically realizable and by not violating the principle of causality, as the invariance Eq.$\,$(9) in \cite{Oberlack22} clearly does. Second, since the universal structure of parallel lines considered in this subsection is just trivial Taylor asymptotics around channel center, and not some special turbulent flow property, it is obvious that such a trivially predictable scaling does not need to be modelled. We do so nevertheless, to show, for completeness, how next to the already implemented statistical translation also a statistical scaling invariance can be consistently added to the analysis.

Now, to model the universal quadratic scaling of the streamwise fluctuation moments $\overline{u_1^n}$ in channel center with invariant functions, we consider the following statistical scaling invariance
of \eqref{220126:0948} (up to the same order as the foregoing statistical translation invariance \eqref{220126:2319}):
\begin{equation}
\left.
\begin{aligned}
\!\!\!\!\mathsf{Q}\!:&\quad\! t^*=t, \quad x_i^*=x_i,\quad \overline{U}_{\!i}^*=\overline{U}_{\!i},\quad \overline{P}^*=P,
\\[0.5em]
&\quad\!\overline{u_iu_j}^{\,*}=e^{q_{ij}}\overline{u_iu_j},\quad \overline{u_iu_ju_k}^{\,*}=e^{q_{ijk}}\overline{u_iu_ju_k},
\quad\overline{u_iu_ju_ku_l}^{\,*}=e^{q_{ijkl}}\overline{u_iu_ju_ku_l},
\\[0.5em]
&\quad\overline{u_i\partial_jp}^{\,*}=e^{q_{ij2}}\overline{u_i\partial_jp},
\quad \overline{u_iu_j\partial_kp}^{\,*}=e^{q_{ijk2}}\overline{u_iu_j\partial_kp},
\\[0.5em]
&\quad\!\text{with the constraints:}\;\; q_{12}=0,\;\;\, q_{1112}=q_{11},\;\;\, q_{1222}=q_{22},\;\;\, q_{1332}=q_{33},
\\[0.0em]
&\quad\!\text{and for any $n$-th order index-permutation $\sigma$\hspace{0.25mm}:}\;\; q_{i_1i_2\,\cdots\, i_n}=q_{i_{\sigma(1)}i_{\sigma(2)}\,\cdots\, i_{\sigma(n)}},\;\; n=1,\ldots ,4.
\end{aligned}
~\right\}
\label{220410:1818}
\end{equation}
However, in contrast to the statistical translation invariance \eqref{220126:2319}, the above transformation \eqref{220410:1818}\linebreak[4] is not admitted as an invariance globally for all spatial coordinate values, but only locally for values close to $x_2=0$, i.e., asymptotically close to channel center, the region we are interested in here.\linebreak[4] Under this asymptotic constraint (see Appendix~\ref{SecD}), it can then be easily verified that \eqref{220410:1818} leaves invariant the full system of moment equations \eqref{220126:0948} (when Reynolds-decomposed) and,~hence, is a valid invariance in channel center for statistically stationary flow.

In the same way as discussed for \eqref{220126:2319}, the above invariance \eqref{220410:1818} is also a statistical equivalence of the governing equations \eqref{220125:2229}, when averaged to \eqref{220126:0948}, that can be realized by transforming the fluctuating fields as follows
\begin{equation}
t^*=t,\quad x_i^*=x_i,\quad u_i^*=\lambda_{i}\cdot u_i,\quad p^*=\lambda_p\cdot p,
\label{220410:2026}
\end{equation}
where $\vlambda=(\lambda_1,\lambda_2,\lambda_3,\lambda_p)$ is a correlated multivariate random variable and, like the univariate random variable $\eta$ \eqref{220410:2051}, again with zero mean and statistical independence of the governing fluctuation fields $u_i$ and $p$. The realizability conditions in channel center are:
\begin{equation}
\overline{\lambda_{i_1}\lambda_{i_2}\cdots\lambda_{i_n}}=e^{q_{i_1i_2\,\cdots\, i_n}},\quad\;
\overline{\lambda_1\lambda_2}=1,\quad\;
\overline{\lambda_i\lambda_p}=\overline{\lambda_i^2\lambda_2}\, ,\quad\;
\overline{\lambda_i^2\lambda_p}=\overline{\lambda_1\lambda_i\lambda_p}=\overline{\lambda_1\lambda_i^2\lambda_2}=\overline{\lambda_i^2}\, .
\label{220410:2028}
\end{equation}
When adding now the above Lie-group $\mathsf{Q}$ \eqref{220410:1818} to $\mathsf{S}_1\circ\mathsf{S}_2\circ\mathsf{G}\circ\mathsf{T}$ \eqref{220126:2221}-\eqref{220126:2319},
we then obtain to \eqref{220130:1025} the\linebreak[4]
following\hfill extended\hfill characteristic\hfill system,\hfill valid\hfill in\hfill the\hfill
asymptotic\hfill region\hfill of\hfill channel\hfill center\hfill ($x_2\rightarrow 0$),
\begin{equation}
\frac{dx_2}{a_2x_2}
=\frac{d\overline{U}_{\!1}}{(a_2-a_1)\overline{U}_{\!1}+c_1}
=\frac{d\overline{u_1^2}}{\big[2(a_2-a_1)+q_{11}\big]\overline{u_1^2}+c_2}
=\frac{d\overline{u_1^3}}{\big[3(a_2-a_1)+q_{111}\big]\overline{u_1^3}+c_3},
\label{220411:0009}
\end{equation}
which then trivially achieves the aimed universal quadratic scaling in channel center for the turbulence-relevant {\it fluctuation} moments (up to order $n=3$),
by fixing $a_1=-a_2$, $q_{11}=-2a_2$, $q_{111}=-4a_2$.

\label{p14}
\subsection{Reiterating the causality principle for statistical symmetries\label{Sec42}}

It is worthwhile to reiterate once again the fact that the two new statistical symmetries proposed herein, the statistical streamwise translation symmetry $\mathsf{T}$ \eqref{220126:2319} (valid from channel center down to the inertial sublayer) and the statistical scaling symmetry $\mathsf{Q}$ \eqref{220410:1818} (valid only in channel center), do {\it not} result from a deterministic symmetry of the Navier-Stokes or Euler equations. They are symmetries that only result from the (unclosed) {\it averaged} equations.\footnote{When being mathematically precise, the statistical invariances $\mathsf{T}$ \eqref{220126:2319} and $\mathsf{Q}$ \eqref{220410:1818} have to be
identified as equivalences and not as symmetries, which is also true for the nonphysical invariances in \cite{Oberlack22}, simply because the underlying equations are unclosed. Although $\mathsf{T}$ \eqref{220126:2319} and $\mathsf{Q}$ \eqref{220410:1818} only act as equivalences, we nevertheless refer to them here in this section as symmetries in order to be in line with the vocabulary in \cite{Oberlack22}. For further details on the fine but important distinction between equivalences and symmetries in unclosed systems, see again footnote \hyperref[fn13]{13}, as well as upcoming~Sec.$\,$\ref{SecA} and the references therein.} Therefore they are in distinct contrast to the statistical symmetries $\mathsf{S_1}$, $\mathsf{S_2}$ \eqref{220125:2134}, and $\mathsf{G}$ \eqref{220126:0950}, which all three have their origin in the corresponding deterministic symmetries of the non-averaged equations.

Now, although $\mathsf{T}$ \eqref{220126:2319} or $\mathsf{Q}$ \eqref{220410:1818} do not originate from any symmetry of the underlying deterministic (non-averaged and thus closed) dynamic equations, here the Euler equations,
they nevertheless dynamically result from a specific {\it non-invariantly} mapped stochastic motion, $\mathsf{T}$ from a univariate~\eqref{220129:1727} and~$\mathsf{Q}$ from multivariate~\eqref{220410:2026} non-Gaussian white-noise process, which the Euler equations can either realize or not. The level of confidence that the Euler equations may realize this kind of motion in the statistically stationary regime, either intermittently or globally over the whole time, rests on the fact that the stochastic mappings for $\mathsf{T}$ and $\mathsf{Q}$ are not some arbitrary mappings, but very specific ones, which leave the (unclosed) averaged Euler equations to {\it all} orders$\,$\footnote{The symmetries $\mathsf{T}$  and $\mathsf{Q}$ are not restricted to the moment order as specified in \eqref{220126:2319} and \eqref{220410:1818}, respectively. They can be readily extended to any higher order in the unclosed hierarchy.} \hspace{-0.45mm}of the infinite hierarchy invariant.

But unfortunately, since this hierarchy is unclosed, even when taking along all infinite orders,\footnote{An infinite hierarchy of elements which does not converge, {\it irrespective of its representation}, cannot be considered as closed, even when taking along all its infinite elements. The unclosed statistical hierarchy of the Navier-Stokes or Euler equations is of such a category. A detailed discussion on this issue can found e.g. in Sec.$\,$1.1 in \cite{Frewer21.5}.} there simply is no absolute guarantee that the deterministic Euler equations will permanently or intermittently realize these particular transformations $\mathsf{T}$ and $\mathsf{Q}$, although they are fully admitted as symmetries to all orders by the statistical Euler equations. In other words, although the invariant functions associated to the symmetries $\mathsf{T}$ and $\mathsf{Q}$ will solve the stationary moment equations to all orders and reduce them to the identity $0=0$ in the regimes where the symmetries apply, there simply is no guarantee for unclosed systems that such a reduction will then automatically imply these invariant functions as realizable solutions of the underlying deterministic equations. Because, particularly for the unclosed systems resulting from the statistical Euler or Navier-Stokes equations, there still is an infinite pool of other possible symmetries the deterministic equations can choose from. Hence, as already said in the beginning of Sec.$\,$\ref{Sec4}, all invariant functions that follow from the symmetries $\mathsf{T}$ and~$\mathsf{Q}$ are thus only possible but not guaranteed solutions of turbulent channel flow. And exactly for this reason it is so important to recognize what the invariances $\mathsf{T}$ and $\mathsf{Q}$ really are, namely only being equivalences that map (unclosed) equations to different (unclosed) equations of the same class, and not as being true symmetries that map solutions to different solutions of the same set of (closed)~equations. This careful distinction will avoid making fundamental mistakes, as it happened in \cite{Oberlack22} and also in all other previous publications of M. Oberlack since he first published on this topic in 2001.

After this clarification we can now turn to the subject of this section. The decisive difference between the statistical symmetries $\mathsf{T}$ and $\mathsf{Q}$ proposed herein, and the statistical symmetries Eqs.$\,$(8-9) proposed in \cite{Oberlack22}, is that the latter ones have no cause at all from which they can emerge. As can be rigorously  proven,\footnote{See e.g. Sec.$\,$I in \cite{Frewer15}, Sec.$\,$3 and Appendix A in \cite{Frewer17},
and Appendix B in \cite{Frewer18.2}.} no cause exists or can be constructed such that they can dynamically emerge from the underlying deterministic equations. Hence, the confidence level that the statistical symmetries Eqs.$\,$(8-9) proposed in \cite{Oberlack22} can be realized by the Euler or Navier-Stokes equations is exactly zero. This is shown by us through the red lines in Fig.$\,$\hyperref[Fig3]{3(a)-(b)}, a severe mismatch between the invariant functions induced by the symmetries Eqs.$\,$(8-9) and the DNS data, thus proving that these invariant functions are not realized by the deterministic equations that the DNS solves. In particular, the inconsistency in these invariant functions already starts at order $n=2$, and then systematically infects all higher orders with increasing intensity. Another clear and independent indication that the symmetries Eqs.$\,$(8-9) in \cite{Oberlack22} are not realizable and thus nonphysical is that the matching to the moments will improve by several orders of magnitude as soon as one discards these symmetries or puts them to zero, as shown e.g. in Sec.$\,$5 in \cite{Frewer14.1}, or by Oberlack {\it et al.} themselves~in~\cite{Sadeghi20}.

The explanation why the symmetries Eqs.$\,$(8-9) in \cite{Oberlack22} are not realizable and therefore fail is simple: They violate the classical principle of cause and effect. There simply is no cause for these symmetries on the deterministic (fine-grained) level such that they can emerge as an effect on the averaged\linebreak[4] (coarse-grained) level. In other words, no cause {\it at all} exists from which Eq.$\,$(8) or (9) can emerge as a symmetry transformation. In clear contrast of course to the statistical symmetries $\mathsf{T}$ \eqref{220126:2319} and $\mathsf{Q}$~\eqref{220410:1818}\linebreak[4] proposed herein, or the ones presented in \cite{Frewer18.2}, or in other third party studies (see the discussion and the references in Sec.$\,$1~in~\cite{Frewer16.1}), which all have a dynamical cause on the deterministic (fine-grained) level.\linebreak[4] The cause is mostly a non-invariant fine-grained collective motion organized such that when viewed on a larger space-time scale a symmetry on the coarse-grained level is observed. In the very same~way as for example the course-grained (macroscopic) diffusion equation acquires a scaling symmetry from its underlying fine-grained (microscopic) motion of a random walk which itself does not admit this scaling symmetry (see \cite{Frewer16.1} for a detailed analysis and discussion). In other words, the cause itself on the fine-grained level need not to be a symmetry in order to induce a symmetry as an effect on the coarse-grained level.

Therefore, if a dynamical symmetry on a large space-time scale is observed, then a cause in form of a specific motion on a smaller space-time scale must exist (at least in classical physics),\linebreak[4] simply because the large-scale symmetry needs to emerge or to be generated from something, where, as already said, the cause itself need not to be a symmetry in order to generate a symmetry on a higher dynamical level. And exactly such a necessary cause-effect relationship does not exist for the statistical symmetries Eqs.$\,$(8-9) in \cite{Oberlack22}. They are causeless and therefore nonphysical. In the end they are just mathematical artefacts of the unclosed statistical equations.

This necessary cause-effect relationship that need to exist for statistical symmetries (at least in classical physics), can now be used as a guiding modelling principle whenever symmetries get determined from unclosed systems that result from a course-graining process of an underlying dynamical set of closed equations that can be directly simulated or measured. Because, for as soon as such obtained symmetries violate this principle of cause and effect, for example as the symmetries Eqs.$\,$(8-9) in \cite{Oberlack22} globally do in having no cause at all from which they can originate, then they can be immediately ruled out as possible candidates, simply because the level of confidence is exactly zero that they can be realized by the underlying dynamical equations.

\newpage\noindent

\appendix

\label{p16}
\section{On the usefulness of a Lie-group symmetry analysis in turbulence\label{SecA}}

The method of Lie symmetry groups is a successful tool to either model dynamical rules that should admit a certain given set of symmetries, or to provide deep insight into the structure of the solution space for a given but closed set of dynamical equations, including the possibility to even allow for their full integration (see e.g. \cite{Ovsiannikov82,Stephani89,Bluman89,Fushchich93,Olver93,Ibragimov94,Andreev98,Hydon00,Cantwell02}).

The (statistical) equations of turbulence, however, are different, both conceptually and practically. These equations are mathematically unclosed and need to be modelled empirically. Hence, caution has to be exercised when extracting new (statistical) symmetries from the unclosed and unmodelled theory itself, not to run into any circular arguments. For example, to derive new symmetries from unclosed equations to then use them in order to close those same equations again, is such a circular argument \cite{Khujadze20}. Also, to explore the solution structure of unclosed equations with new symmetries only\linebreak[4] admitted by those unclosed equations, turns out to be inconclusive, not only because the admitted set of symmetries is unclosed by itself, but also, once a particular choice from such an infinite (unclosed) set of possible symmetries is made, there is a high chance that a nonphysical symmetry will be chosen which is not reflected by experiment or numerical simulation.

All these well-known and crucial facts are not mentioned in \cite{Oberlack22}, nor in any of the first author's previous publications ever since his first paper \cite{Oberlack01} on turbulence and symmetries appeared more than two decades ago --- a key paper of his which is even technically flawed \cite{Frewer14.2} in that the Lie-group symmetry analysis has been misapplied (see Appendix \ref{SecF}).

Among one of the basic facts not understood by Oberlack {\it et al.} ever since is that for unclosed systems the concept of symmetries breaks down and gets replaced by the weaker concept of equivalences. It is not a semantic sophistry to carefully distinguish for differential equations between symmetry and equivalence transformations, because a symmetry transformation maps a solution of a specific (closed) equation to a new solution of
the {\it same} equation, while an equivalence transform acts in a weaker sense in that it only maps an (unclosed) equation to a new (unclosed) equation of the same class --- and  therefore, since equivalence transformations map equations and not solutions, they do not allow for the same insight into the solution structure of differential equations as symmetry transformations do.\footnote{Of course, this does not mean that equivalence transformations are not useful. For example, they can be successfully applied to classify unclosed differential equations according to the number of symmetries they admit when specifying the unclosed terms (see e.g. \cite{Khabirov02.1,Khabirov02.2,Meleshko02,Chirkunov12,Meleshko15,Bihlo17,Vaneeva20}). A typical task in this context sometimes is to find a specification of the unclosed terms such that the maximal symmetry algebra is gained. Once the equation is closed by a such a group classification, invariant solutions can be determined. But in how far these equations and their solutions are physically relevant and whether they can be matched to empirical data is not clarified {\it a priori} by this approach, in particular if such a pure Lie-group-based type of modelling is carried out completely detached from empirical findings.}

In particular, when generating invariant functions from equivalence transformations, as constantly done and argued by Oberlack {\it et al.} for the non-modelled and unclosed equations of turbulence, they do {\it not} constitute solutions of the unclosed system, but only possible candidate functions for a possible solution. In other words, they perform as invariant functions which only possibly but not necessarily can serve as useful turbulent scaling functions. Moreover, since their invariance analysis also never makes any choice or specification on which differential and integral variables the unclosed terms may depend, it always results into an infinite-dimensional equivalence group. Thus, the admitted set of equivalences is never closed by such an approach, when properly and correctly performed. Therefore, within the Lie-group invariance analysis itself an own closure problem is generated, with the result that any thinkable invariance can be derived and not only those few reported by Oberlack {\it et al.}\linebreak[4] Ultimately this means that the choice of an invariance is made by the user and not dictated by theory.

Referring again specifically to \cite{Oberlack22}, the search for new ``symmetries" from the unclosed equations of~Eq.$\,$(4), even when considering the entire infinite and non-modelled set, inevitably leads to an infinite dimensional and thus unclosed Lie-algebra, where (nearly) any invariant transformation and hence (nearly) any desirable scaling law can be generated. The simple reason for this is that at each order of the infinite hierarchy almost any change due to a variable transformation can always be balanced or compensated by an unclosed term of the next higher order.\footnote{\label{fn27}For explicit examples, see for instance the recent invariance analysis
in \cite{Frewer21.4} (Appendix A), or \cite{Frewer14.2,Frewer16.3,Frewer18.2}.} Ultimately one has an infinite set of invariant possibilities to choose from when performing a full and correct Lie-group symmetry analysis for unclosed equations. A crucial information which is not shared with the reader~in~\cite{Oberlack22}.

Hence, the Lie-group symmetry method in turbulence is not free of any assumptions. It is an {\it ad~hoc}\linebreak[4] method too, not in the same but in a similar way as the classical self-similarity method used by von K\'arm\'an and Prandtl a century ago: Instead of using an {\it a priori} set of scales, the Lie-group method has to make use of an {\it a priori} set of symmetries, namely to select the correct relevant symmetries from an infinite (unclosed) set. In other words, the particular selection of the additionally chosen symmetries Eqs.$\,$(8-9) in \cite{Oberlack22} is an assumption and not a result that comes from theory, as the authors try to convey. Because, as just referenced in the previous footnote \hyperref[fn27]{27}, when correctly performing a complete and systematic Lie-group symmetry analysis on the considered set of unclosed equations in the untruncated form of Eq.$\,$(4), one gets an infinite set of functionally independent invariances, and not only those few as first reported in \cite{Oberlack10} and presented here again through Eqs.$\,$(8-9) in \cite{Oberlack22} --- to note is\linebreak[4] that all ``new" invariances in \cite{Oberlack10}, or equally in \cite{Rosteck13}, were obtained only through heuristics and a trial-and-error ansatz, and {\it not} through a complete and systematic Lie-group analysis, which would have given an unclosed set of invariances and thus an overall different conclusion, namely that the Lie-group method alone, like any other analytical method, cannot bypass the closure problem of~turbulence.

Another basic fact to be understood before applying the Lie-group symmetry method to unclosed equations is to know whether they are based or induced by a more fundamental closed equation. If this\linebreak[4]
is the case then additional invariant modelling rules apply. For Euler or Navier-Stokes turbulence a critical modelling rule is to not violate the classical principle of cause and effect between the
fluctuating\linebreak[4] and the mean fields (see e.g.$\,$\cite{Frewer15,Frewer16.1,Frewer17,Frewer18.2,Frewer21.4}). The reason for this restriction is clear: Since~the deterministic\linebreak[4] set of Navier-Stokes equations naturally defines a causal structure on the statistically induced equations, that is, since the deterministic (fine-grained) equation implies its statistical (coarse-grained) equations and not opposite, a strict principle of cause and effect is formulated by this asymmetric relation which should be respected and not violated during any modelling process. For turbulence, the following cause-effect relations between the fluctuations (cause) and their correlations (effect) can be formulated: (1) Every statistical invariance need to have a cause at the fine-grained fluctuating level from which it can emerge, where (2) the cause itself need not to be an invariant in order to induce an invariance as an effect on the coarse-grained averaged level, but (3) if the cause is an invariant, then the induced effect is automatically also an invariant, but which, however, can be intermittently or globally broken in certain flow processes.

Therefore, to unravel the complexity of Navier-Stokes turbulence, not only the unclosed statistical equations, but also their defining deterministic equations, the instantaneous Navier-Stokes equations themselves, should be considered and taken into account in any modelling and solution finding process\linebreak[4] --- and not to be ignored, as done in \cite{Oberlack22}, with the consequence then that two non-realizable and thus nonphysical invariances Eqs.$\,$(8-9) get proposed, which are even falsely elevated to two very special\linebreak[4] symmetries that apparently should {\it ``reflect the two well-known characteristics of turbulent flows:\linebreak[4] non-Gaussianity and intermittency"} [p.$\,$1]. Both invariances clearly violate the causality principle, since no cause on the fluctuating level exists such that the invariances Eqs.$\,$(8-9) can result as an
effect~\cite{Frewer15,Frewer16.1,Frewer17,Frewer18.2,Frewer21.4}. This violation then clearly shows itself as the matching failure in Fig.$\,$\hyperref[Fig3]{3}.

What we know and can say so far, when scanning the literature also beyond Oberlack {\it et al.}, is that for Euler or Navier-Stokes turbulence no breakthrough has yet been achieved when using the invariant function method of Lie-group symmetries. Up to date, all systematic results to predict the statistical scaling behaviour of turbulent flows with Lie-group symmetries, are either not rigorous to convince or are not correct to be adopted. In the former case, the Lie-group symmetry results are standardly based on strong low-order assumptions which typically turn out to be incompatible to associated higher-order relations in showing an increasing mismatch to empirical results the higher the statistical order gets, while in the latter case, the Lie-group symmetry results are already inconsistent from the outset in violating certain immutable constraints already on the lowest statistical order. One reason for this prominent failure and the missing breakthrough is that the classical Navier-Stokes theory does not allow for a local space symmetry, in strong contrast, for example, to the theory of quantum fields, which is based on such a symmetry, the local gauge symmetry, which successfully predicts the unknown functional structure of the interacting fields between the various elementary particles.

\newpage

\label{p18}
\section{Parameter values for the figures shown\label{SecB}}

\paragraph*{Fig.$\,$1a:}
The basis profile ($n=1$) is that of laminar channel flow $\overline{U}_{\!1}=U_{\scriptscriptstyle\! L}=\rho(1-x_2/h)(1+x_2/h)$, with $\rho=u_\tau\text{Re}_\tau/2$, and values $u_\tau=0.034637$ and
$\text{Re}_\tau=10049$ taken from the database of \cite{Oberlack22}.

The symbols in the figure are the increasing powers of $U_{\scriptscriptstyle\! L}$ (in deficit form), i.e.
$(U_{\scriptscriptstyle\! L}^{\scriptscriptstyle{(0)\hspace{0.3mm}}\scriptstyle n}-U_{\scriptscriptstyle\! L}^n)/u_\tau^n$, up~to order $n=6$ (from bottom to top), where $U_{\scriptscriptstyle\! L}^{\scriptscriptstyle{(0)}}=\rho$ is the value at channel center $x_2/h=0$. The laminar profile $U_{\scriptscriptstyle\! L}$ was sampled at discrete points exactly at those locations as given in Fig.$\,$3 in \cite{Oberlack22}.

The solid lines in the figure are the best-fit using the turbulent scaling law Eqs.$\,$(19-20) from~\cite{Oberlack22}, with the fitted values: $\sigma_1=2.0$, $\sigma_2=1.98$, $C_1^\prime=\text{Re}_\tau/2$,
$C_2^\prime\approx 4.5823\text{e}{7}$, $C_3^\prime\approx 3.0693\text{e}{11}$, $C_4^\prime\approx 1.9094\text{e}{15}$, $C_5^\prime\approx 9.6579\text{e}{18}$, $C_6^\prime\approx 5.7244\text{e}{22}$,
$\alpha^\prime\approx 1.0213$, $\beta^\prime\approx 8.8096$.

\paragraph*{Fig.$\,$1b:}
The symbols in the figure display the turbulent full-field correlations $\overline{U_1^n}$ in deficit form up to order $n=6$ (from bottom to top), taken from Fig.$\,$3 in \cite{Oberlack22}.

The solid lines in the figure show in deficit form the increasing powers of the fitted mean velocity profile $\overline{U}_{\!1}$ ($n=1$, bottom curve), the only profile that was fitted in this arrangement. Using the defining turbulent scaling law Eq.$\,$(19) from~\cite{Oberlack22} for $n=1$, the only fitted values are: $\sigma_1=1.95$ and $C_1^\prime\approx6.43$, coinciding with the values in \cite{Oberlack22}. All solid lines above the lowest one ($n=1$) are then just obtained by
$(\overline{U}_{\!1}^{\scriptscriptstyle{(0)\hspace{0.3mm}}\scriptstyle n}-\overline{U}_{\!1}^{\,n})/u_\tau^n$, without any further fitting needed.

\paragraph*{Fig.$\,$2a:}

The symbols in the figure display the turbulent full-field correlations $\overline{U_1^n}$ in wall-units up to order $n=6$ (from bottom to top), taken from Fig.$\,$1(a) in \cite{Oberlack22}.

The solid lines in the figure show the increasing powers of the fitted mean velocity profile $\overline{U}_{\!1}^{\,+}$ ($n=1$, bottom curve), the only profile that was fitted in this arrangement.
Using the turbulent scaling law Eq.$\,$(15) from~\cite{Oberlack22}, the only fitted values are: $\kappa\approx 0.3909$, $B\approx 5.1251$. All solid lines above the lowest one ($n=1$) are then obtained by just evaluating $\overline{U}_{\!1}^{\,+\hspace{0.3mm} n}$, without any further fitting needed.

\paragraph*{Fig.$\,$2b:}

The symbols in the figure display the turbulent full-field correlations $\overline{U_1^n}$ in wall-units up~to order $n=6$ (from bottom to top), taken from Fig.$\,$1(a) in \cite{Oberlack22}.

The solid lines in the figure show the increasing powers of the fitted mean velocity profile $\overline{U}_{\!1}^{\,+}$ ($n=1$, bottom curve), the only profile that was fitted in this arrangement. But, instead of a log-law, the mean velocity profile was fitted here as a power law, $\overline{U}_{\!1}^{\,+}=C_1(y^+)^\omega+B_1$, which is obtained also as a valid scaling law in the inertial sublayer when solving Eq.$\,$(10) in \cite{Oberlack22} without the symmetry breaking constraint, i.e. for $a_{Sx}-a_{St}+a_{Ss}\neq0$. Since the translation group parameter for the mean velocity field was put to zero to also invariantly map the wall-boundary conditions, i.e., since $a_{1{\scriptscriptstyle \{1\}}}^H=0$, which implies $B_1=0$, the only fitted parameters are: $\omega\approx 0.1179$, $C_1\approx 9.9991$. All solid lines above the lowest one ($n=1$) are then obtained by just evaluating $\overline{U}_{\!1}^{\,+\hspace{0.3mm} n}$, without any further fitting~needed.

\paragraph*{Fig.$\,$3a:}

The symbols in the figure display the fluctuation correlations $\overline{u_1^n}$ (in deficit form) of the even orders $n=2,4,6$ (from bottom to top), taken from Fig.$\,$3 in \cite{Oberlack22} for the full-field moments and then transformed to the fluctuation moments using the unique relationship \eqref{220122:1231}. Hence, the discrete points shown (connected with a thin line) correspond exactly to those points shown in~Fig.$\,$3~in~\cite{Oberlack22} when transforming from the full-field to the fluctuation correlations.

The red solid line shows the failure already at lowest level $n=2$, when matching the data according to the prescribed scaling law Eq.$\,$(19) in~\cite{Oberlack22}, which in the transformed representation of the fluctuation correlation reads
\begin{equation}
\frac{\overline{u_1^2}^{(0)}-\overline{u_1^2}}{u^2_\tau}=C_2^\prime\left(\frac{x_2}{h}\right)^{\sigma_2}-\frac{\overline{U}_{\!1}^{(0)\hspace{0.3mm}2}-\overline{U}_{\!1}^{\hspace{0.3mm}2}}{u_\tau^2},
\label{220130:2304}
\end{equation}
where the mean velocity field $\overline{U}_{\!1}$ is also given by Eq.$\,$(19) in~\cite{Oberlack22}, but now for $n=1$, which trivially is equivalent to the full-field form
\begin{equation}
\frac{\overline{U}_{\!1}^{(0)}-\overline{U}_{\!1}}{u_\tau}=C_1^\prime\left(\frac{x_2}{h}\right)^{\sigma_1}.
\label{220130:2305}
\end{equation}

\newpage\noindent
The fitting procedure is as follows: First the mean velocity is fitted via \eqref{220130:2305}, where the result is then explicitly solved for $\overline{U}_{\!1}$, and then plugged into \eqref{220130:2304} to fit the second moment. While the scaling law for the first-moment \eqref{220130:2305} can be well fitted, with $\sigma_1=1.95$ and $C_1^\prime\approx6.43$, the fit for the second-moment~\eqref{220130:2304}~fails, and is shown as the red line for the best-fitted values $\sigma_2\approx 1.94$ and $C_2^\prime\approx3.3683\text{e}2$.

To note is that since the left-hand side \eqref{220130:2304} is negative, the whole equation has to be multiplied by $-1$ in order to display it in a log-log-plot.

\paragraph*{Fig.$\,$3b:}

The symbols show the second moment (squares) and the third moment (diamonds) of the streamwise fluctuation field in wall units, taken from Fig.$\,$1(a) in \cite{Oberlack22} for the full-field moments and then transformed to the fluctuation moments using the unique relationship~\eqref{220122:1231}. Hence, the discrete points shown correspond exactly to those points shown in~Fig.$\,$1(a)~in~\cite{Oberlack22} when transforming from the full-field to the fluctuation correlations.

The red solid line shows the failure for the moment $n=3$, when matching the data according to the prescribed scaling law Eq.$\,$(16) in~\cite{Oberlack22}, which in the transformed representation of the fluctuation correlation reads
\begin{equation}
\overline{u_1^3}^{\,+}=C_3\big(y^+\big)^{2\omega}-B_3-\left(\overline{U}_{\!1}^{\,+\hspace{0.3mm}3}+3\hspace{0.3mm}\overline{U}_{\!1}^{\,+\,}\overline{u_1^2}^{\,+}\right),
\label{220131:1012}
\end{equation}
where the streamwise Reynolds-stress $\overline{u_1^2}^{\,+}$ and the mean velocity field $\overline{U}_{\!1}^{\,+}$ are prescribed by Eq.$\,$(16) and Eq.$\,$(15) in~\cite{Oberlack22}, respectively, where only the latter is trivially equivalent again to the full-field form
\begin{equation}
\left.
\begin{aligned}
&\overline{u_1^2}^{\,+}=C_2\big(y^+\big)^{\omega}-B_2-\overline{U}_{\!1}^{\,+\hspace{0.3mm}2},
\\[0.5em]
&\overline{U}_{\!1}^{\,+}=\frac{1}{\kappa}\ln\big(y^+\big)+B.
\end{aligned}
~~~\right\}
\label{220131:1013}
\end{equation}
The fitting procedure is as follows: First the mean velocity and the second moment are fitted via~\eqref{220131:1013}, where each result is then plugged into \eqref{220131:1012} to fit the third moment. While the scaling law for the first- and second-moment \eqref{220131:1013} can be well fitted, with $\kappa\approx 0.39087$, $B\approx 4.5251$, $\omega\approx 0.11529$, $C_2\approx 4.3877\text{e}2$, and $B_2\approx 4.7533\text{e}2$,\footnote{Note that while the second moment $\overline{u_1^2}^{\,+}$ \eqref{220131:1013} can be well fitted with the scaling approach of \cite{Oberlack22}, as shown in Fig.$\,$\hyperref[Fig3]{3(b)},\linebreak[4] it fits very unnaturally, since parameters of order 100 are needed to fit a quantity that only varies by order 1 (see also Sec.$\,$\ref{SecB2}). This~problem is then solved in Fig.$\,$\hyperref[Fig4]{4(b).}}
the fit for third-moment~\eqref{220131:1012}~fails, and is shown as the red line for the best-fitted values $C_3\approx 3.2918\text{e}3$ and $B_3\approx4.9263\text{e}3$.

\paragraph*{Fig.$\,$4a:}

The symbols show the second moment of the streamwise fluctuation field in deficit form, presented in the same arrangement as before in Fig.$\,$3a. The solid line shows the best-fit according to the scaling law \eqref{220131:1155} in the range $0<x_2/h<0.15$, with parameters $2\sigma\approx 2.0481$ and $C_2^\prime\approx 1.0242\text{e}1$. The dashed line shows the best-fit for the same scaling law, but for the range
$0.15<x_2/h<0.7$, resulting in $2\sigma\approx 1.5314$ and $C_2^\prime\approx 4.0867$.

\paragraph*{Fig.$\,$4b:}

The symbols show the second moment (squares) and the third moment (diamonds) of~the streamwise fluctuation field in wall units, presented in the same arrangement as before in Fig.$\,$3b.
Based~on the best-fit of the mean velocity field $(n=1)$ according to the scaling law \eqref{220131:1156}, which yields~$\gamma\approx 0.1153$ in the whole range $400<y^+<2500$,
the solid line through the square symbols shows the best-fit of the second moment according to the same scaling law, but for $n=2$. For $\gamma\approx 0.1153$,
the~fitted values are $C_2\approx -1.1355$ and $B_2\approx 1.0544\text{e}1$, valid also in the whole range $400<y^+<2500$.

The left solid line through the diamond symbols shows the best-fit of the third moment $(n=3)$ in the fitted range $400<y^+<600$, based on the already fixed scaling exponent $\gamma\approx 0.1153$. As shown in the figure, the fitted range can then be extended to the lower end $150<y^+<600$. The fitted values are: $C_3\approx -0.7345$ and $B_3\approx 4.4188$.

The right solid line through the diamond symbols shows the best-fit of the third moment $(n=3)$ in the fitted range $1500<y^+<2500$, based again on the same fixed scaling exponent $\gamma\approx 0.1153$. As shown in the figure, the fitted range can then be extended to the upper end $1500<y^+<5000$. The fitted values are: $C_3\approx 0.1659$ and $B_3\approx -4.9267$.

\newpage

\label{p20}
\subsection{Indicator functions to \texorpdfstring{Fig.$\,$\hyperref[Fig2]{2}}{Fig.2}\label{SecB1}}

\begin{figure}[t!]
\centering
\begin{minipage}[c]{.49\linewidth}
\includegraphics[width=0.90\textwidth]{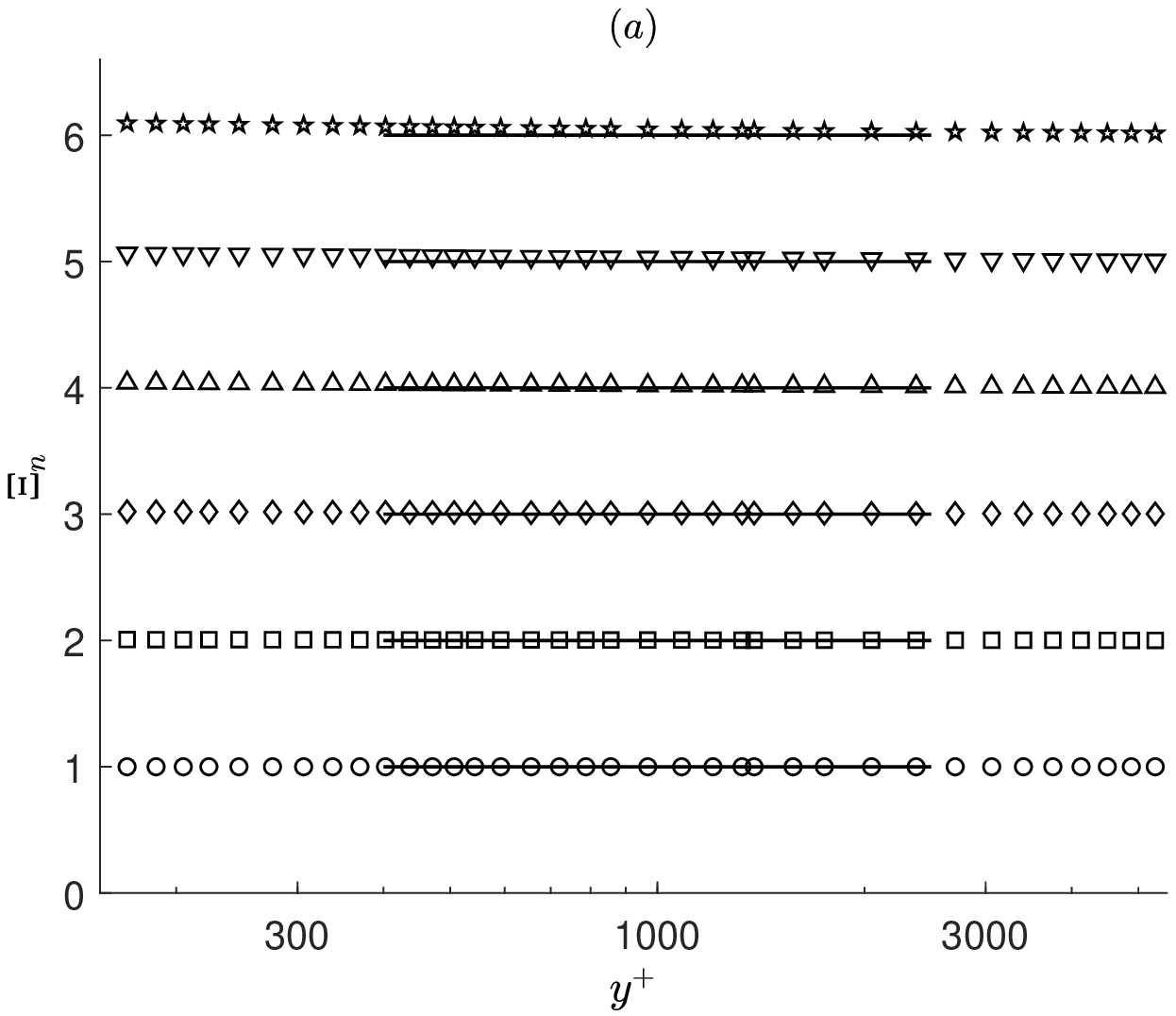}
\end{minipage}
\hfill
\begin{minipage}[c]{.49\linewidth}
\includegraphics[width=0.90\textwidth]{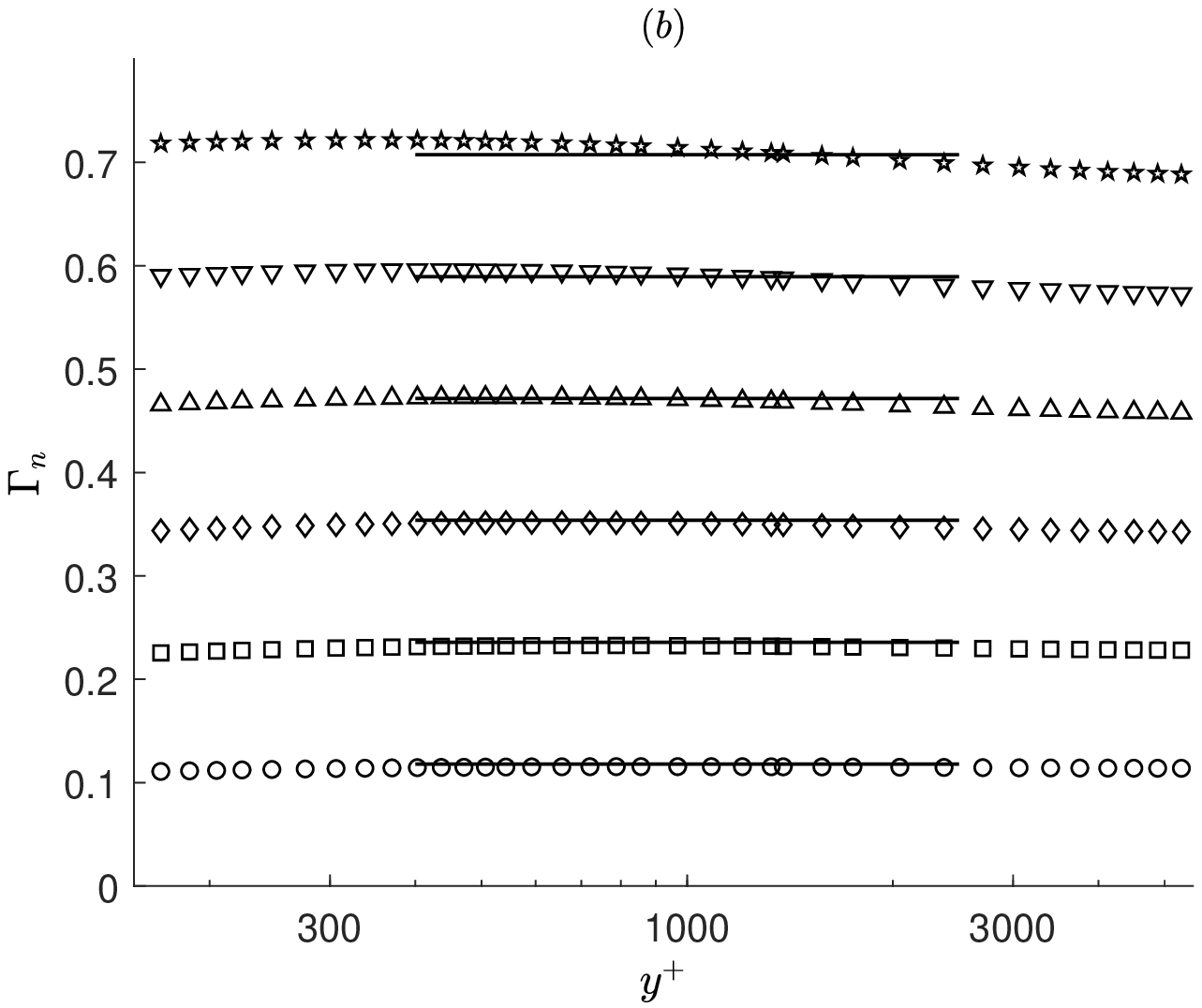}
\end{minipage}
\caption{
Comparison plots to Fig.$\,$1(b) in \cite{Oberlack22}. The symbols display the data side (left-hand side) and the solid horizonal lines the modelling side (right-hand side) of the indicator functions: in (a) for
the model-free and thus\linebreak[4] {\it unbiased} indicator function~$\Xi_n$~\eqref{220216:1102}, in (b) for $\Gamma_n$ \eqref{220216:1053} with $\omega\approx 0.12$ and data modifying
parameter~$C_1\approx 10$.
}
\label{Fig5}
\end{figure}

Normally, the indicator function to detect a power-law or a log-law in the data is respectively defined~as
\begin{equation}
\Gamma=\frac{y^+}{\mathcal{F}}\frac{d\mathcal{F}}{dy^+},\qquad \Xi=y^+ \frac{d\mathcal{F}}{dy^+},
\label{220216:0904}
\end{equation}
where $\mathcal{F}$ is some statistical correlation function that can be built from the data alone, i.e., a function that should not involve any modelling parameters. The power-law indicator function Eq.$\,$(18)
in~\cite{Oberlack22}, however, is not of this type. It makes use of the modelling parameter $B_n$,
\begin{equation}
\Gamma_n=\frac{y^+}{\overline{U_1^n}^{\,+}+B_n}\frac{d\overline{U_1^n}^{\,+}}{dy^+},
\label{220216:0920}
\end{equation}
i.e. $\mathcal{F}=\overline{U_1^n}^{\,+}+B_n$, which modifies the data $\overline{U_1^n}^{\,+}$ in bias towards the modelling function used. Therefore, in order to have a fair comparison to what is shown in Fig.$\,$1(b) in \cite{Oberlack22} and the power-law used herein in Fig.$\,$\hyperref[Fig2]{2(b)},
\begin{equation}
\overline{U_1^n}^{\,+}=\Big(\overline{U}_{\!1}^{\,+}\Big)^n,\;\; \text{with $\:\overline{U}_{\!1}^{\,+}$}=C_1\big(y^+\big)^\omega,
\label{220216:1142}
\end{equation}
we will not use the model-free indictor function $\Gamma$ \eqref{220216:0904}, but instead use the function
\begin{equation}
\Gamma_n=\frac{\ln\Big(\overline{U_1^n}^{\,+}/\,\,C_1^n\Big)}{\ln\big(y^+\big)}=n\cdot\omega,\quad n\geq 1,
\label{220216:1053}
\end{equation}
which, like the indicator function \eqref{220216:0920} from \cite{Oberlack22} with its parameters $B_n$, modifies the data, but here only with a single modelling parameter $C_1$. The indicator $\Gamma_n$ \eqref{220216:1053} is shown above in Fig.$\,$\hyperref[Fig5]{5(b)}.

In Fig.$\,$\hyperref[Fig5]{5(a)}, the indicator function
\begin{equation}
\Xi_n=\frac{\ln\Big(\overline{U_1^n}^{\,+}\Big)}{\ln\Big(\overline{U}_{\!1}^{\,+}\Big)}=n,\quad n\geq 1,
\label{220216:1102}
\end{equation}
is shown, based on the log-law model of Fig.$\,$\hyperref[Fig2]{2(a)}, given by
\begin{equation}
\overline{U_1^n}^{\,+}=\Big(\overline{U}_{\!1}^{\,+}\Big)^n,\;\; \text{with $\:\overline{U}_{\!1}^{\,+}$}=\frac{1}{\kappa}\ln\big(y^+\big)+B.
\label{220216:1143}
\end{equation}
To note is that \eqref{220216:1102} is a model-free and thus {\it unbiased} indicator function, i.e., the data shown by the symbols in Fig.$\,$\hyperref[Fig5]{5(a)} is not modified by any modelling parameter. Certainly, \eqref{220216:1102} also applies to the power-law model \eqref{220216:1142}. Hence, Fig.$\,$\hyperref[Fig5]{5(a)} demonstrates once again that the scaling of the {\it full-field} correlations $\overline{U_1^n}^{\,+}$ is simply driven by the scaling of the mean velocity~$\overline{U}_{\!1}^{\,+}$, independent of the specific model used.

\newpage
\newgeometry{left=2.0cm,right=2.0cm,top=2.0cm,bottom=1.2cm,headsep=1em}

\label{p21}
\subsection{The unnatural scaling in \cite{Oberlack22} for the Reynolds-stress \texorpdfstring{$\boldsymbol{\overline{u_1^2}}$}{u1sq} in the log-layer\label{SecB2}}

What we know already from Fig.$\,$\hyperref[Fig3]{3(b)} is that although the Reynolds stress $\overline{u_1^2}$ in the log-layer can be fitted with the scaling law from \cite{Oberlack22}, it fits very unnaturally. Because, a quantity as $\overline{u_1^2}$, which varies in the fitted region of order~$1$, has to be fitted with parameters of order $100$. Under this scaling, given~by

\vspace{-0.75em}
\begin{equation}
\overline{u_1^2}^{\,+}=C_2\big(y^+\big)^{\omega}-B_2-\overline{U}_{\!1}^{\,+\hspace{0.3mm}2},
\;\;\text{with}\;\;
\overline{U}_{\!1}^{\,+}=(1/\kappa)\ln\big(y^+\big)+B,
\label{220425:2101}
\end{equation}

\begin{figure}[t!]
\centering
\begin{minipage}[c]{.47\linewidth}
\includegraphics[width=.90\textwidth]{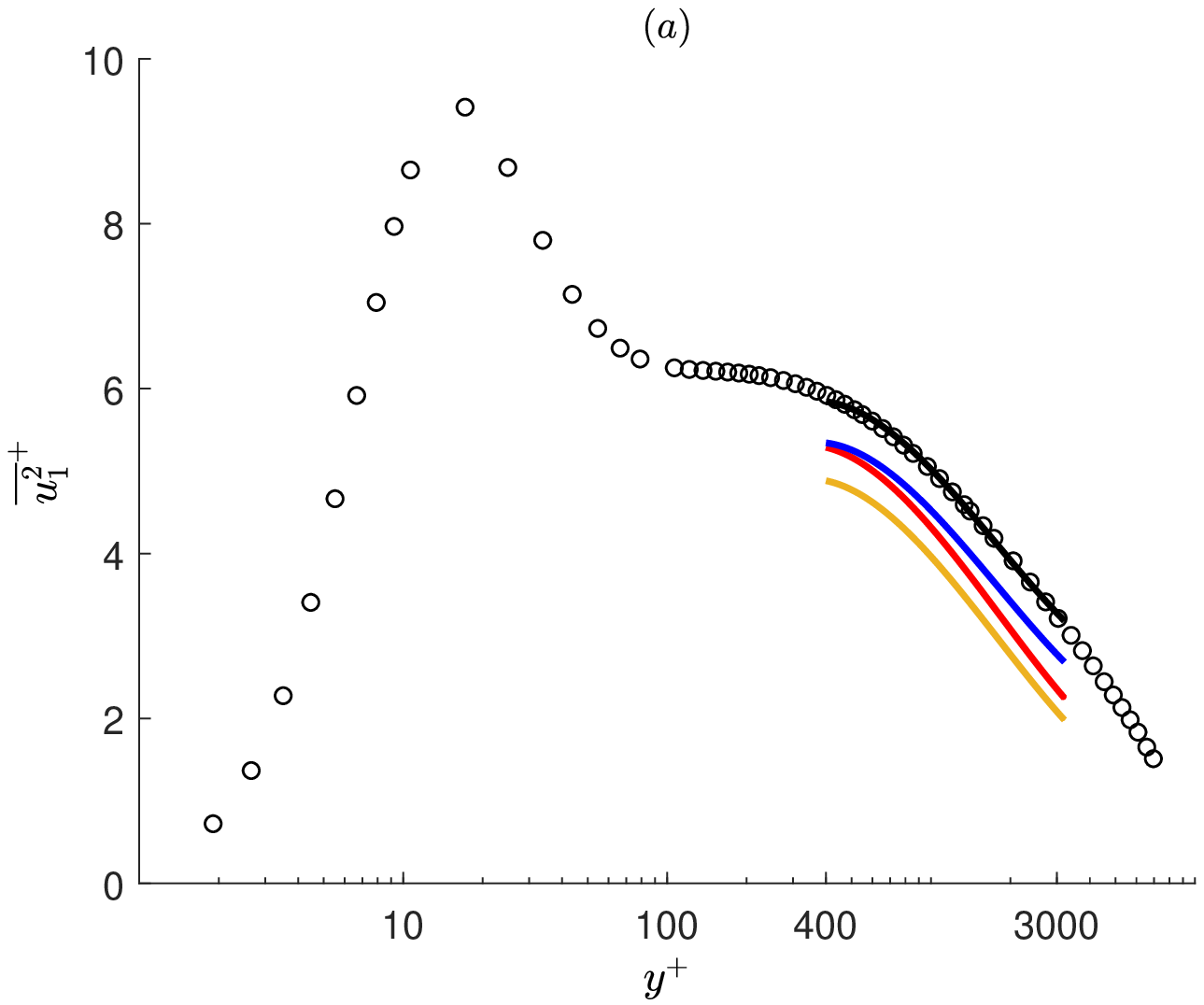}
\end{minipage}
\hfill
\begin{minipage}[c]{.47\linewidth}
\includegraphics[width=.90\textwidth]{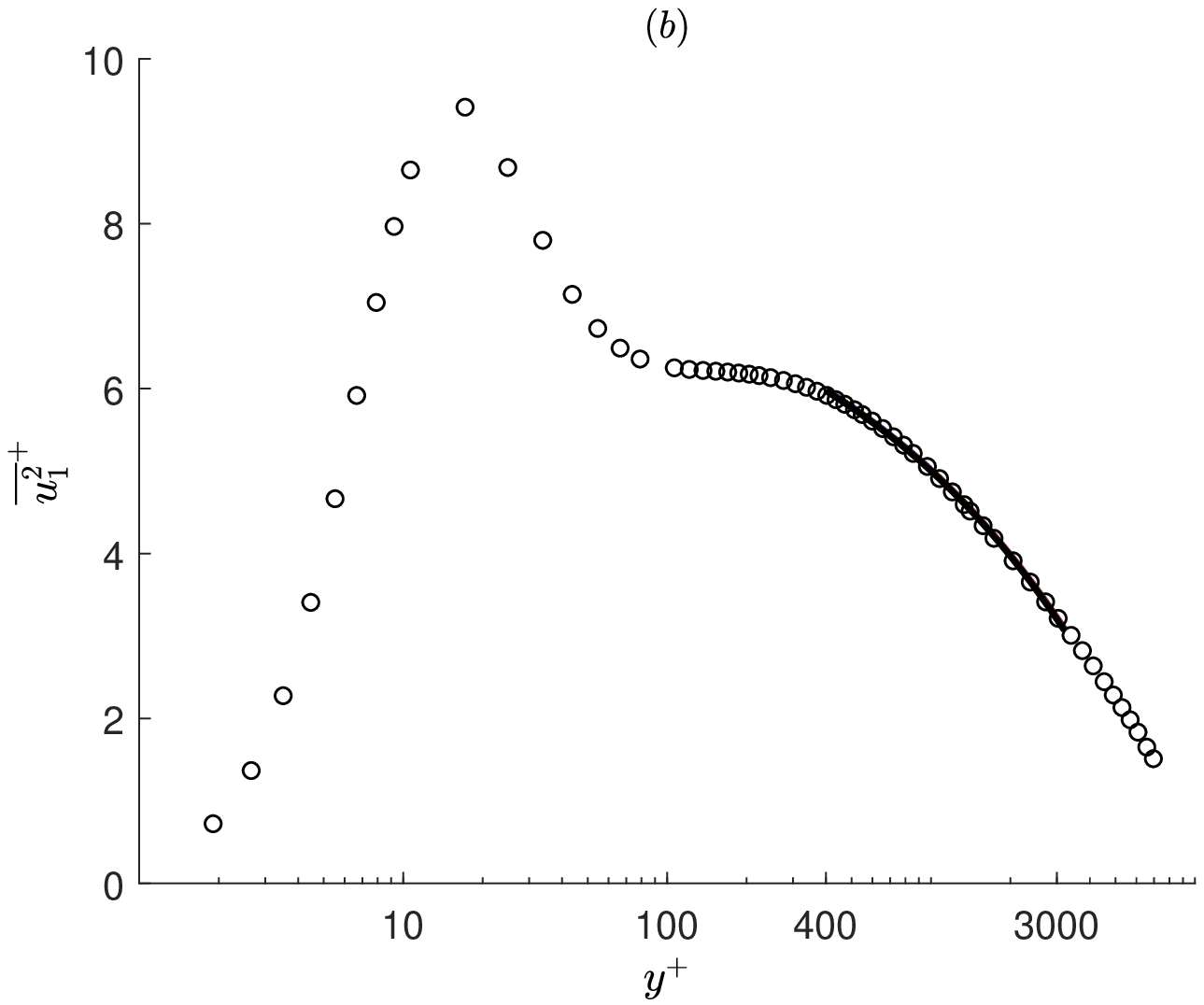}
\end{minipage}
\caption{
Best-fit of $\overline{u_1^2}$ in the log-layer (black solid line), including a tiny $0.1\%$ change in the fitted parameter values (coloured solid lines), once for the structurally inconsistent invariant scaling law \eqref{220425:2101} from~\cite{Oberlack22}, shown in~(a), and once for the consistent invariant scaling law \eqref{220131:1156}, shown in (b). Circles show the DNS data of~\cite{Oberlack22}.\linebreak[4] The values for the black curve in (a) are (to four significant digits): $\omega=0.1087$, $B_2=524.9$, $C_2=482.4$, along\linebreak[4] with the first-moment parameters $\kappa$ and $B$, as given in \cite{Oberlack22}. Changing any second-moment parameter now by~a tiny amount of $0.1\%$, either to $\omega=0.1086$, $B_2=525.5$, $C_2=481.9$, results in each case in a large discrepancy, as shown by the red, blue and yellow curve in (a), respectively. This demonstrates that the fit in this region is highly unstable and thus not natural. In clear contrast to what is shown in (b). There the scaling is stable and~robust. Changing the best-fit values $\gamma=0.2145$, $B_2=7.988$, $C_2=-0.1542$ in (b) by the same amount\linebreak[4]
\vspace{-0.15em}of~$0.1\%$~as in (a), results in complete insensitivity to this change on the plotting scale shown. This unnatural\linebreak[4]
high sensitivity to small perturbations in Oberlack's symmetry-based scaling law to $\overline{u_1^2}$, as shown in (a), is not special to channel flow.
It has also already been demonstrated in turbulent ZPG-flow (see Fig.$\,$4 in \cite{Frewer14.1}).
}
\label{Fig6}
\vspace{-1em}
\end{figure}

\vspace{-0.25em}\noindent
which results directly from Eq.$\,$(16) in \cite{Oberlack22} when reformulated to the equivalent representation of the fluctuation moments,
the values of the shift and normalization parameters, $B_2$ and $C_2$, are both around $400$ to $500$ and therefore not natural. This is particularly expressed
by the fact that the resulting fit is not stable and therefore not robust. Already the smallest change in the parameter values leads to a large discrepancy, as shown in Fig.$\,$\hyperref[Fig6]{6(a)}~above.\footnote{The fitting
procedure in Fig.$\,$\hyperref[Fig6]{6} only focusses on the second moment without searching for an overall best-fit when the next higher-order moment $\overline{u_1^3}$ gets included. This results in higher accuracy, which also allows to extend the fitting region to $400<y^+<3000$, as it was done in Fig.$\,$1(a) in \cite{Oberlack22}, and in figure on slide 60 in \cite{Oberlack22x}, which has also been fact-checked in \cite{Frewer22}. To note is that when combining the best-fit of the second and the first moment, then \eqref{220425:2101} has some advantage over \eqref{220131:1156}, because in the former the set of parameters of the first and second moment do not overlap, while in the latter they do, by sharing the same scaling parameter $\gamma$.
This is not surprising, because \eqref{220425:2101} is based on more invariances than \eqref{220131:1156}, however, the latter with the same number of degrees of freedom as the former due to a forced symmetry breaking in \cite{Oberlack22}. But this drawback can be solved by just searching for more realizable equivalences from the infinite pool of possibilities.
Nevertheless, in Fig.$\,$\hyperref[Fig4]{4(b)} with $\gamma=0.1153$, a combined best-fit of the first two moments is reached, valid even for the third moment, which cannot be reached with the scaling in \cite{Oberlack22}, as shown in Fig.$\,$\hyperref[Fig3]{3(b)}.}
In clear contrast to the insensitive result of the best-fit shown in Fig.$\,$\hyperref[Fig6]{6(b)}, which is based on the structurally consistent invariant scaling law \eqref{220131:1156}.

The unnatural high sensitivity of \eqref{220425:2101} to small perturbations is the consequence of its inconsistent structure, due to being based on two non-realizable and thus nonphysical invariances Eqs.$\,$(8-9) in \cite{Oberlack22}. This inconsistency, which starts at $n=2$, then systematically infects all higher orders with increasing intensity, so much that already at the next higher order, $n=3$, the fluctuation moment $\overline{u_1^3}$ cannot be matched to the data anymore, as shown by the red line in Fig.$\,$\hyperref[Fig3]{3(b)} --- in channel center this mismatch already starts with the Reynolds stress at $n=2$, as shown in Fig.$\,$\hyperref[Fig3]{3(a)}. And since this problem is a structural one and not a convergence issue of the data or a numerical stability issue of the fitting procedure, i.e., since in the space of all norm functions no better fit can be generated than shown in Figs.$\,$\hyperref[Fig3]{3(a)-(b)} and since this problem continues for all higher orders, regardless of how good the DNS data converges, proves that the scaling laws
in~\cite{Oberlack22} are not solutions of the Navier-Stokes equations.

\restoregeometry

\newpage

\label{p22}
\section{On the DNS accuracy and uncertainty of the fluctuation moments\label{SecC}}

One of the tasks in a DNS or experiment is to determine the fluctuation moments, say of the third streamwise velocity moment$\,$\footnote{The brackets $\L\cdot\R$ in \eqref{230108:1054} denote some averaging procedure.}
\begin{equation}
R_3:=\big\L u_1^3\big\R,
\label{230108:1054}
\end{equation}
from the computed or measured random streamwise velocity field $U_1$. This can be done either directly in the DNS or experiment from the random field itself by taking the difference to its mean field $\overline{U}_1$, which basically defines the fluctuation field $u_1$,
\begin{equation}
R_3=\big\L u_1^3\big\R,\quad u_1=U_1-\overline{U}_1,
\label{230108:1153}
\end{equation}
or indirectly, as a post-processing step after the DNS or experiment has concluded, through the relation
\begin{equation}
R_3=H_3-3H_1H_2+2H_1^3,\quad H_k:=\big\L U_1^k\R, \quad H_1:=\overline{U}_1,
\label{230108:1154}
\end{equation}
by using the computed or measured full-field moments up to third order, $H_1$, $H_2$ and $H_3$, the DNS or experiment provides. It is clear that both procedures, \eqref{230108:1153} and \eqref{230108:1154}, yield the very same result for the fluctuation moment $R_3$, if in both procedures the same computational or measuremental precision is used.\footnote{In a DNS the precision is standardly double precision, i.e., the computations are up to 16 decimals precise.} The simple reason is that \eqref{230108:1153} algebraically implies \eqref{230108:1154}, and vice versa. This equality we will also see confirmed in a small experiment later on.

Now, the aim of this whole Sec.$\,$\ref{SecC} is to explicitly show in a small experiment (which will be done in the upcoming subsection) that if the computed or measured full field moments $H_1$, $H_2$ and $H_3$ are accurate, then the fluctuation moment $R_3$ determined via relation \eqref{230108:1154} is reasonably accurate as~well.\linebreak[4] This will refute any claim that the $R_3$ values used to generate Fig.$\,$\hyperref[Fig3]{3(b)},
or Fig.$\,$\hyperref[Fig4]{4(b)} is meaningless, by arguing that relation \eqref{230108:1154} leads to a significant large error or uncertainty when the values of all three summands, $H_3$, $3H_1H_2$ and $2H_1^3$, become large. But, as we prove below, such an argument is simply false here because it is based on wrong assumptions.

In what follows, it is important to carefully distinguish between accuracy and precision. They are two different concepts that need to be kept separate. Precision is how repeatable a series of computations or measurements are, that is, how close the computed or measured values approach each other, while accuracy is how close a computed or measured value approaches an accepted or known value.

Let's discuss the issue of precision for relation \eqref{230108:1154} first: Taking the DNS values from \cite{Oberlack22} at some fixed spatial position, say at $x_2^+\approx 1000$, for the first three full-field moments (to 8 digits precise)
\begin{equation}
H_1\approx 2.2217185\text{e}1,\quad H_2\approx 4.9861022\text{e}2,\quad H_3\approx 1.1297592\text{e}4,
\label{220108:1544}
\end{equation}
we see that we yield significantly different values for $R_3$ from \eqref{230108:1154} when rounding the above\linebreak[4] values~\eqref{220108:1544} down to a too low precision than the DNS provides:
\begin{center}
\begin{tabular}{c|c|c|c|c}
Precision of $H_k$ \eqref{220108:1544} rounded to & 8 digits & 6 digits & 4 digits & 2 digits\\ \hline
Resulting value of $R_3$ from \eqref{230108:1154} & $-2.60$  & $-2.56$  & $4.61$   & $-704$
\end{tabular}
\end{center}
The reason for this strong instability lies of course in the large values of the summands in \eqref{230108:1154}: Values of the order $10^4$ get added and subtracted here, where a sufficient precision is thus necessary to obtain a reasonable result. Hence, when determining $R_3$ from \eqref{230108:1154}, it is always best to use the same precision as the DNS provides.

Now, let's discuss the issue of accuracy for relation \eqref{230108:1154}. Say we have the following relative uncertainties$\,$\footnote{Unfortunately, the DNS database of \cite{Oberlack22} does not provide
the uncertainties $\delta H_i$. Hence,~we guess here some reasonable values. To guide the uncertainties for the first two moments, $\delta H_1$ and $\delta H_2$,
we basically rely on the DNS by M. Lee and R. Moser (2015), taken from their database for the case $\text{Re}_\tau=5200$ (\href{https://turbulence.oden.utexas.edu}{https://turbulence.oden.utexas.edu}).} in the computed values \eqref{220108:1544} for the $H_k$:
\begin{equation}
\delta H_1/H_1=0.05\%,\quad \delta H_2/H_2=0.1\%,\quad \delta H_3/H_3=0.4\%.
\label{221203:1243}
\end{equation}
Then, since \eqref{230108:1154} is a polynomial relation, the uncertainty of the value $R_3\approx -2.6$$\,$\footnote{When using the same precision of 8 digits as in \eqref{220108:1544}, the value of $R_3$ is $-2.6$.} can be determined exactly (no linear approximation is needed here) as
\begin{align}
\delta R_3&=(H_3\pm\delta H_3)-3(H_1\pm\delta H_1)(H_2\pm\delta H_2)+2(H_1\pm\delta H_1)^3-R_3\nonumber\\[0.5em]
&=H_3(1\pm 0.004)-3H_1H_2(1\pm 0.0005)(1\pm 0.001)+2H_1^3(1\pm 0.0005)^3-R_3,
\label{221203:1302}
\end{align}
which, if the $\pm$-signs are given unfavorably (e.g. in the order $++-+$),\footnote{Since $H_1$ appears twice in the relation for $R_3$, there are obviously only $2^3$ independent combinations as how the signs in~\eqref{221203:1302} can be chosen.} leads to an extremely large absolute uncertainty of maximal $\delta R_3\approx 94.7$, where the relative uncertainty then is a whopping
\begin{equation}
\delta R_3/R_3\approx 3650\%.
\label{221203:1346}
\end{equation}
The reason for this large uncertainty lies of course again in the large values of the $H_k$ \eqref{220108:1544}, i.e., despite having small relative uncertainties \eqref{221203:1243}, the uncertainty in $R_3$ \eqref{221203:1302} depends on large terms as $H_3$, $H_1H_2$ and~$H_1^3$, which all are of the order $10^4$. On the other side, however, if the $\pm$-signs are given favorably (e.g. in the order~$+-+-$), then the absolute uncertainty can go as low as $\delta R_3\approx 4.3$, with a minimal relative uncertainty then of
\begin{equation}
\delta R_3/R_3\approx 165\%.
\label{221204:0803}
\end{equation}
So the uncertainty of $R_3$ must lie somewhere between these two extremes
\begin{equation}
165\%\;<\;\delta R_3/R_3\;<\;3650\%,
\label{221204:0813}
\end{equation}
if the prediction of turbulence statistics and its uncertainties were only so simple, but in reality it is~not. The uncertainty \eqref{221204:0813} is a wrong overestimation of the real situation. The mistake is in \eqref{221203:1302}, in trying to determine the uncertainty of $R_3$ deterministically. In turbulence, this approach is wrong and not applicable for the following simple reasons: Firstly, the $H_k$ are stochastic quantities constructed or measured by sampling over time, making it possible that unfavorable uncertainties can cancel over time, where thus extreme unfavorable uncertainties can turn into statistical outliers. Secondly, since the $H_k$ are stochastic quantities, also their uncertainties $\delta H_k$ are stochastic and therefore never certain. And thirdly, and most importantly, the construction or measurement of $H_1$, $H_2$ and $H_3$ are not independent of each other, but depend all on a single field, the full instantaneous streamwise velocity $U:=U_1$, in the following~way:
\begin{equation}
H_1=\big\L U\big\R,\quad H_2=\big\L U^2\big\R,\quad H_3=\big\L U^3\big\R,
\end{equation}
that is, any uncertainty in the construction of $H_1$ will influence the uncertainty when constructing $H_2$ and so~on. Hence, a statistical method has to be used to determine the correct uncertainty of $R_3$, and not a deterministic one that dramatically and falsely overestimates the uncertainty.

Predicting the correct uncertainty of $R_3$ from theory is not straightforward, and we will not pursue it here. Instead, we perform a small and simple experiment that mimics the situation of the problem. It will show that with a confidence level of 95\% the relative uncertainty of $R_3$ will not grow beyond 10\% and with a confidence of even 99\% that it will not grow beyond 12\%. A huge fundamental difference to the purely deterministic prediction~\eqref{221204:0813} that the uncertainty can grow up to 3650\% and that it cannot go below 165\%.

\newpage

\label{p24}
\subsection{An experiment quantifying the uncertainty of \texorpdfstring{$\boldsymbol{R_3}$}{R3}\label{SecC1}}

Let's consider an oversimplified toy-model for the full instantaneous velocity $U$ (at a fixed point in space) as a random variable $x$ with the following continuous probability distribution function~(PDF):
\begin{equation}
F(x):=\frac{1}{3}\cdot\frac{e^{-\frac{1}{2}(x-20.5)^2}}{\sqrt{2\pi}}+\frac{2}{3}\cdot\frac{e^{-\frac{1}{2}(x-23.5)^2}}{\sqrt{2\pi}},
\label{221201:1635}
\end{equation}
being a weighted sum of two normal distributions $\mathcal{N}_1(20.5,1)$, with mean $\mu_1=20.5$ and variance $\sigma_1^2=1$, and, $\mathcal{N}_2(23.5,1)$, with mean $\mu_2=23.5$ and variance $\sigma_2^2=1$. We chose \eqref{221201:1635} for three~reasons: First, to have a simple distribution from which all moments can be determined exactly, second, to have a skewed distribution so that the third central moment (analogous to the fluctuation correlation $R_3$) is not zero, and third, to have for the first three moments (analogous to the full-field correlations $H_k$) more or less the same values with the same magnitude as in \eqref{220108:1544}, in order to have a fair comparison to the values used in the deterministic method~\eqref{221203:1302}. The first three moments of \eqref{221201:1635} are
\begin{equation}
\left.
\begin{aligned}
&H_1=\int_{-\infty}^\infty x\,F(x)\,dx=22.5,\\[0.5em]
&H_2=\int_{-\infty}^\infty x^2F(x)\,dx=509.25,\\[0.5em]
&H_3=\int_{-\infty}^\infty x^3F(x)\,dx=11591.125,
\label{221201:1720}
\end{aligned}
~~~\right\}
\end{equation}
with the third central moment as
\begin{equation}
R_3=\int_{-\infty}^\infty (x-H_1)^3F(x)\,dx=H_3-3H_1H_2+2H_1^3=-2.
\label{221201:1859}
\end{equation}
Now, let's run an experiment where we have $N\gg 1$ samples of the above random variable $x$, which effectively represent at a fixed spatial location the time samples of the full instantaneous velocity $U$ from the DNS in \cite{Oberlack22}. Since the sample size $N$ is finite, we obtain a natural sampling error when determining the moments $H_k$ \eqref{221201:1720} from these $N$ samples. The numerical discretization error of the DNS is not mimicked here because it is usually small (seen at the fast convergence rate and the small error of the first moment) and, if small enough, is not as relevant as the sampling error, which predominately plays the key role in the statistical quality of higher order moments.\footnote{The sampling error is so dominant and prevailing that it even prevents the correct prediction of the discretization error, as discussed e.g.~in \cite{Oliver14}.} Hence, we only focus on the sampling error in the following to determine the uncertainty in the~moments.

If we now run the experiment, say for $N=10^4$, i.e., by generating $N=10^4$ random numbers $X_i$ from the PDF \eqref{221201:1635}, then we obtain, for example, the following realization for the three moments $H_k$ and the central moment $R_3$ (another run, with the same number of samples $N$, will give of course another realization, since the experiment itself is~random)
\begin{equation}
\left.
\begin{aligned}
\!\!\!\!H_1^r=\frac{1}{N}\sum_{i=1}^N X_i\approx 22.5362,\quad\;\; H_2^r=\frac{1}{N}\sum_{i=1}^N X^2_i\approx 510.897,\quad\;\; H_3^r=\frac{1}{N}\sum_{i=1}^N X^3_i\approx 11647.583,\\[0.0em]
\!\!\!\!R_3^r=H^r_3-3H^r_1H^r_2+2H^{r3}_1\approx -2.06,\quad\;\; R_3^{r}=\frac{1}{N}\sum_{i=1}^N (X_i-H_1^r)^3\approx -2.06,\hspace{1.15cm}
\end{aligned}
~~\right\}
\label{221201:2001}
\end{equation}
where it is important to note here that we evaluate the moment $R_3$ in two different ways and both give the very same result, once via the relation $R_3=H_3-3H_1H_2+2H_1^3$, and once directly from the random field $X$ itself.
Therefore, in the end, it does not matter which construction method for $R_3$ one prefers.

When comparing now \eqref{221201:2001} to the exact values \eqref{221201:1720}-\eqref{221201:1859}, the first and most important thing that can be seen from this experiment, already from a single but common realization, is that the uncertainty in the $H_k$  does not dramatically change the uncertainty in the central moment~$R_3$, as the deterministic uncertainty propagation \eqref{221203:1302}-\eqref{221204:0813} would wrongly predict.

\newpage\noindent
Fact is, by explicitly comparing the uncertainty of $R_3$ from this single realization \eqref{221201:2001}, which is
\begin{equation}
\delta R^r_3/R_3=(R_3^r-R_3)/R_3\approx 3\%,
\label{221206:2329}
\end{equation}
with the  deterministic method \eqref{221203:1302}, based here on the uncertainties $H_k$ from \eqref{221201:2001},
\begin{equation}
\delta H^r_1/H_1\approx 0.16\%,\quad \delta H^r_2/H_2\approx 0.32\%,\quad \delta H^r_3/H_3\approx 0.49\%,
\end{equation}
it will give, for this particular realization, the following minimal and maximal uncertainty
\begin{equation}
50\%\;<\;\delta R^r_3/R_3\;<\;11000\%,
\label{221206:2009}
\end{equation}
which obviously is not a realistic estimate. The problem is that result \eqref{221206:2009}, calculated by the method~\eqref{221203:1302}, wrongly assumes that the $H_k$ are constructed independently, but which of course is not~true. It ignores the fact that the sampling errors of $H_k$ in \eqref{221201:2001} are coupled. They are finely tuned quantities constructed over a long sampling interval from a single field $X$, and thus lead to a verifiably meaningful result for $R_3$ \eqref{221201:2001}-\eqref{221206:2329}, and not to the absurd result \eqref{221206:2009}.

To obtain a probability distribution for the values of $R_3$, we now run this experiment, say, for $M=1000$ times (for fixed~$N=10^4$). From the obtained sample set $\{R^r_3\,\}_{r=1,\dotsc,M}$ we then numerically construct the PDF, from which we then can extract additional statistical quantities.

For instance, the standard deviation from the mean $\mu_{R_3}\approx -2$ is $\sigma_{R_3}\approx 0.1$, which gives quite a large relative spread of 5\% from the mean, when compared with the relative spreads of the $H_k$ from their means, which in this experiment are around 0.08\%, 0.16\%, 0.23\%, for $H_1$, $H_2$ and $H_3$, respectively.

Nevertheless, the distribution of $R_3$ gives a 95\% confidence that the uncertainty will {\it not} grow beyond~10\%, and even a 99\% confidence that it will {\it not} grow beyond 12\%, which clearly is in stark contrast to the statement made by the deterministic  method in \eqref{221206:2009}, where the uncertainty in a specific but common realization can grow up to 11000\% and not below 50\%, which certainly is not~true.

Hence, from the above finding we can conclude that in a DNS, if sampled over a sufficiently long time in the statistically stationary regime, the chances are almost certain that the sampled values $H_k$ have favorable uncertainties so that the fluctuation moment $R_3$ can be determined with confidence, either indirectly via $R_3=H_3-3H_1H_2+2H_1^3$, or directly via $R_3=\big\L (U-H_1)^3\big\R$ --- both evaluations will yield the very same result in the end.

\vspace{1.5em}\noindent
\PRLsep

\vspace{1em}\noindent
{\it Note:} The above experiment was performed in Mathematica and the key commands used were:\\[0.75em]
{\small\verb|> dist=MixtureDistribution[{1,2},{NormalDistribution[41/2,1],NormalDistribution[47/2,1]}];|}\\
{\small\verb|> NN=10^4; M=10^3; (* NN: number of samples, M: number of realizations *)|}\\
{\small\verb|> For[r=1,r<=M,r++,|}\\
\phantom{> For[}
{\small\verb|X=RandomVariate[dist,NN];|}\\
\phantom{> For[}
{\small\verb|H1[r]=Total[X]/NN; H2[r]=Total[X^2]/NN; H3[r]=Total[X^3]/NN;|}\\
\phantom{> For[}
{\small\verb|R3[r]=H3[r]-3*H1[r]*H2[r]+2*H1[r]^3;|\\
\phantom{> For[ }
{\small\verb|R3p[r]=Total[(X-H1[r])^3]/NN; (* alternative method to determine R3 *)|\\
\phantom{> For }
{\small\verb|];|\\
{\small\verb|> (* constructing the PDF of R3 *)|\\
{\small\verb|> dataR3=Table[R3[r],{r,1,M}]; distR3=SmoothKernelDistribution[dataR3]; pdfR3=PDF[distR3,z];|\\
{\small\verb|> (* extracting statistical quantities *)|\\
{\small\verb|> mu=Mean[distR3]; sigma=Sqrt[Variance[distR3]];|\\
{\small\verb|> (* constructing the range of 10% error relative to the exact value R3E=-2 *)|\\
{\small\verb|> zL=Solve[(z-R3E)/R3E==0.1,z][[1,1,2]]; zR=Solve[(z-R3E)/R3E==-0.1,z][[1,1,2]];|\\
{\small\verb|> (* computing the probability that the error is within this range *)|\\
{\small\verb|> NProbability[zL<z<zR,Distributed[z,distR3]];|
\normalsize

\newpage

\label{p26}
\section{System of moment equations for statistically stationary flow in channel center\label{SecD}}

The 1-point moment equations in their general form (up to order $n=3$) are given by \eqref{220126:0948}, which will now be reduced to a system of equations for stationary channel flow in the asymptotic limit of channel center ($x_2\rightarrow 0$). Up to the considered moment-order, the {\it asymptotic expansion} in $x_2$ will be consistently of {\it first order} only here. Beginning with the moment equations of first order, i.e. with the moment equations of mass and momentum
\begin{equation}
\partial_k \overline{U}_{\!k}=0, \qquad \partial_t \overline{U}_{\!i}+\partial_k\overline{U_iU_k}=-\partial_i \overline{P},
\end{equation}
in the asymptotic leading-order limit around the center $x_2\rightarrow 0$ of stationary channel flow, they simply reduce (in Reynolds-decomposed form) to
\begin{equation}
\partial_2 \overline{u_1u_2}=-\partial_1\overline{P},\qquad
\partial _2 \overline{u_2^2}=0,\qquad \partial_2\overline{P}=0,
\label{220415:2150}
\end{equation}
where we used the fact that in channel flow the fluctuation moments $\overline{u_1^nu_2^m}$ for all $n\geq 0$ are even functions when $m\geq 0$ is even, and odd functions when $m$ is odd. This implies that in channel center to leading order in $x_2\rightarrow 0$ we have$\,$\footnote{In the asymptotic limit $x_2\rightarrow 0$, it is clear that if $\overline{u_1^nu_2^m}=0$, then this does not imply $\partial_2\overline{u_1^nu_2^m}=0$, and vice versa. For example, in this limit $\overline{u_1u_2}=0$, while $\partial_2\overline{u_1u_2}\neq 0$.}
\begin{equation}
\left.
\begin{aligned}
&\text{$\forall m\geq 0$ even:}\quad \partial_2\overline{u_1^nu_2^m}=0,
\\[0.5em]
&\text{$\forall m\geq 1$ odd:}\quad\; \overline{u_1^nu_2^m}=0,
\end{aligned}
~~~\right\}\quad\forall n\geq 0,\quad n+m\geq 2.
\label{220415:1550}
\end{equation}
The second-order moment equations \eqref{220126:0948}
\begin{equation}
\partial_t\overline{U_iU_j}+\partial_k\overline{U_iU_jU_k}=-\overline{U_i\partial_jP}-\overline{U_j\partial_iP},
\end{equation}
then reduce to$\,$\footnote{The mean-field $\overline{U}_{\!1}$-dependence in \eqref{220415:2151} was eliminated by using the lower-order equations \eqref{220415:2150} along with the condition $\partial_2\overline{U}_{\!1}=0$,
valid in the limit $x_2\rightarrow 0$. This elimination is also done later in \eqref{220415:2153}, by then using \eqref{220415:2151}.}
\begin{equation}
\partial_2\overline{u_1^2u_2}=-2\overline{u_1\partial_1p},\qquad\partial_2\overline{u_2^3}=-2\overline{u_2\partial_2p},\qquad
\partial_2\overline{u_2u_3^2}=-2\overline{u_3\partial_3p},
\label{220415:2151}
\end{equation}
where we used the leading-order limits in $x_2\rightarrow 0$ ($\forall n\geq 0$):
\begin{equation}
\partial_2\overline{U}_{\!1}=0, \qquad \overline{u_1^nu_2^m\partial_k p}=
\begin{cases}
\, 0,\;\; \text{for $k=1$, and $\forall m\geq 1$ odd},
\\
\, 0,\;\; \text{for $k=2$, and $\forall m\geq 0$ even},
\end{cases}
\label{220415:2152}
\end{equation}
based on the same odd-even argument as before in \eqref{220415:1550}, as well as the fact that in stationary channel flow the following moments are {\it globally} zero ($\forall x_2$, $\forall n_1,n_2\geq 0$):
\begin{equation}
\overline{u_1^{n_1}u_2^{n_2}u_3^{n_3}}=0,\;\,\text{$\forall n_3\geq 1$ odd};\qquad
\overline{u_1^{n_1}u_2^{n_2}u_3^{n_3}\partial_k p}=
\begin{cases}
\, 0,\;\; \text{for $k\neq 3$, and $\forall n_3\geq 1$ odd},
\\
\, 0,\;\; \text{for $k=3$, and $\forall n_3\geq 0$ even}.
\end{cases}
\end{equation}
The third-order, and thus the last considered system of moment equations in \eqref{220126:0948}
\begin{equation}
\partial_t\overline{U_iU_jU_k}+\partial_l\overline{U_iU_jU_kU_l}=-\overline{U_iU_j\partial_kP}-\overline{U_jU_k\partial_iP}-\overline{U_kU_i\partial_jP},
\end{equation}
then reduces to
\begin{equation}
\left.
\begin{aligned}
&\partial_2 \overline{u_1^3u_2}=-3\left(\overline{u_1^2}\partial_1\overline{P}+\overline{u_1^2\partial_1p}\right),\qquad
\partial_2\overline{u_2^2u_3^2}=0,\qquad \overline{u_2u_3\partial_3p}=0,\qquad \overline{u_3^2\partial_2p}=0,
\\[0.5em]
&\partial_2\overline{u_1u_2u_3^2}=-\overline{u_3^2}\partial_1\overline{P}-2\overline{u_1u_3\partial_3p}-\overline{u_3^2\partial_1p},
\qquad
\partial_2\overline{u_1u_2^3}=-\overline{u_2^2}\partial_1\overline{P}-2\overline{u_1u_2\partial_2p}-\overline{u_2^2\partial_1p}.
\end{aligned}
~~~\right\}
\label{220415:2153}
\end{equation}
Important to note here again is that the reduced equations \eqref{220415:2150}, \eqref{220415:2151}, \eqref{220415:2152}  and \eqref{220415:2153} are not globally valid equations in $x_2$, but local equations valid only in the leading-order limit $x_2\rightarrow 0$.

\newpage

\label{p27}
\section{The non-existent universal log-law in \cite{Avsarkisov14}\label{SecE}}

Based on our investigation \cite{Frewer16.2} concerning the reproducibility of Fig.$\,$9(a) in \cite{Avsarkisov14}, a key figure which should prove the sensational claim of a universal log-law in turbulent channel flow with wall transpiration, the authors provided a correction in \cite{Avsarkisov21} upon request of the journal. However, this Corrigendum is still flawed because the corrected figure shown is based on erroneous data. There is simply
no~universal log-law for different transpiration rates as claimed, and, as we will prove below, there is even a simple physical explanation~for~it.

To have an orientation of the situation already at this point, here a brief summary of the facts when comparing the results of the Corrigendum with those of the original article:
\begin{itemize}
\item
The original article \cite{Avsarkisov14}: The data that was used therein was self-consistent and did not lead to any contradictions. The main problem therein, however, Fig.$\,$9(a) still can not be reproduced with this data.
\item
The Corrigendum \cite{Avsarkisov21}: Fig.$\,$1(a) can indeed be reproduced with the new data provided. The problem now, however, is that the new data is not self-consistent anymore and leads to fundamental contradictions (see below). All plots shown in Fig.$\,$1 are therefore not valid, particularly the approximate collapse of curves as shown in Fig.$\,$1(a) can not be observed when using consistent data. What the authors want to suggest with this figure is not true and also does not~exist.
\end{itemize}

\noindent
The proof that the new data set in \cite{Avsarkisov21} is inconsistent and contradictive is based on the definition of the mean bulk velocity $U_B$, which is given explicitly by Eq.$\,$(2.4) in the original
article \cite{Avsarkisov14}. Therein it is defined as a universal constant for all Reynolds numbers and transpiration rates and has the fixed value $U_B=0.89$. This universal feature of $U_B$ is again confirmed in the Corrigendum (see the caption of Fig.$\,$1, where $U_B=0.89$ for all configurations).

The data used in the original article \cite{Avsarkisov14} is consistent with the definition of $U_B$. We checked this relation Eq.$\,$(2.4) for all considered cases of different Reynolds numbers $\text{Re}_\tau$ and transpiration rates~$v_0^+$. Indeed, when evaluating the integral Eq.$\,$(2.4) by using the original data from 2014, we get the value $U_B=0.89$. We also performed a different but mathematically equivalent test, by evaluating the integral of $(\bar{U}-U_B)/u_\tau$ which should give zero, and indeed it gives zero (up to numerical accuracy). So, in the original article \cite{Avsarkisov14} the data is self-consistent.

In the Corrigendum \cite{Avsarkisov21}, however, this is not the case anymore. All plots shown in Fig.$\,$1 are not consistent to Eq.$\,$(2.4) anymore and lead to fundamental contradictions. The integral
of $(\bar{U}-U_B)/u_\tau$ for the new data set is not zero. For example, for the cases shown in Fig.$\,$1(c)-(d) we get exactly $-1$ (up to numerical accuracy), which is wrong, but not a surprising result, because, as claimed in the Corrigendum, the original data was shifted in wall-normal direction by $-1$:
\begin{center}
{\it ``The [old] plots in figure 9(a-d) were erroneously shifted vertically by $1$."} \cite{Avsarkisov21}
\end{center}
However, this was and is not the case, because their old original plots in \cite{Avsarkisov14} were and still are fully correct in this regard. Hence, the Corrigendum is even worse than the original publication, as consistent DNS data has now been changed with this shift into inconsistent data.

From our point of view, the modus operandi is clear: The authors merely made a change to bring their data in line with the incorrect symmetry-based scaling law of Eq.$\,$(1).
Instead of changing or adapting the scaling law, they modified the correct DNS data by shifting everything in the wall-normal direction by $-1$. In our opinion, this demonstrates the blind faith in their obtained symmetry-based scaling laws: If the symmetry-based scaling law cannot be matched to the data, then in their reasoning the experiment or the numerical simulation must be wrong and need to be changed.

Due to the Corrigendum we are thus dealing here with two different data sets. One that is self-consistent (the original data from 2014), and one that is not (the new data from 2021). However, in both cases the authors present a plot where the curves for a fixed Reynolds number with different transpiration rates collapse into a single curve: Fig.$\,$9(a) in the original article and Fig.$\,$1(a) in the Corrigendum, based on the original and modified data, respectively. This immediately raises the question, how is that possible?

The problem clearly lies with Fig.$\,$9(a) in \cite{Avsarkisov14}, since it can neither be reproduced with the original 2014\hspace{0.4mm}-data, nor with the new 2021-data. With the 2014\hspace{0.4mm}-data, the curves do not collapse into a single curve, as incorrectly shown in Fig.$\,$9(a) and correctly shown here in Fig.$\,$\hyperref[Fig7]{7(b)}, and with the 2021-data, the curves are not only shifted downward by $1$, but also do not collapse as perfectly in channel center as shown in Fig.$\,$9(a). So, the question is: How did the authors manage to create this perfect Fig.$\,$9(a) in their original article? What was the mathematical rule behind it? How was the data changed? It's surely not the way as described in the Corrigendum, because, as already said, Fig.$\,$1(a) is strikingly different to Fig.$\,$9(a), not only in the vertical shift, but also in the quality of the~collapse.

\begin{figure}[t!]
\centering
\begin{minipage}[c]{.49\linewidth}
\vspace{-1em}
\includegraphics[width=.90\textwidth]{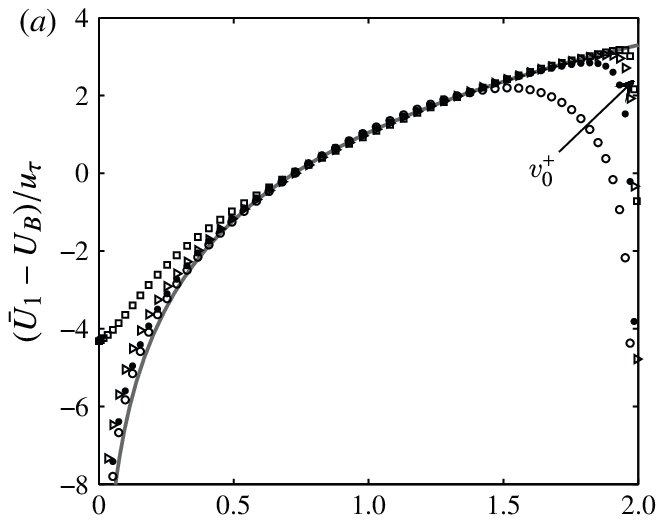}
\end{minipage}
\hfill
\begin{minipage}[c]{.49\linewidth}
\includegraphics[width=.90\textwidth]{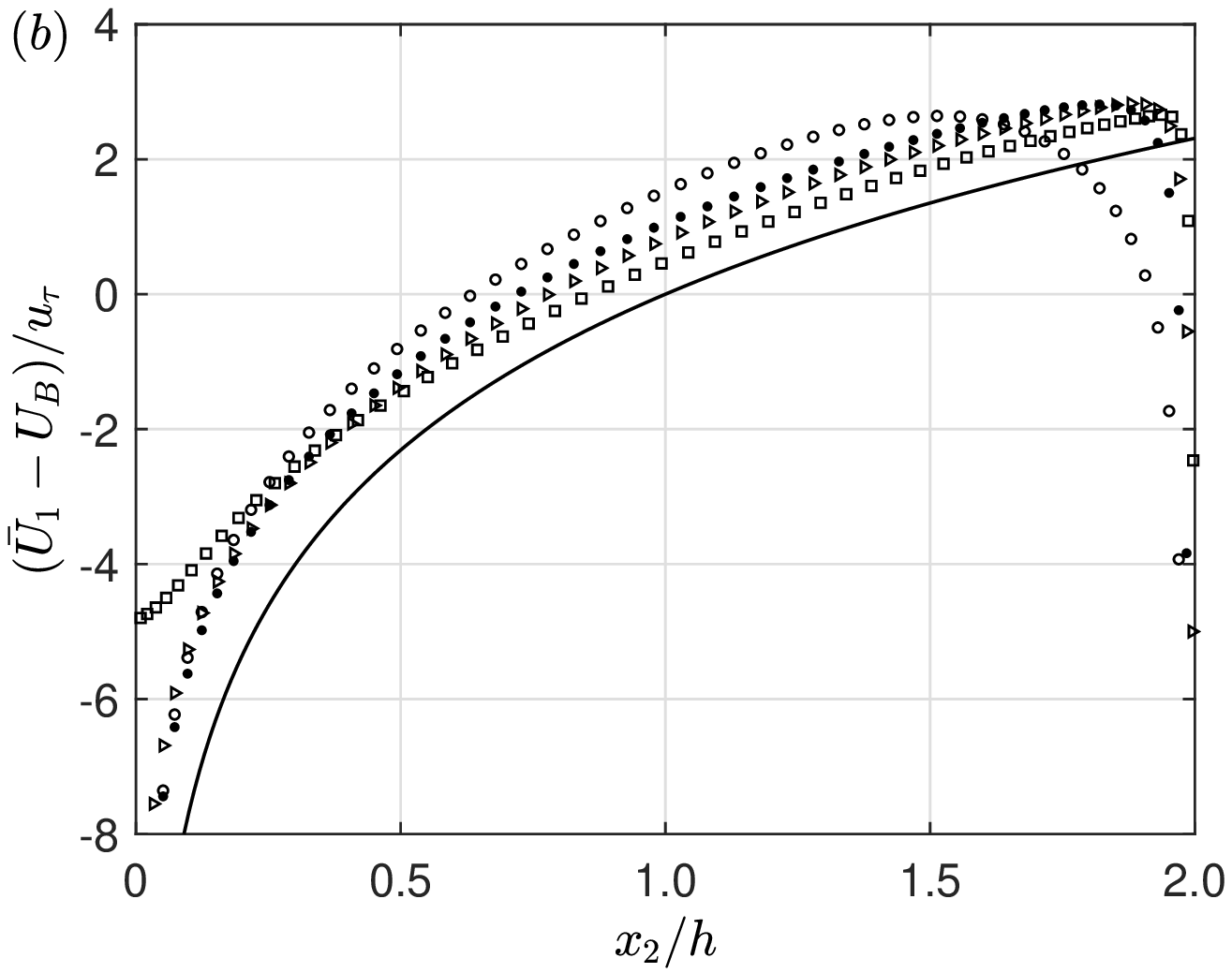}
\end{minipage}
\caption{
For constant Reynolds number $Re_\tau=480$, the mean velocity profile in deficit form is displayed for different transpiration rates:
$v_0^+=0.05$ ($\circ$);$\:$ $v_0^+=0.10$~($\bullet$);$\:$ $v_0^+=0.16$ ($\,\triangleright$);$\:$ $v_0^+=0.26$ (${\scriptstyle \square}$).
The left plot~(a) is an exact copy of Fig.$\,$9(a) from \cite{Avsarkisov14} (used with permission of JFM), while the right plot (b) is taken from our comment~\cite{Frewer16.2}, displaying the correct version
of Fig.$\,$9(a), as it should be, when taking their consistent DNS data~(from 2014) and plot it correctly.
In both plots, the solid line represents the ``new logarithmic scaling~law" of \cite{Avsarkisov14}, also presented as Eq.$\,$(1) in \cite{Avsarkisov21}, as the so-called universal log-law
$(\bar{U}_1-U_B)/u_\tau=(1/\gamma)\ln(x_2/h)$, containing only
a single free parameter $\gamma$, and being even globally constant, i.e., independent of both Reynolds number and transpiration rate. But, as figure (b) reveals, there can be no such universal law, and there is a simple
explanation for it (see text).
}
\label{Fig7}
\vspace{-1em}
\end{figure}

As the final step in our proof, we now show from a physical point of view why the curves for a fixed Reynolds number and varying transpiration rate cannot collapse into a single one for the specific
normalization $(\bar{U}-U_B)/u_\tau$, thus showing that a universal log-law cannot exist.

Fact is, the proposed normalization $(\bar{U}-U_B)/u_\tau$ in \cite{Avsarkisov14} only works for Fig.$\,$9(c), but not for Fig.$\,$9(a). The reason is that the curves in Fig.$\,$9(a) are driven by a different physical process than those in Fig.$\,$9(c).
The curves in Fig.$\,$9(c) show the mean velocity profiles at a constant transpiration rate but with different Reynolds numbers. There the normalization $(\bar{U}-U_B)/u_\tau$ makes sense and indeed leads to an approximate collapse of the curves. The curves for Fig.$\,$9(a), however, represent the mean velocity profiles at a fixed Reynolds number but with different transpiration rates. Obviously, the physical process in this case is completely different, however, the very same normalization $(\bar{U}-U_B)/u_\tau$ is used. And since this normalization is independent of the transpiration rate $v_0^+$, there is no reason for why the curves should collapse in this case. Hence, as correctly shown by us in Fig.$\,$\hyperref[Fig7]{7(b)}, a collapse of curves and with it a universal log-law does not exist.

The last point we would like to make on this matter is that the authors' acknowledgement in \cite{Avsarkisov21} is misleading and sarcastic, as it gives the impression that we have taken note of their Corrigendum and therefore agreed with its content before publication. But this is not the case. Although we initiated this Corrigendum, neither the authors nor the editor gave us the chance to see the Corrigendum before it was accepted for publication. For then we could have raised our objection in time and a revision could have taken place. Now the Corrigendum is a publication that is worse than the original article. Moreover, our names are now associated with an article which is clearly wrong and whose content we reject. For the journal, the case is closed.

\newpage

\label{p29}
\section{The importance for a correction, or if necessary, for a retraction of \cite{Oberlack01}\label{SecF}}

Our comment \cite{Frewer14.2} from 2014 was the first to prove that the paper {\it ``A unified approach for symmetries in plane parallel turbulent shear flows"}, by M. Oberlack (2001) \cite{Oberlack01}, is seriously flawed.
The Lie-group based symmetry approach to turbulence, as first initiated, developed and presented therein, and used ever since by Oberlack, is not only methodologically but also technically flawed. Moreover, besides several serious errors, it also contains a non-reproducible result.

Unfortunately, we cannot publish our comment in Journal of Fluid Mechanics (JFM), where Oberlack's work was published back in 2001, because of JFM's publication policy, which does not allow for Comments. We also tried to force a Corrigendum from M. Oberlack through the journal, once in 2014 and once recently because of a new proof that does not require any knowledge in Lie-group theory, but both times the editor declined, with the final argument that the paper is old and not worth to be corrected. We see it differently, because, as can be seen with the current paper \cite{Oberlack22}, the community is again confronted with a highly misleading paper based on this 2001 work, a work from which all past and current problems emanate and therefore requires an official correction or a retraction in order to avert even further damage.

In the following we will present the new proof we shared with the journal. Knowledge in Lie-group symmetry analysis is not required to follow and understand this proof. An alternative proof, which requires some knowledge in Lie's symmetry method, is also provided, taken from \cite{Frewer14.2} to obtain a complete picture.

\label{p29}
\subsection{Introduction\label{SecF1}}

The study by Oberlack (2001) \cite{Oberlack01} is an influential paper that is still cited as a successful study today.
However, the key analytical result in \cite{Oberlack01}, on which the overall conclusion of the paper is based, is in error and cannot be reproduced. A rigourous proof for this claim is given in Sec.$\,$\ref{SecF2}. Knowledge in Lie-group symmetry analysis is not required to follow and understand this proof. An alternative proof, which requires some knowledge in Lie's symmetry method, is given in Sec.$\,$\ref{SecF4}, summarizing the full proof in our comment \cite{Frewer14.2}, which is also assisted by using a computer algebra system~(CAS) to validate on the results given by several different third-party symmetry software~packages.

What stands out here is not only the mere fact that the author has made a severe technical mistake which, when corrected, affects the overall conclusion of his study, but the fact that the author makes a clear statement about his result, in how he obtained it by performing a specific mathematical step, but which ultimately can not be realized, namely that from equation (3.15) the term \emph{``$\mathscr{N}_iu_j+\mathscr{N}_ju_i$ may be factored out''} [p.$\,$308] to get to the central result (3.16).

Important to realize here is that even if the author can provide a detailed mathematical explanation\linebreak[4] in how he managed to obtain the key crucial result (3.16) by just `factoring out' the specific term\linebreak[4]
$\mathscr{N}_iu_j+\mathscr{N}_ju_i$ from (3.15), the result (3.16) is still in error.

In our opinion, the paper \cite{Oberlack01} should be retracted. The reason is that result (3.16) is key to~the\linebreak[4] overall conclusion of the paper. Correcting this result not only changes the overall conclusion to the~opposite as to what is claimed, it also turns numerous statements throughout the paper into false\linebreak[4] statements, as shown and described in detail in Sec.$\,$\ref{SecF5}. Therefore, this paper is uncorrectable and should be retracted according to the standards of the Publication Ethics Committee (COPE)\href{https://publicationethics.org/retraction-guidelines}{$\;$\Large\ExternalLink}, on~the grounds that the results and conclusion of the paper cannot be relied upon due to a major error.

Independent of this major error, the paper faces a second major technical error, namely when going from (3.16) to (3.17). This is proven and discussed separately in Sec.$\,$\ref{SecF3}, which should make it clear that this error is really independent of the error discussed before in Sec.$\,$\ref{SecF2}. Also, both errors do not cancel each other, but only amplify the error in the final result (3.19).

Critical and important to know here is that the paper's central and final result (3.19), which directly results from the two errors in (3.16) and (3.17), represents an incomplete result that is highly misleading and which already caused a great deal of damage in the turbulence community, and still continues to do so. The problem with (3.19) is that it gives the misleading impression that the Lie-group\linebreak[4] symmetry method is able to generate first-principle solutions for the mean velocity field $\bar{u}_1$ without making any prior assumptions, but which in reality is not the case.

Because when correcting the two above mentioned derivation errors leading to (3.19), it leads to a fundamentally more general result in the symmetry solution $\eta_{\bar{u}_1}$ for the mean velocity than what is claimed in (3.19). The correct solution for $\eta_{\bar{u}_1}$ was carefully derived in \cite{Frewer14.2}, and is shown therein on the right-hand side of Table~1 [p.$\,$7], while on its left-hand side (3.19) is shown.

The difference between both solutions is striking: The left-hand side shows an algebraically reduced result for $\eta_{\bar{u}_1}$, which gives the misleading impression that is was possible to circumvent the closure problem of turbulence since a closed expression in the shear-normal coordinate $x_2$, the only independent coordinate for the mean velocity, has been derived, while the right-hand side shows the complete opposite: It shows the correct solution, obtained when correctly performing the Lie-group  analysis, with the result that it picks up an arbitrary function $F(x_2)$, which again shows that the set of symmetries is thus not closed --- because this arbitrary function can now be used to generate {\it any} desirable invariant scaling law for the mean velocity profile $\bar{u}_1(x_2)$, and not only those specific few mentioned in \cite{Oberlack01}.

The same problem we also face with the invariant scaling solutions of the two-point velocity correlations in \cite{Oberlack01}. This is discussed in the second point of Sec.$\,$\ref{SecF4}, which shows that also the analysis for the two-point correlations in \cite{Oberlack01} is flawed, simply due to the overall fact that the Lie-group symmetry method has been misapplied by the author.
Its correct symmetry solution is given by (A.8) in \cite{Frewer14.2}.

{\it Conclusion:} The main message here is that the Lie-group method alone, like any other analytical method, cannot bypass the closure problem of turbulence. When applied correctly, the Lie-group method only shifts the closure problem from the equations to its admitted set of symmetries. Without modelling, nothing is gained here by using the Lie-group method. For a more detailed discussion, see~Sec.$\,$\ref{SecF5} and \cite{Frewer14.2}.

\label{p30}
\subsection{Proof that result (3.16) in \cite{Oberlack01} is in error and cannot be reproduced\label{SecF2}}

The following layout will prove that the analytical result (3.16) cannot be derived from equation~(3.15) as claimed in \cite{Oberlack01}. Crucial for the proof is to know and understand what the individual symbols in (3.15) and (3.16) stand for:
\begin{itemize}
\item[$u_i$\hspace{0.3mm}:]
Components of the fluctuating velocity field, which globally$\,$\footnote{Of course, locally for certain space-time points the fluctuating velocity components $u_i$ can map to zero, but not globally for {\it all} space-time points.}$\,$are non-zero fields, $u_i\not\equiv 0$.
\item[$ \mathscr{N}_i$\hspace{0.3mm}:]
All collected terms that form componentwise the fluctuating Navier-Stokes momentum equations; see (2.11). Thus, when putting $\mathscr{N}_i$ to zero, they directly represent these Navier-Stokes equations, i.e., $\mathscr{N}_i=0$ ultimately represents a differential equation.
\item[$X$\hspace{0.3mm}:]
A scalar differential operator which acts on functions and their derivatives, see (3.11).\footnote{The prolonged part $X_p$ up to the second derivative order is shown explicitly by (2.6) in \cite{Frewer14.2}.}
\item[$\eta_{u_i}$\hspace{0.3mm}:]
Functions that represent componentwise the infinitesimal generator for the symmetries of the fluctuating velocity field; see (3.14) for a possible realization.
\end{itemize}
Now, in \cite{Oberlack01} it is said that equation (3.15) is the direct result of (3.12c) by {\it ``carrying~out the\linebreak[4] differentiations in (3.12c)''} [p.$\,$307]. When carrying out this step by applying the product rule of differentiation for the differential operator $X$ in (3.12c), we obtain the correct result
\begin{align}
0&\underset{(3.12c)}{=}X\Big(\mathscr{N}_iu_j+\mathscr{N}_ju_i\Big)\Big\vert_{(\mathscr{N}_iu_j+\mathscr{N}_ju_i)=0}\nonumber\\[0.5em]
&\underset{\phantom{(3.12c)}}{=}\Big(\mathscr{N}_iXu_j+\mathscr{N}_jXu_i+u_jX\mathscr{N}_i+u_iX\mathscr{N}_j\Big)\Big\vert_{(\mathscr{N}_iu_j+\mathscr{N}_ju_i)=0},
\label{210123:1725}
\end{align}

\pagebreak[4]\noindent
which is identical to (3.15), up to the preassigned evaluation constraint which got suppressed in (3.15). As we will see now, it is important not to suppress this constraint here because it leads to information
which can be used to show that \eqref{210123:1725} is not an equation but a null identity. Indeed, by having a closer look at this evaluation constraint
\begin{equation}
\mathscr{N}_iu_j+\mathscr{N}_ju_i=0,
\label{210123:1834}
\end{equation}
which in \cite{Oberlack01} is also defined as the {\it ``instantaneous velocity product \underline{equation}"} (2.12), it represents nothing less than a differential equation that can be solved.

Now, since equation \eqref{210123:1834}, as set by its own construction rule on p.$\,$302 in \cite{Oberlack01}, has to be valid for \underline{\textbf{all}} possible flow configurations $u_i$, which again have to satisfy the underlying fluctuating Navier-Stokes momentum equation $\mathscr{N}_i=0$, it consistently follows that the general solution of equation~\eqref{210123:1834} can only be
\begin{equation}
\mathscr{N}_i=0, \qquad \forall u_i,\;\: \forall i,
\label{210123:1923}
\end{equation}
which also is clearly seen when equivalently rewriting \eqref{210123:1834} as
\begin{equation}
\mathscr{N}_j=-\frac{u_j}{u_i}\cdot\mathscr{N}_i,\qquad \forall u_i,\;\: \forall i.
\end{equation}
That $\mathscr{N}_i=0$ \eqref{210123:1923} is the only possible solution to \eqref{210123:1834}, is consistent with the fact that the very existence of equation \eqref{210123:1834} is only due to the underlying existence of
$\mathscr{N}_i=0$ itself, because \eqref{210123:1834} originally resulted from equation $\mathscr{N}_i=0$ in the first place. Hence, equation \eqref{210123:1834} can be identified as the equivalence$\,$\footnote{The equivalence \eqref{210124:1711} can also be read as that the solution space of  $\mathscr{N}_iu_j+\mathscr{N}_ju_i=0$ is identical to that of $\mathscr{N}_i=0$.}
\begin{equation}
\mathscr{N}_iu_j+\mathscr{N}_ju_i=0\quad\Leftrightarrow\quad \mathscr{N}_i=0,\qquad  \forall u_i,\;\: \forall i,
\label{210124:1711}
\end{equation}
and therefore relation \eqref{210123:1725} can be equivalently written as
\begin{equation}
\Big(\mathscr{N}_iXu_j+\mathscr{N}_jXu_i+u_jX\mathscr{N}_i+u_iX\mathscr{N}_j\Big)\Big\vert_{\mathscr{N}_i=0}=0,\qquad \forall u_i,\;\: \forall i.
\end{equation}
When evaluating the above relation,
\begin{equation}
\underbrace{\mathscr{N}_i\Big\vert_{\mathscr{N}_i=0,\,\forall i}}_{=0}\cdot\; Xu_j\;+\;\underbrace{\mathscr{N}_j\Big\vert_{\mathscr{N}_i=0,\,\forall i}}_{=0}\cdot\; Xu_i
\quad +\quad
u_j\cdot \underbrace{X\mathscr{N}_i\Big\vert_{\mathscr{N}_i=0,\,\forall i}}_{\underset{(3.12b)}{=}0}\;+\;
u_i\cdot \underbrace{X\mathscr{N}_j\Big\vert_{\mathscr{N}_i=0,\,\forall i}}_{\underset{(3.12b)}{=}0}\;\; =\;\;0,
\label{210123:2024}
\end{equation}
we see that the first two terms do not contribute, while the last two terms vanish because of the constraint equation given by (3.12b). Hence, in reality, equation \eqref{210123:1725} is not an equation but just a null identity
$0=0$, which directly transfers to (3.15) and it's prior results (3.12b-c), when based on the constraint (2.12) as given and constructed in \cite{Oberlack01}.\footnote{Or, put in other words: If (3.15) is seen as an equation, then, along with (2.12), it is fully redundant to (3.12b), i.e., it does not provide any new information than what (3.12b) already gave --- for more details on this redundancy issue, please see \cite{Frewer14.2}. Therein we even performed a full and complete symmetry analysis (assisted by~CAS), which offers a further independent proof of this redundancy.}

But the above null-result \eqref{210123:2024} is clearly at odds with the non-null result (3.16) given in \cite{Oberlack01}. Important to recognize here is the explanation by the author how this result (3.16) from (3.15) was obtained. It is said that

\vspace{1em}
\emph{``The last two terms [of (3.15)] do not contribute to the constraints for the infinitesimals since\linebreak[4]
\phantom{x}\hspace{0.6cm}$\mathscr{N}_iu_j+\mathscr{N}_ju_i$ may be factored out. Due to (2.12) this term cancels''} [p.$\,$308].

\vspace{1em}\noindent
It is exactly the content of this sentence which cannot be reproduced. In fact, there is no possibility that $\mathscr{N}_iu_j+\mathscr{N}_ju_i$ can be factored out from the last two terms in (3.15). The author should provide a clear mathematical explanation how he managed to do so.

In addition it should be noted that it's not true that the last two terms do not contribute. As already shown above in \eqref{210123:2024}, it is exactly the opposite: It are the first two terms in (3.15) which do not contribute when applying (2.12), and not the last two. This can be clearly seen again also from a different perspective, by choosing, for example, the simple (diagonal) specification $i=1$ and $j=1$ in~(3.15), which then simplifies to:
\begin{equation}
2\mathscr{N}_1\cdot Xu_1+2u_1\cdot X\mathscr{N}_1=0.
\label{210123:2141}
\end{equation}
Now, when using equation (2.12) as done and explained by the author, which for the chosen specification $i=1$, $j=1$ then reads
\begin{equation}
2\mathscr{N}_1u_1=0,
\end{equation}
and which then again is equivalent to
\begin{equation}
\mathscr{N}_1u_1=0\quad\Leftrightarrow\quad \mathscr{N}_1=0,\qquad  \forall u_1\not\equiv 0,
\end{equation}
we see that due to this result $\mathscr{N}_1=0$, it are the first two (here additively combined) terms in \eqref{210123:2141} which do not contribute, and not the last terms as falsely claimed, as these are still bound to the differential operator $X$ and therefore cannot be directly evaluated through the constraint $\mathscr{N}_1=0$. Hence, for the specification $i=1$, $j=1$, the constraint equation (2.12) reduces relation (3.15) to the form
\begin{equation}
2u_1\cdot X\mathscr{N}_1 \Big\vert_{\mathscr{N}_1=0}=0,\qquad  \forall u_1\not\equiv 0,
\end{equation}
which is equivalent to
\begin{equation}
X\mathscr{N}_1 \Big\vert_{\mathscr{N}_1=0}=0,
\end{equation}
being thus finally identical to result (3.12b), and also consistent again to the result \eqref{210123:2024} as it was derived above. This demonstration just shows again that (3.15), along with (2.12), is fully redundant to (3.12b).\footnote{Or, equivalently: (3.12c) is fully redundant to (3.12b), since obviously (3.15)$\,$\&$\,$(2.12) is equivalent to (3.12c).} It does not lead to any new information as it's falsely proposed through (3.16).

\label{p32}
\subsection{Proof that result (3.17) in \cite{Oberlack01} is in error\label{SecF3}}

The following layout will prove that if we assume (3.16) to be a correct result, then equation (3.17) cannot be derived from (3.16). In other words, in this proof we will ignore all previous findings
of Sec.$\,$\ref{SecF2} above and pretend that the result (3.16) is correct, exactly just as the author did in \cite{Oberlack01} to get from (3.16) to (3.17). But, as will be shown now, equation (3.17) does not follow from~(3.16). We start with (3.16) and then follow the approach as it's described below (3.16) in \cite{Oberlack01}, namely that

\vspace{1em}
\emph{``The~term $\mathscr{N}_iu_j+\mathscr{N}_ju_i$ may again be separated out from (3.15) [sic].\footnote{Typo mistake: Instead of (3.15) it should read (3.16).}\! Since this term also cancels\linebreak[4]
\phantom{x}\hspace{0.6cm}out due to~(2.12) we obtain the remaining restrictions''} [p.$\,$308].

\vspace{1em}\noindent
Thus the first step is to separate out $\mathscr{N}_iu_j+\mathscr{N}_ju_i$ from (3.16), which indeed is a true observation that can be done:
Inserting into (3.16) the result for $\eta_{u_i}$, which is correctly given by (3.14) as$\,$\footnote{For the sake of better readability, we suppress all dependencies in the solution functions of $\eta_{u_i}$.}
\begin{equation}
\eta_{u_i}=[a_1-a_4]u_i+\Big(a_2u_3+\dot{f}_1-g_1\Big)\delta_{i1}+\Big(-a_2[u_1+\bar{u}_1]+\dot{f}_2\Big)\delta_{i3},
\end{equation}
equation (3.16) can be reformulated as
\begin{align}
0\;=&\;\;\mathscr{N}_i\eta_{u_j}+\mathscr{N}_j\eta_{u_i}\nonumber\\[0.5em]
\;=&\;\;[a_1-a_4]\Big(\mathscr{N}_iu_j+\mathscr{N}_ju_i\Big)\nonumber\\[0.5em]
&\hspace{0.71cm}
+\,\Big(a_2u_3+\dot{f}_1-g_1\Big)\Big(\mathscr{N}_i\delta_{j1}+\mathscr{N}_j\delta_{i1}\Big)
\,+\,\Big(-a_2[u_1+\bar{u}_1]+\dot{f}_2\Big)\Big(\mathscr{N}_i\delta_{j3}+\mathscr{N}_j\delta_{i3}\Big),
\label{210124:1630}
\end{align}
where $\mathscr{N}_iu_j+\mathscr{N}_ju_i$ has been separated out as the first term. The second step now is to make use of the constraint (2.12), namely that $\mathscr{N}_iu_j+\mathscr{N}_ju_i$ is zero. This then sets the first term in \eqref{210124:1630} to zero, reducing this equation and therefore (3.16) thus to
\begin{equation}
\Big(a_2u_3+\dot{f}_1-g_1\Big)\Big(\mathscr{N}_i\delta_{j1}+\mathscr{N}_j\delta_{i1}\Big)
\,+\,\Big(-a_2[u_1+\bar{u}_1]+\dot{f}_2\Big)\Big(\mathscr{N}_i\delta_{j3}+\mathscr{N}_j\delta_{i3}\Big)=0.
\label{210124:1640}
\end{equation}
The final argument in \cite{Oberlack01}, then, is that this equation \eqref{210124:1640} can only be fulfilled if the terms in the first and third bracket add to zero, i.e., if
\begin{equation}
\left.
\begin{aligned}
a_2u_3+\dot{f}_1-g_1=0,\\[0.5em]
-a_2[u_1+\bar{u}_1]+\dot{f}_2=0,
\end{aligned}
~~~\right\}
\label{210124:1652}
\end{equation}
which exactly is result (3.17). But such a deduction is not valid. Equation \eqref{210124:1640} does not necessarily imply \eqref{210124:1652}, because the second and fourth bracket can also become zero, which, in fact, are exactly zero by construction!

Because, as already explained in the case of Sec.$\,$\ref{SecF2}, when applying the constraint equation~(2.12),\linebreak[4] i.e., $\mathscr{N}_iu_j+\mathscr{N}_ju_i=0$, it's equivalent to applying $\mathscr{N}_i=0$, $\forall i$ (see explanation to \eqref{210124:1711}).
Hence, when applying (2.12) to~\eqref{210124:1630}, it correctly reduces to
\begin{equation}
0\,=\,[a_1-a_4]\cdot 0
\,+\,\Big(a_2u_3+\dot{f}_1-g_1\Big)\cdot 0
\,+\,\Big(-a_2[u_1+\bar{u}_1]+\dot{f}_2\Big)\cdot 0,
\end{equation}
which then just turns into the identity relation $0=0$ from which no information can be extracted. In other words, when applying (2.12) to (3.16), as done in \cite{Oberlack01}, it just turns into a null-identity
and {\it not} to the additional constraint equations (3.17).

The restrictions (3.17) simply do not exist, and therefore no symmetry breaking is triggered as falsely claimed by the author from (3.17) onwards. The final result (3.19) in \cite{Oberlack01} is thus not only wrong but also seriously misleading since it gives the misleading impression that the Lie-group symmetry method in turbulence is able to \underline{analytically} circumvent the closure problem of turbulence, because~(3.19) shows that in the symmetry finding process no arbitrary space-dependent functions for the mean-flow symmetry generator $\eta_{\bar{u}_1}$ were picked up, but which in reality is not true!

The correct result for (3.19) is given on the right-hand side of Table~1 on p.$\,$7 in \cite{Frewer14.2}. Important to note is the difference in the result for $\eta_{\bar{u}_1}$, where the correct result on the right-hand side includes an arbitrary function in the shear-normal direction $x_2$, representing thus a non-closed result which ultimately just states that the closure problem in turbulence cannot be bypassed through only using the Lie-group symmetry method.

\label{p33}
\subsection{Further points for correction in \cite{Oberlack01}\label{SecF4}}

Although the two errors pointed out in Secs.$\,$\ref{SecF2} and \ref{SecF3} are central errors, in that they change the overall conclusion of the paper once corrected, these are not the only errors that can be found in \cite{Oberlack01}. There are three more technical errors, two major ones (Pt.~1 and 2 below) and one minor one (Pt.~3), all being independent of each other and which all need to be corrected.

\vspace{1em}\noindent
\hypertarget{SecF4P1}{{\bf 1.}} The first technical mistake is in (3.12a-c) and relates to the Lie-group symmetry method itself. In our comment \cite{Frewer14.2} we claimed this to be the key mistake in \cite{Oberlack01} as it lies at the heart of that study. For the investigation here, however, the strategy was changed in that (3.16) and (3.17) are declared as the key mistakes, in order to make this investigative review as simple and easy as possible, since no knowledge in Lie-group theory is required to understand the issues raised by the cases in Secs.$\,$\ref{SecF2} and \ref{SecF3}. But the origin of those two errors is already hidden in~(3.12a-c). Because instead of
\begin{equation}
\left.
\begin{aligned}
X\mathscr{C}\vert_{\mathscr{C}=0} &=0,\\[0.5em]
X\mathscr{N}_i\vert_{\mathscr{N}_i=0} &=0,\\[0.5em]
X(\mathscr{N}_iu_j+\mathscr{N}_ju_i)\vert_{(\mathscr{N}_iu_j+\mathscr{N}_ju_i)=0} &=0,
\end{aligned}
~~~\right\}
\label{140618:0007}
\end{equation}
the correct determining system of equations, in order to perform a correct and in particular a complete Lie-group symmetry analysis, is {\it not} given by \eqref{140618:0007}$\equiv$(3.12a-c), but by
\begin{equation}
X\mathscr{T}_{n}\vert_{\mathscr{T}_{n}=0}=0,\;\; \text{with}\;\;
\mathscr{T}_{n}:=(\mathscr{C},\mathscr{N}_i,\mathscr{N}_iu_j+\mathscr{N}_ju_i),
\label{131228:1947}
\end{equation}
with the key difference to be noted in the evaluation constraint, where $\mathscr{C}=0$, $\mathscr{N}_i=0$, and as well as $\mathscr{N}_iu_j+\mathscr{N}_ju_i=0$ have to be evaluated simultaneously and {\it not} separately as in \eqref{140618:0007}. The reason is that these three system of equations form a strongly coupled system in which each cannot be solved\pagebreak[4] on its own. A correctly applied Lie-group symmetry analysis then tells us that in order to get the complete and full set of symmetries for such a coupled system, one has to evaluate the $X$-derivative of this system by all available constraints simultaneously. Hence, the symmetry ansatz \eqref{140618:0007}, and thus (3.12a-c) in \cite{Oberlack01}, is a misapplication of the Lie-group symmetry method.

At first sight it is surprising that ansatz \eqref{140618:0007} leads to a more restrictive set of symmetries than~\eqref{131228:1947}. One would think that since \eqref{140618:0007} contains fewer evaluating constraints than
\eqref{131228:1947}, that the former will lead to a more general solution than the latter. But it is just the other way around. The reason is that $X\mathscr{T}_{n}=0$ without any evaluations induces a highly overdetermined system which thus can only be made less restrictive the more equations get inserted into this system through additional evaluation~constraints, as only this reduces the number of equations to make the system less overdetermined and therefore less restrictive.

In other words, the symmetry condition $X\mathscr{T}_{n}=0$ without any evaluations is more restrictive than $X\mathscr{T}_{n}\vert_{\{E\}=0}=0$ when including a set of evaluations $\{E\}=0$, and the more elements this evaluation set has, the less restrictive the symmetry solution is. And this is exactly the reason why the analysis in~\cite{Oberlack01} leads to such a highly restricted result (3.19), which is falsely claimed as the result of a symmetry breaking, but which in reality is not true, since the Lie-group method was not correctly~applied.

As already explained before, the correct application of the Lie-group symmetry method leads to a far less restrictive symmetry result. For instance, for the mean-flow symmetry generator $\eta_{\bar{u}_1}$ it leads to
the inclusion of an arbitrary space-dependent function as shown on the right-hand side of Table~1~[p.$\,$7] in \cite{Frewer14.2}. This correct result just states that the Lie-group symmetry method alone, like any other analytical method, cannot bypass the closure problem of turbulence. A statement opposite as to what is falsely claimed in \cite{Oberlack01}.

\vspace{1em}\noindent
\hypertarget{SecF4P2}{{\bf 2.}} As for the symmetry solution of the mean flow, also the solutions for all velocity correlations are less restrictive when performing the symmetry analysis correctly. Under the constraint (B8), the symmetry solution (B7) for the two-point correlations in \cite{Oberlack01} is not complete. The correct solution is given by (A.8) on p.$\,$13 in \cite{Frewer14.2}, which now involves arbitrary functions in both $x_2$ and the correlation distance $\vr=(r_1,r_2,r_3)$.

Hence for the correlations too, a Lie-group symmetry analysis does not provide any clue how they should scale, since through their arbitrary functions any desirable scaling law can be generated. Once~more this result just shows again that without modelling to numerical or experimental data, the closure problem cannot be bypassed, irrespective of whether we consider the mean flow or any correlation.

\vspace{1em}\noindent
\hypertarget{SecF4P3}{{\bf 3.}} The results (B7) and (B8) as they stand are only true for the non-rotating case ($\Omega_k=0$).\linebreak[4] For the rotating case ($\Omega_k\neq 0$) additional terms arise. Also, the symmetry analysis gets more involved since the correlation equations do {\it not} decouple anymore, as falsely claimed by saying that in order to obtain the results (B7) and (B8) \emph{``only the equations for $R_{22}$ in (B1), $\overline{pu_2}$ in (B3) and $\overline{u_2p}$~in~(B4) need to be examined, because these equations decouple from the other components in the tensor equations''} [p.$\,$325]. This statement is only true for the simple non-rotating case.

\label{p34}
\subsection{List of all false statements made in \cite{Oberlack01}\label{SecF5}}

{\bf 1a.} \textit{``the purpose of (2.12) [the velocity product equations] regarding the symmetry properties of plane shear flows is quite different"} [p.$\,$302]

\vspace{1em}\noindent
{\bf 1b.} \textit{``(2.12) is crucial to find self-similar mean velocity profiles consistent with the second moment and all higher-order correlation equations"} [p.$\,$307]

\vspace{1em}\noindent
{\bf 1c.} \textit{``the major difference between the classical turbulence modelling approach and the present procedure is the treatment of equation (2.12)"} [p.$\,$309]

\vspace{1em}\noindent
All these statements regarding the constraint equation (2.12) are not true, because, as was shown Secs.$\,$\ref{SecF2} and \ref{SecF3}, the equation (2.12) is fully equivalent or redundant to $\mathscr{N}_i=0$,
the fluctuating\linebreak[4]

\newgeometry{left=2.0cm,right=2.0cm,top=2.0cm,bottom=1.45cm,headsep=1em}

\noindent
Navier-Stokes equations themselves. Thus equation (2.12) has no impact on the symmetry properties of the flow. In particular the crucial symmetry-breaking property which is attributed to (2.12) on~p.$\,$308 does not exist.

\vspace{1em}\noindent
{\bf 2a.} \textit{``In the case of the logarithmic law of the wall, the scaling with the distance from the wall arises as a \underline{result} of the analysis and has \underline{not been assumed} in the derivation."} [p.$\,$299]

\vspace{1em}\noindent
{\bf 2b.} \textit{``important to note that group theoretical arguments very much guide the finding ... where the mean velocity profiles are applicable."} [p.$\,$306]

\vspace{1em}\noindent
{\bf 2c.} \emph{``This is an \underline{assumption} in the classical derivation [von-K\'arm\'an-derivation] of the log law of the~wall \underline{but is a result} of the present analysis."} [p.$\,$312]

\vspace{1em}\noindent
{\bf 2d.} \emph{``The theory is fully algorithmic and \underline{no intuition} is needed to find a self-similar mean velocity profile."} [p.$\,$321]

\vspace{1em}\noindent
All these statements regarding the use of the Lie-group symmetry method in turbulence and its generated solutions are wrong and seriously misleading, because they all claim that this method can analytically generate first-principle solutions for the scaling problem of turbulence without modelling or making any assumptions, and therefore implying that the Lie-group method is able to analytically circumvent the closure problem of turbulence. But this is not true.

First of all, the above statements in \cite{Oberlack01} are the result of the fact that the Lie-group method itself has been misapplied by the author. Because when correctly applying this method, the exact opposite conclusions are found. Arbitrary space-dependent functions in the symmetry solutions arise, which allow to generate any desirable scaling law for the statistically stationary flow configurations being considered. Without prior modelling or assumptions in the analysis, the Lie-group method alone does not give any information as how turbulence should scale.

A key aspect to be recognized here is that since the statistical equations of turbulence are unclosed, so is their admitted set of Lie symmetries. Unclosed equations, as those considered in \cite{Oberlack01},
inevitably lead to infinite dimensional and therefore unclosed Lie-algebras, which means that any invariant transformation can be generated, and thus also any desirable scaling law. Ultimately one has an infinite set of invariant possibilities to choose from when performing a full and correct Lie-group symmetry analysis for unclosed equations --- see e.g. \cite{Frewer16.3}, or \cite{Frewer18.2}, for further case examples in turbulence.

For example, if we consider for the mean velocity field the \emph{``log-law condition''} as described in Sec.$\,$3.3.2 on p.$\,$312 in \cite{Oberlack01}, then the correct corresponding scaling law which will be obtained from a correctly performed Lie-group symmetry analysis, is not given by (3.29), but by (see~(2.23) in \cite{Frewer14.2} for its derivation)
\vspace{-0.5em}
\begin{equation}
\bar{u}_1=\int \frac{F(x_2)}{a_1x_2+a_3}dx_2+C,
\label{210125:1252}
\end{equation}
where $F$ is a completely arbitrary function in the integration variable $x_2$. The log-law, for instance, is then obtained by the \underline{assumption} that $F$ should be a globally constant function in $x_2$. But also any other choice for $F$ can made, with the result that {\it any} desirable scaling law for the mean flow can be generated. The Lie-group analysis itself does not tell us how this function $F$ should be chosen. The correct result \eqref{210125:1252} is just an alternative yet more complicated representation for the unknown mean velocity function $\bar{u}_1$. One thus gained nothing in using the method of Lie-groups, one just shifted the problem from one unknown function $\bar{u}_1$ to another unknown function $F$, explicitly showing that the closure problem of turbulence thus cannot be bypassed when using this method.

Hence, the Lie-group symmetry method in turbulence is {\it not} free of assumptions. It is an {\it ad~hoc}\linebreak[4] method too, not in the same but in a similar way as the classical self-similarity method as first used by von K\'arm\'an and Prandtl a century ago: Instead of using an {\it a priori} set of scales, the Lie-group method has to make use of an {\it a priori} set of symmetries, namely to select the correct relevant symmetries from an infinite (unclosed) set. Even when including all unclosed higher order correlation equations, one still gets an infinite and therefore unclosed set of functionally independent invariances if the analysis is properly performed, and not only those few as always reported by M. Oberlack up to this day. In~other words, the Lie-group method in turbulence is effectively no different to the classical invariance method of von K\'arm\'an and Prandtl.

\restoregeometry
\nocite{apsrev42Control}
\bibliographystyle{apsrev4-2}
\bibliography{References}

\end{document}